\documentclass[]{aa} % for two-column version
\pdfoutput=1 
\usepackage{times}
\usepackage{graphicx}
\usepackage{hyperref}
\usepackage{multirow}
\usepackage{float}	
\usepackage{subfigure}
\usepackage{gensymb} % for \circ
\usepackage{natbib}
\usepackage{float}
\usepackage[T1]{fontenc} % if needed
\newcommand{\ba}{\begin{eqnarray}}
\newcommand{\ea}{\end{eqnarray}}
\newcommand{\be}{\begin{equation}}
\newcommand{\ee}{\end{equation}}
\newcommand{\dd}{{\rm d}}
\usepackage{array}
\newcolumntype{P}[1]{>{\centering\arraybackslash}p{#1}}
\newcolumntype{M}[1]{>{\centering\arraybackslash}m{#1}}
\usepackage{xcolor}

\begin{document}
\title{Cosmic-ray propagation around the Sun: investigating \\
  the influence of the solar magnetic field on the cosmic-ray \\ Sun shadow}
\titlerunning{Cosmic-ray propagation around the Sun}

\author{J. Becker Tjus\inst{1} \and P. Desiati\inst{2} \and N. D\"opper\inst{1}
  \and H. Fichtner\inst{1} \and J. Kleimann\inst{1} \and M. Kroll\inst{1} \and F. Tenholt\inst{1}}

\institute{
  Theoretische Physik IV, Ruhr-Universit\"at Bochum, 44780 Bochum, Germany \\
  \email{frederik.tenholt@rub.de}
  \and
  Wisconsin IceCube Particle Astrophysics Center, Madison, WI 53703, USA
}

\abstract{
  %% State of the art:
  The cosmic-ray Sun shadow, which is caused by high-energy charged cosmic rays being blocked and deflected by the Sun and its magnetic field, has been observed by various experiments, such as Argo-YBJ, HAWC, Tibet, and IceCube.
  Most notably, the shadow's size and depth was recently shown to correlate with the 11-year solar cycle. 
  The interpretation of such measurements, which help to bridge the gap between solar physics and high-energy particle astrophysics, requires a solid theoretical understanding of cosmic-ray propagation in the coronal magnetic field.
  %% Aim:
  It is the aim of this paper to establish theoretical predictions for the cosmic-ray Sun shadow in order to identify observables that can be used to study this link in more detail.
  %% Methods:
  To determine the cosmic-ray Sun shadow, we numerically compute trajectories of charged cosmic rays in the energy range of 5--316~TeV for five different mass numbers. We present and analyze the resulting shadow images for protons and iron, as well as for typically measured cosmic-ray compositions.
  %% Results:
  We confirm the observationally established correlation between the magnitude of the shadowing effect and both the mean sunspot number and the polarity of the magnetic field during the solar cycle.
  We also show that during low solar activity, the Sun's shadow behaves similarly to that of a dipole, for which we find a non-monotonous dependence on energy. 
  In particular, the shadow can become significantly more pronounced than the geometrical disk expected for a totally unmagnetized Sun.
  For times of high solar activity, we instead predict the shadow to depend monotonously on energy, and to be generally weaker than the geometrical shadow for all tested energies.
  These effects should become visible in energy-resolved measurements of the Sun shadow, and may in the future become an independent measure for the level of disorder in the solar magnetic field.
}
\maketitle

\section{Introduction}
\label{sec:intro} 
The blocking of cosmic rays by the Sun, as first proposed in \cite{clark}, is an interesting subject in the investigation of cosmic-ray properties.
In particular, it has been predicted that interactions of cosmic rays with the Sun would result in gamma- and neutrino-production that would eventually be detectable \citep[see][]{1991ApJ...382..652S}. 
Even cosmic-ray electrons were predicted to produce a halo component in gamma-rays. 
Today, gamma-ray signatures can indeed be measured: the halo component has been verified at an intensity level as expected by the prediction in \citet{1991ApJ...382..652S}; also see \citet{Orlando:2008uk} and \citet{2011ApJ...734..116A}.
The disk component, however, was found to be an order of magnitude higher than expected \citep{Ng:2015gya}, a discrepancy that could still not be resolved in the literature.
The neutrino signal that accompanies the disk component has been recalculated in the past few years \citep{2017JCAP...07..024A, Ng+2017_PhRvD, 2017JCAP...06..033E} as it is now an important background for dark-matter searches with neutrino telescopes from the direction of the Sun.
\citet{whitepaper2020} also provides a general discussion of the possibilities of using the Sun as a laboratory for astroparticle physics.

The proper modeling of the cosmic-ray component itself only became important in recent years when ground-based detectors started detecting the shadow of the Sun in cosmic rays produced by their interactions with the solar surface and corona. 
The Tibet AS-Gamma experiment was the first to demonstrate that the cosmic-ray Sun shadow varies with time, in correlation with the variation of the sunspot number \citep{2013PhRvL.111a1101A}. 
As the latter is, in turn, correlated with the magnetic field strength and structure, these measurements were first to prove an effect of cosmic-ray deflection in the inner heliospheric field on the Sun shadow.
Most recently, a similar study at higher energies $\left<E_{\rm CR}\right>\sim 40$~TeV has been performed with the IceCube detector \citep{ICECUBE_SUNSHADOW_PAPER} and a general temporal variation of the cosmic-ray Sun shadow was verified. 
A correlation with the sunspot number, however, could not be validated yet because the solar cycle during the time of the measurements (2010/2011--2014/2015) was not variable enough to allow for a statistically significant statement \citep{Bos2017}. 
The HAWC collaboration \citep{Enriquez:2015nva, hawc_aps2017} also observed a deficit of cosmic rays from the direction of the Sun for the years 2013 and 2014 between $1$~TeV and $142$~TeV. 
A study of the variability of the Sun shadow could not yet, however, be presented. 
In the coming years, different experiments are expected to be able to probe the variation of the cosmic-ray Sun shadow with the solar magnetic field across a broad energy range and with a better time resolution. 

It is the aim of this paper to investigate which effects are present and how the cosmic-ray shadow varies with time in the $\sim 5$\,TeV to $\sim 316$\,TeV energy range based on numerical calculations of particle propagation.
Our studies are specifically aimed at a quantification of the energy behavior of the Sun shadow. 
While the Tibet, IceCube, and HAWC measurements present results at different energies, they are not comparable at this point for two central reasons: (1) the Sun shadow is presented on the event level and therefore, the individual measures for the shadow cannot be compared directly; (2) measurements are not necessarily performed during the same year and solar cycle. 
Thus, this study is aimed at carrying out the first theoretical investigation of the energy behavior in order to make a prediction on whether and how observatories could study cosmic-ray propagation in even greater detail. 
Given the high statistics level reached with Tibet, IceCube, and HAWC, this should certainly be possible in the future.

The propagation of cosmic rays can generally be modeled by using either an ensemble- or a single-particle approach. 
The many-particle approach is typically applied in diffusive regimes in which the gyro-radius $r_{\mathrm{g}}$ of the particles is small compared to the system size $L$, such that a high number of interactions with the turbulent magnetic field for particle trajectories in the considered volume is expected \citep[see, e.g.,][]{Schlickeiser2002}.
For Galactic cosmic-ray propagation, with a magnetic field strength of $\sim \mu \mathrm{G}$ and assuming a particle energy of $10\,\mathrm{TeV}$, the resulting gyroradius of $r_{\mathrm{g}} \approx 10^{-4}\,\mathrm{pc}$ is, for example, much smaller than the diameter of the Milky Way of roughly $30\,\mathrm{kpc}$.
As a consequence, particles will complete a high number of gyrations and a diffusive description is appropriate.
Over recent decades, sophisticated simulation schemes have been developed for Galactic cosmic-ray propagation.
In particular, these efforts have resulted in the propagation tool GALPROP \citet[see, e.g.,][]{0004-637X-509-1-212, 2041-8205-722-1-L58, 2011CoPhC.182.1156V} and newer propagation software, such as DRAGON \citep{1475-7516-2008-10-018}, which is capable to address questions of anisotropy, along with PICARD \citep{2014APh....55...37K}, and CRPropa 3.1 \citep{2017JCAP...06..046M}. 
For the problem studied in this paper, due to the resulting small number of gyrations within the solar corona, a diffusive description of the propagation of $>10\,\mathrm{TeV}$ cosmic rays around the Sun would not be appropriate:
the diffusive regime has not been reached in that case and a different approach needs to be applied.
In this context, the ballistic (single-particle) approach, in which the equation of motion is solved explicitly for each particle, is suitable.
This approach works well for cases in which the gyro-radii are comparable to or larger than the system size, allowing for the computation of the actual trajectories of TeV cosmic rays.
A well-structured and comprehensive state-of-the-art tool originally designed for extragalactic propagation is CRPropa \citep{2013arXiv1307.2643A}.
As we applied different scales in our simulations, we developed a dedicated code for the simulations.
In the solar environment, the gyro-radius of $r_{\mathrm{g}} \approx 0.5\,R_{\odot}$ ($E=10\,\mathrm{TeV}$, $B=1\,\mathrm{G}$) is on the same order of magnitude as the extent of the studied region of approximately $1\,R_{\odot}$ to $10\,R_{\odot}$.
For the purpose of efficiency, a back-tracing scheme has been set up in which anti-particles are propagated toward the Sun.

This paper is organized as follows:
In Sect.~\ref{ssec:smf}, some general properties and the modeling of the solar magnetic field are discussed.
Afterwards, we describe the details of the numerical propagation, the weighting scheme, and the final analysis quantifying the strength of the Sun shadow (Sect.~\ref{sec:description}).
The results obtained are then presented in Sect.~\ref{sec:results}, followed by a summary (Sect.~\ref{sec:conclusions}) and an outlook (Sect.~\ref{sec:outlook}).

\section{The solar magnetic field}
\label{ssec:smf}
In this section, we summarize some general properties of the solar magnetic field (see Sect.~\ref{ssec:general}), followed by the details of the implementation used in this work (see Sect.~\ref{ssec:implement}).

\begin{figure}[t]
  \centering
  \includegraphics[width=\linewidth]{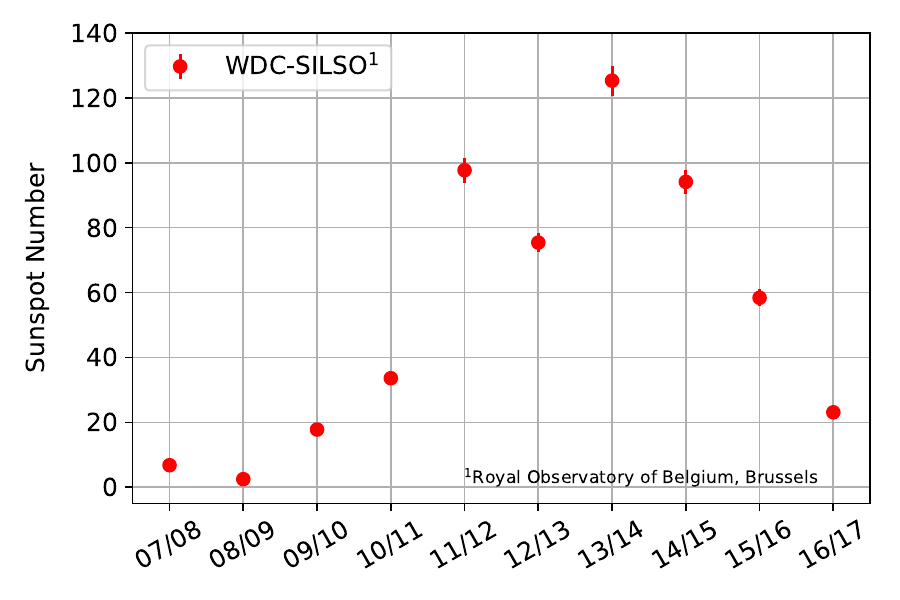}
  \caption{Sunspot numbers taken from \cite{sidc} and averaged over the four months of each season are shown for a ten-year time period.
    The seasons 2007-08 and 2008-09 report a low sunspot number, while in seasons 2011-12, 2012-13, 2013-14, and 2014-15 a higher number is detected, before decreasing again afterwards.
    This approximately describes the solar activity cycle of 11~years \citep[see, e.g.,][]{2006RPPh...69..563S}.
    \label{fig:SunSpots}}
\end{figure}

\subsection{General properties}
\label{ssec:general}
The Sun goes through a cycle of approximately 11~years during which the solar magnetic field varies in strength and structure.
At the beginning of each cycle, the magnetic field strength is minimal and the magnetic field near the solar surface can approximately be described by a dipole field which, with increasing radial distance, gets deformed into a radial field by the solar wind. 
After five to six years, the magnetic field strength reaches its maximum and the dipolar structure disappears. 
During this period of time, the structure is highly irregular and also changes on the timescale of solar rotation \citep[e.g.][]{Priest1982}.
After a total of approximately 11~years, the structure returns to a dipole structure with reversed polarity. 
Measurements show that maxima and minima of solar magnetic activity correlate with maxima and minima of the sunspot number \citep[see, e.g.,][]{1998SoPh..179..189H, 2006RPPh...69..563S}.
Figure~\ref{fig:SunSpots} shows the sunspot number from  2007 through 2017 over the analyzed time frames. 
Since our aim is to qualitatively compare our results to the IceCube measurements reported by \cite{ICECUBE_SUNSHADOW_PAPER}, we define a season as the period consisting of the months November through February of each Antarctic summer.
The magnetic field strength at the solar surface is measured using observatories on Earth or in space. 
Such measurements are called magnetograms and are produced for each Carrington rotation (CR).

\subsection{Implementation}
\label{ssec:implement}
\begin{figure*}[th]
  \centering
  \includegraphics[width=\linewidth]{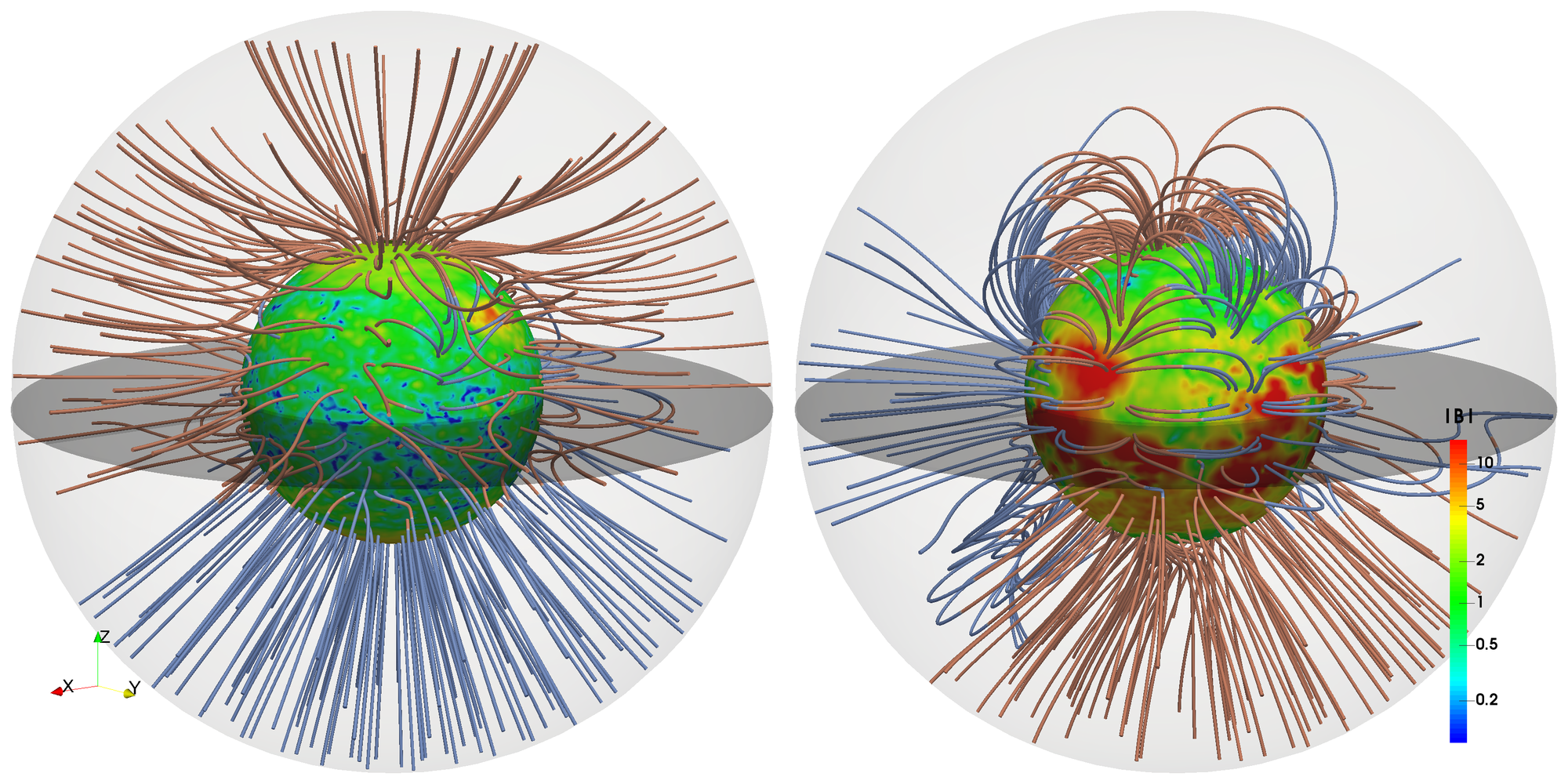}
  \caption{Magnetic field strength on the solar surface and selected magnetic field lines for Carrington rotations 2077 (December 2008 with low solar activity, \textit{left}) and 2158 (December 2014 close to the solar activity maximum, \textit{right}), based on GONG magnetogram data and the PFSS model.
    Field lines are shown between the solar surface and the source surface at 2.5 solar radii, with blue indicating outgoing ($B_r > 0$), brown indicating incoming ($B_r < 0$) polarity.
    Magnetic field strength on the solar surface is shown using a rainbow color scale.
    It can clearly be seen that the dipolar structure which typically prevails during solar minimum turns into a more irregular field structure towards solar maximum.
  }
  \label{fig:bfield}
\end{figure*}

The Global Oscillation Network Group \citep[GONG,][]{1996Sci...272.1284H} provides integral synoptic magnetograms, which we used in order to extrapolate to a full three-dimensional magnetic field using different models.
Here we use the Potential Field Source Surface (PFSS) model \citep{1969SoPh....6..442S, 1969SoPh....9..131A}.

The magnetic field is computed using the FDIPS software \citep{2016ascl.soft06011T}, which allows the user to define parameters, such as the grid resolution at which the field will be computed or the so-called source surface radius $R_{\mathrm{SS}}$ at which the magnetic field is assumed to be purely radial. 
More details can be found in such works as \cite{2014ApJ...788...80W}. 
This procedure results in field components on a spherical grid equidistant in the radius $r$, longitude $\varphi$, and the sine of the co-latitude $\vartheta$. 
Here, we choose $R_{\mathrm{SS}} = 2.5 R_{\odot}$. 
A typical result for the coronal magnetic field between the solar surface and source surface can be seen in Fig.~\ref{fig:bfield}.
The magnetic field data used for Fig.~\ref{fig:bfield} had already been employed in \citet{2014ApJ...788...80W}, where also a different visualization of the magnetic field can be found (page 8, Fig.~6).
As Carrington rotation 2060 was comparably quiet in terms of solar activity, most field lines appear in an ordered, large-scale structure.
It is only close to the surface, where the magnetic field strength becomes significantly larger in some places, that more unordered, small-scale structures are visible.
For times of higher solar activity, as, for example, between 2011 and 2015, these unordered, small-scale structures become more relevant regarding cosmic-ray propagation through the solar magnetic field.
For $r > R_{\mathrm{SS}}$ the magnetic field is implemented as purely radial and with a magnitude decreasing proportional to the inverse square of the distance. 
This implementation omits angular components of the magnetic field as present in the Parker spiral model \citep{parker}. 
However, we restricted our study to a region close to the Sun in which these angular components do not contribute significantly to the deflection of the tracked particles (compare Sect.~\ref{ssec:computation}).
We also refrained from modeling the turbulent component of the heliospheric magnetic field since it is not relevant for the $>10\,\mathrm{TeV}$ particles studied in this work; it would only become important in the case of a diffusive behavior and as argued in the introduction, we are investigating scales that are comparable to the gyroradii of the particles.
Thus, propagation is assumed to be ballistic.
An alternative approach for extrapolating from magnetograms to a three-dimensional solar magnetic field is the Current Sheet Source Surface (CSSS) model \citep{csss}, which additionally uses a so-called cusp surface located at a cusp radius $R_{\mathrm{cp}}$.
Beyond this radius, all field lines are assumed to be open.
In both the PFSS and the CSSS model, field lines are assumed to be purely radial at $r>R_{\mathrm{SS}}$.

\section{Simulation of the Sun shadow}
\label{sec:description}
The basic setup of our simulation framework includes the solar magnetic field and anti-particles that are traced back in the vicinity of the Sun. 
The back-tracing ensures that only those particles that propagate in the direction of Earth after they have traversed the magnetic field are considered (those particles, in turn, are relevant for the cosmic-ray Sun shadow observed at Earth).

The Sun shadow is determined by defining a quadratic, planar injection window with a side length of four solar radii, located at a distance of five solar radii from the Sun's center and arranged such that it is parallel to the $yz$-plane of the Heliocentric Earth equatorial (HEEQ) coordinate system, meaning that it is parallel to the Sun's rotation axis and perpendicular to the intersection of the solar equator and the solar central meridian as seen from Earth \citep[compare][]{hapgood}.
For simplicity, we assumed the solar rotational axis to be perpendicular to the ecliptic, thus neglecting the Sun's obliquity of $7.25^{\circ}$ throughout this paper.

\begin{figure}[t]
  \begin{center}
    \includegraphics[width=\linewidth]{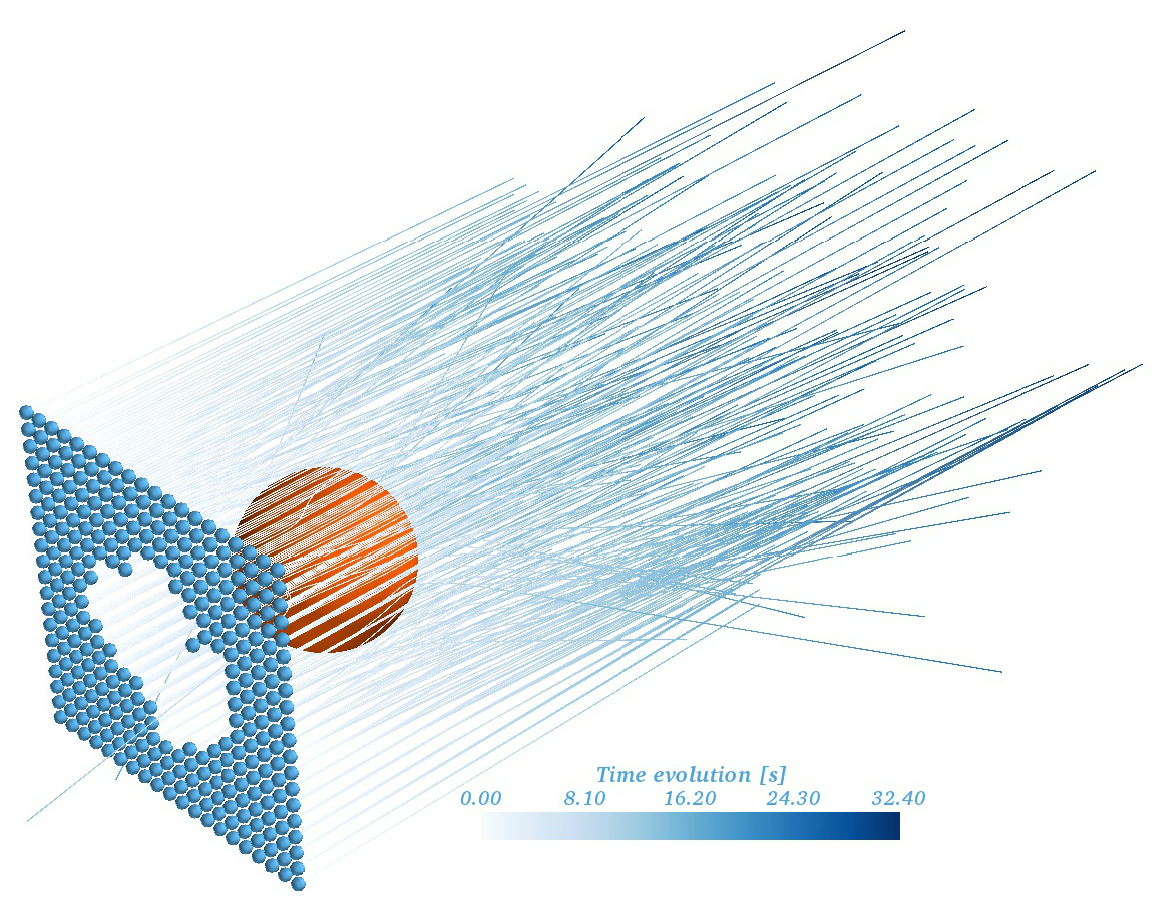}
  \end{center}
  \caption{Schematic view of the simulation setup (distances not to scale).
    Starting points of anti-particles as indicated by the blue bubbles are arranged in the injection window.
    Timing information for the trajectories of the anti-particles is given in color code.
    The starting points of those anti-particles hitting the Sun are removed from the plot.
    In this way, the cosmic-ray deficit caused by the Sun is simulated using back-tracing.}
  \label{fig:injection_plane}
\end{figure}

This injection window is uniformly filled with an equidistant, rectangular grid of $N=100^2 = 10^4$ starting points of anti-particles.
The initial momentum vector of a given anti-particle $i$ points along a straight line from a virtual point on the $x$-axis at $x=1$~AU to the anti-particle's starting point $\vec{x}_i(0)$, where $1 \leq i \leq N$. 
In other words, the point of divergence of the initial directions of all anti-particles is at the position of the Earth.
An example of the shadow picture of an anti-proton ($|Z|=1$) with an energy $E = 40\,\mathrm{TeV}$ for December 2014 (Carrington rotation CR2158) is shown in Fig.~\ref{fig:injection_plane}.

\subsection{Computation of particle trajectories}
\label{ssec:computation}
In general, the relativistic equation of motion for a particle of rest mass $m$ under the influence of an external force $\vec{F}$ is given by
\begin{align}
  \vec{F} = \frac{\dd}{\dd t} \left( \gamma(v) \, m \, \vec{v} \right) .
\end{align}
In this equation, $\vec{v}$ is the velocity and $\gamma$ is the Lorentz factor.
The motional electric field at radius $r$ is of order $r \Omega B$, with the solar angular rotational speed $\Omega$. 
Comparing this to the $(\vec{v} \times \vec{B}$) term due to the Lorentz force shows a ratio of $r \Omega/c$.
Since this ratio is on the order of $10^{-5}$, we assume that the particles passing through the solar magnetic field are only affected by the Lorentz force. 
And since the latter cannot change $\vec{v}$ in magnitude because it is perpendicular to it, the Lorentz factor is constant and thus unaffected by the time derivative.
With that, the equation of motion becomes
\begin{align}
  \frac{\dd \vec{v}(t)}{\dd t} = \frac{q}{\gamma m} \left(\vec{v}(t)\times\vec{B}\right)
  \quad \textrm{with} \quad
  \frac{\dd \vec{x}(t)}{\dd t} = \vec{v}(t) ,
  \label{eq:dgl_lorentz_force}
\end{align}
with $\vec{x}$ and $q$ denoting the location and the charge of the particle. 
The equation of motion \eqref{eq:dgl_lorentz_force} is solved numerically using the so~called Boris Push method \citep{boris}.
The main advantage of the Boris Push method as compared to other standard methods like the Runge-Kutta-based Cash-Karp method is the conservation of energy for the case that there are no electric fields (which is shown in \cite{why_is_boris_so_good}). 
Also its relatively short computing time compared to Runge-Kutta algorithms with comparable accuracy is beneficial.
Given the initial conditions $\vec{x}(t_0)$ and $\vec{v}(t_0)$, $\vec{v}(t_0+\Delta t)$ and $\vec{x}(t_0+\Delta t)$ can be determined as follows:
\begin{align}
  \vec{l} &= \frac{q\vec{B}\Delta t}{2\gamma m} , \\
  \vec{s} &=  \frac{2\vec{l}}{1+|\vec{l}|^2} , \\
  \vec{v}^{\prime} &= \vec{v}_{\perp}(t_{0}) + \vec{v}_{\perp}(t_{0}) \times \vec{l} , \\
  \vec{v}(t_{0} + \Delta t) &= \vec{v}(t_{0}) + \vec{v}^{\prime} \times\vec{s} , \\
  \label{eq:from_v_to_x}
  \vec{x}(t_{0} + \Delta t) &= \vec{x}(t_{0}) + \vec{v}(t_{0} + \Delta t) \; \Delta t
\end{align}
(and analogously for all following time steps), where $\vec{v}_{\perp}$ denotes the component of $\vec{v}$ perpendicular to $\vec{B}$.

\subsection{Simulation tests}
\label{sec:dipolar_test}
Various tests of increasing complexity have been conducted prior to the actual runs presented in the ensuing sections in order to confirm the proper operation and reliability of our code.
From those tests, we present the noteworthy example of a dipolar magnetic field.
This exercise is useful because its results will help us to understand and interpret the ones of the actual solar magnetic field, for which a dipole is often used as a crude approximation.

\begin{figure}[t]
  \begin{center}
    \includegraphics[width=\linewidth]{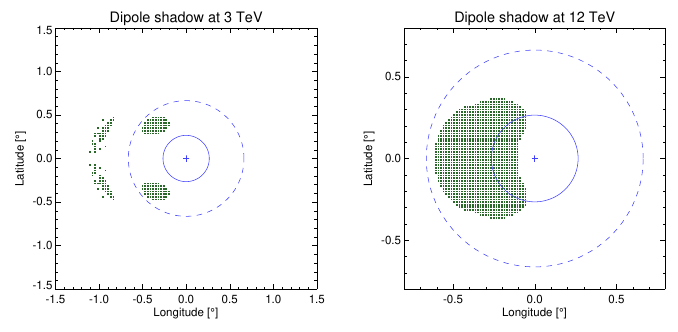}
  \end{center}
  \caption{Shadow maps for mono-energetic protons in the magnetic field of a dipole, using the example of \mbox{$E_\textrm{kin} = 3$\,TeV} (\textit{left}) and $E_\textrm{kin} = 12$\,TeV (\textit{right}), and \mbox{$100 \times 100$} particles in each case.
    Green squares represent the starting points of particle trajectories intersecting the solar surface.
    Shadow-to-disk ratio amounts to 0.787 for the left case and 1.516 on the right.
    To guide the eye, blue circles indicate projected radii of 1~$R_{\odot}$ (solid) and $R_{\rm SS} = 2.5 \ R_{\odot}$ (dashed).
    We note that for the right plot, the lateral half-length $d$ of the quadratic injection window is decreased from $d = 1.5^{\circ}$ to $d = 0.8^{\circ}$ for improved resolution.
  }
  \label{fig:greenshadow_2pack}
\end{figure}

Our test setup consists of an ensemble of mono-energetic protons (charge $+e$, mass $m_{\rm p}$) being deflected in the magnetic field
\begin{equation}
  \label{eq:dipole}
  \vec{B} = -\nabla \left(\frac{\vec{p} \cdot \vec{r}}{r^3} \right) =
  \frac{3 \, \vec{r} \, (\vec{p} \cdot \vec{r})}{r^5} - \frac{\vec{p}}{r^3}
\end{equation}
of a dipole with magnetic moment $\vec{p} = p \, \vec{e}_z$ and strength $p = B_{\rm dip} \, R_{\odot}^3$, such that the magnetic field strength at the equator ($x^2+y^2=R_{\odot}^2, z=0$) equals exactly \mbox{$B_{\rm dip} = 1$\,G}. 
Normalizing all lengths to $R_{\odot}$ and speeds to light speed $c$, we obtain %
\begin{equation}
  \label{eq:eom_norm}
  \frac{\dd \vec{\bar{v}}}{\dd {\bar{t}}} =
  \eta \ \vec{\bar{v}} \times \vec{\bar{B}}
  \quad \mbox{with} \quad
  \eta = \frac{Z \, e \, B_0 \, c \, R_{\odot}}{\gamma \, m_{\rm p} \, c^2}
  \approx 20.88 \ \frac{Z \, B_0 \, [\mbox{G}]}{E_{\rm kin} \, \mbox{[TeV]}}
\end{equation}
\newpage
\noindent
(in which dimensionless quantities are denoted by a bar and normalization constants with a lower index ``0'') as the dimensionless form of Eq.~(\ref{eq:dgl_lorentz_force}).
We may, thus, easily identify $\eta$, which is proportional to the normalization field strength divided by the particle's (unchanging) rigidity, as the only external parameter of relevance in this case.
For protons in a static field, it is sufficient to vary only the particle energy between different simulations.

\begin{figure}
  \begin{center}
    \includegraphics[width=\linewidth]{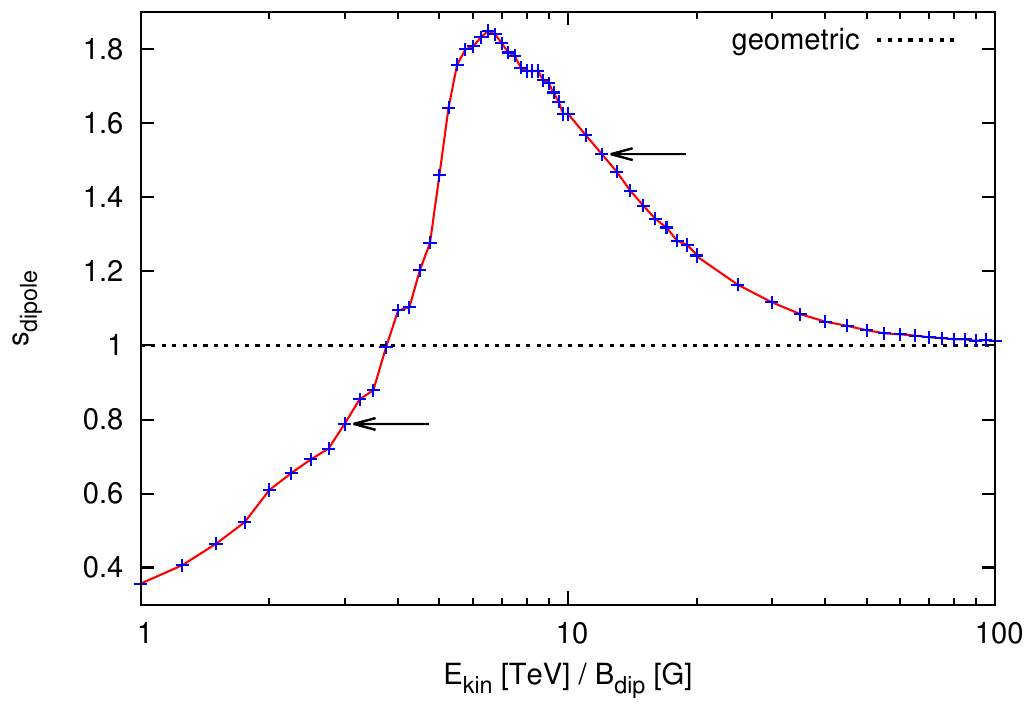}
  \end{center}
  \caption{
    Total shadow area $s$ as a function of particle energy for dipolar field (\ref{eq:dipole}), normalized to the area of the geometrical disk.
    Two arrows indicate the respective energy values of 3 and 12\,TeV, for which shadow images are shown in Fig.~\ref{fig:greenshadow_2pack}.
    The respective side length $2 d$ of the injection window was chosen as $3.0^{\circ}$ ($1.6^{\circ}$, $1.0^{\circ}$) for $E \in [1,10]$ ($[10,20]$, $[20,100]$)\,TeV to optimize spatial resolution, while at the same time keeping the field of view large enough not to lose any ``shadow particles.'' 
    \vspace*{-5mm}
  }
  \label{fig:plot_shadowsize}
\end{figure}

Figure~\ref{fig:greenshadow_2pack} shows shadow maps obtained in this way for two specific energies.
Somewhat surprisingly, we see that, as opposed to what might have been expected in view of previous observations showing the Sun's shadow diminishing with increased magnetic activity, the shadow's total area may, in fact, exceed that of the solar disk considerably at certain energies.
Figure~\ref{fig:plot_shadowsize} summarizes the relative shadow size $s$ as a function of the particle energy employed in each simulation.
Here the shadow size is determined as the total shadow area as seen in green color in Fig.~\ref{fig:greenshadow_2pack}, divided by the area of the solar disk when projected onto the injection window.
We see that as expected, this ratio approaches unity for very large energies (or very small field strengths) because trajectories are then almost straight lines tracing the geometrical, circular shape of the projected Sun.
Interestingly, we may also confirm that the shadow size does in fact increase toward smaller energies, reaching a maximum of about 185 percent of the geometrical disk area before declining again.
But it is only at comparatively low energies that this value decrease below unity as the shape of the shadow starts to break apart to ultimatively dissolve beyond recognition as it fragments into smaller and smaller ``islands.''

This simple example may serve to illustrate that at least for large-scale magnetic fields, the shadow may not necessarily become less pronounced with increasing energy, but may, in fact, get amplified considerably compared to the case of a non-magnetized shadow-casting obstacle.
We will come back to this phenomenon in Sect.~\ref{sec:results:rigidity}.

\subsection{Models of the primary cosmic-ray spectrum and composition}
\label{sec:crmodels}
For the energy spectrum and composition of the simulations, we used the model from \citet{HGm} with a mixed extragalactic component (hereinafter, HGm model) and from \citet[][hereinafter, GH model]{GaisserHonda}.

\subsection{Passing probability and weighting scheme}
\label{section:Passing Probability and Weighting Scheme}
In order to mimic the rotation of the solar magnetic field within a Carrington rotation, which has a synodic rotation period of $\sim 27.3$ days, a total of 36 calculations was performed, revolving the injection plane (instead of the solar magnetic field) in steps of $10^{\circ}$ around the rotation axis of the Sun.
To quantify how likely it is for a particle with energy $E$ and atomic number $Z$ coming from direction $i$ to pass the Sun without hitting its surface, we then defined the passing probability $p$ of a specific Carrington rotation CR as the number $n_{\mathrm{pass}}$ of angular steps for which the particle passes the Sun, divided by the total number $n_{\mathrm{tot}}=36$ of angular steps:
\begin{equation}
  p_{i, Z, E, \mathrm{CR}} = \frac{n_{\mathrm{pass}}}{n_{\mathrm{tot}}}.
  \label{eq:p_i_Z_E_CR}
\end{equation}
This is done for the nuclei of hydrogen, helium, nitrogen, aluminum, and iron with particle energies of $10^{1.0}$, $10^{1.6}$, and $10^{2.2}$~TeV (corresponding to $10$, $\sim 40$ and $\sim158$~TeV).
Particles with higher energies were omitted due to their lower particle flux. 
We run our computations for the Carrington rotations approximately over the time window between November and February of each season, since it is only between these months that the IceCube detector observes the cosmic-ray flux from a window around the Sun, considering the seasons 2007-08 through 2016-17.

The result for each set of atomic number and energy was weighted according to its abundance based on the HGm model (see Sect.~\ref{sec:crmodels}).  
Assuming that the calculated Sun shadow does not vary considerably within an energy range of $\pm0.3$ in $\log(E)$, we obtained the weighting factors $g_{Z, E}$ by integrating the energy spectra in an integration interval from $10^{x-0.3}$~TeV to $10^{x+0.3}$~TeV for all mentioned energies and atomic numbers.
Using these weighting factors, the weighted arithmetic mean of all sets of energy and atomic number was determined for each Carrington rotation (CR) according to 
\begin{equation}
  \bar{p}_{i, \mathrm{CR}} = \frac{\sum\limits_{Z, E}p_{i, Z, E, \mathrm{CR}}\cdot g_{Z, E}}{\sum\limits_{Z, E}g_{Z, E}} .
\end{equation}
In the last step, the mean of $\bar{p}_{i, \mathrm{CR}}$ for all Carrington rotations and the corresponding standard deviation were determined.

\subsection{A quantitative measure for the shadow depth}
\label{sec:numerical}
In order to quantify the calculated shadowing effect of the Sun for a particular Carrington rotation, we compared the computed shadow size with that of an unmagnetized object with the Sun's size and shape.
Since each particle would hit the solar surface with average probability $1-\bar{p}_{i, \mathrm{CR}}$, we expect a total of $N-\sum_{i} \bar{p}_{i, \mathrm{CR}}$ particles to contribute to the shadow.
With $R_{\mathrm{S}} \equiv \arctan[R_{\odot}/(\mbox{1 AU})]$ $\approx 0{.}27^\circ$ as the angular radius of the Sun as seen from Earth, the angular area (solid angle) covered by the solar disk amounts to $\pi R_{\mathrm{S}}^2$. 
Since the injection area of size $\omega \equiv (4 R_{\mathrm{S}})^2$ contains $N$ evenly distributed starting positions, $(\pi R_{\mathrm{S}}^2/\omega) N = (\pi/16) N$ shadowed events are expected from an unmagnetized object with the Sun's size and shape. 
The normalized net shadowing effect for one Carrington rotation can therefore be quantified using the ratio
\begin{equation}
  s_{\mathrm{CR}} = \frac{N-\sum\limits_{i} \bar{p}_{i, \mathrm{CR}}} {(\pi R_{\mathrm{S}}^2/\omega) N}
  = \frac{16}{\pi} \bigg( 1- \frac{1}{N} 
  \sum\limits_{i} \bar{p}_{i, \mathrm{CR}}
  \bigg) .
  \label{verhaeltnis}
\end{equation}
In particular, a ratio of $s_{\mathrm{CR}}=1$ would mean that the shadow's size is equivalent to a geometrical shadowing by an unmagnetized Sun.
As the last step, the mean of those ratios
\begin{equation}
  s = \frac{1}{n_{\mathrm{CR}}} \sum_{\mathrm{CR}} s_{\mathrm{CR}}
  \label{verhaeltnis2}
\end{equation}
for the Carrington rotations considered was calculated.
The uncertainty of $s$ was calculated as the standard deviation of the ratios of the different Carrington rotations. 

\begin{figure}[t]
  \centering
  \includegraphics[trim = 36mm 4mm 4mm 3mm, clip, width=0.8\linewidth]{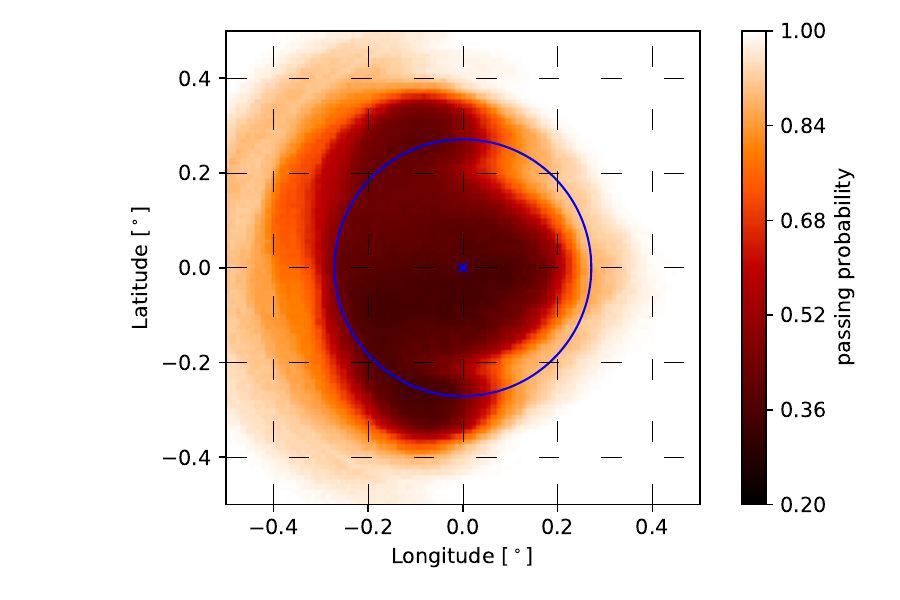} 
  \centering
  \includegraphics[trim = 36mm 4mm 4mm 3mm, clip, width=0.8\linewidth]{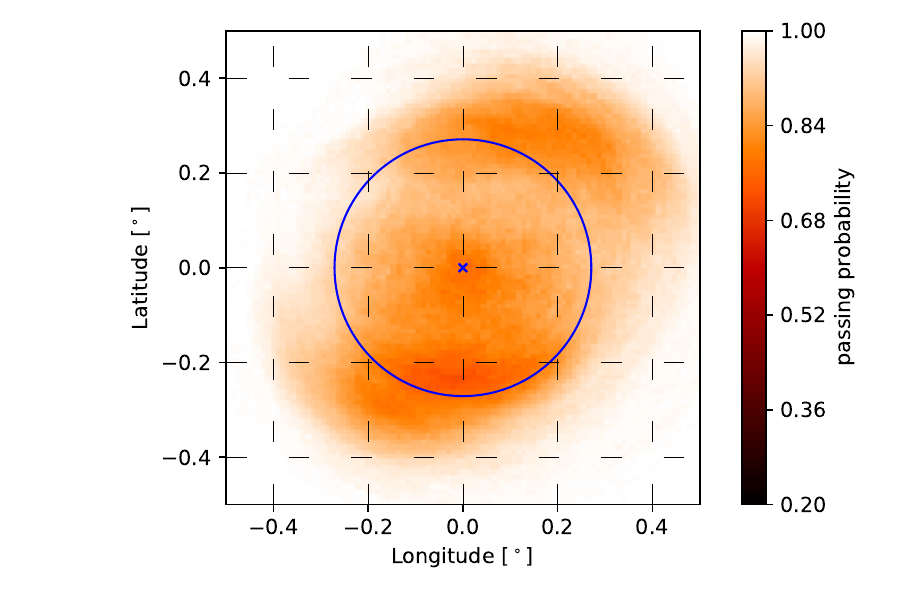}
  \caption{Examples of visualizing the calculated Sun shadow, where 2008-09 (\textit{top}; year of least solar activity) and 2014-15 (\textit{bottom}; year of most solar activity) seasons are shown, using the energy spectrum and composition according to the GH model.}
  \label{fig:shadow_example}
\end{figure}

\section{Results}
\label{sec:results}
The passing probabilities (compare Sect.~\ref{section:Passing Probability and Weighting Scheme}) are visualized on a two-dimensional grid in longitude and latitude.
The passing probability for each bin is represented by a color, with darker colors representing smaller and lighter colors representing larger passing probabilities. 
A passing probability of 1.0 in a specific bin means that every particle from that bin passed the Sun unimpeded, while a passing probability of 0.0 means that every particle hit the Sun.
Two examples for resulting shadow plots can be seen in Fig.~\ref{fig:shadow_example}. 
For the purposes of readability, shadow plots related to the following sections are presented in Appendix~\ref{sec:appendix:shadow}.
For numerical studies of the shadow, the shadow size ratio $s$ as described in Sect.~\ref{sec:numerical} was used.

\subsection{Sun shadow as a function of mass number}
\label{sec:results:mass}
\begin{figure*}[htbp]
  \begin{center}
    \includegraphics[width=0.95\linewidth]{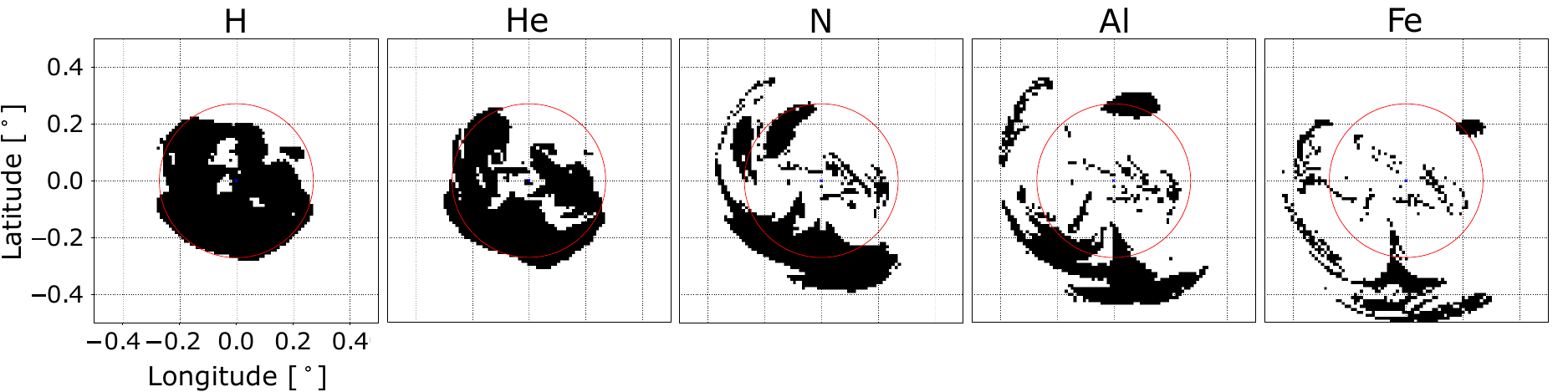}
  \end{center}
  \centering
  \caption{Calculated shadow at $E=40\,\mathrm{TeV}$ for the different nuclei simulated.}
  \label{fig:Dez15_40TeV_alle_teilchensorten}
\end{figure*}

\begin{figure}[htbp]
  \begin{center}
    \includegraphics[width=0.95\linewidth]{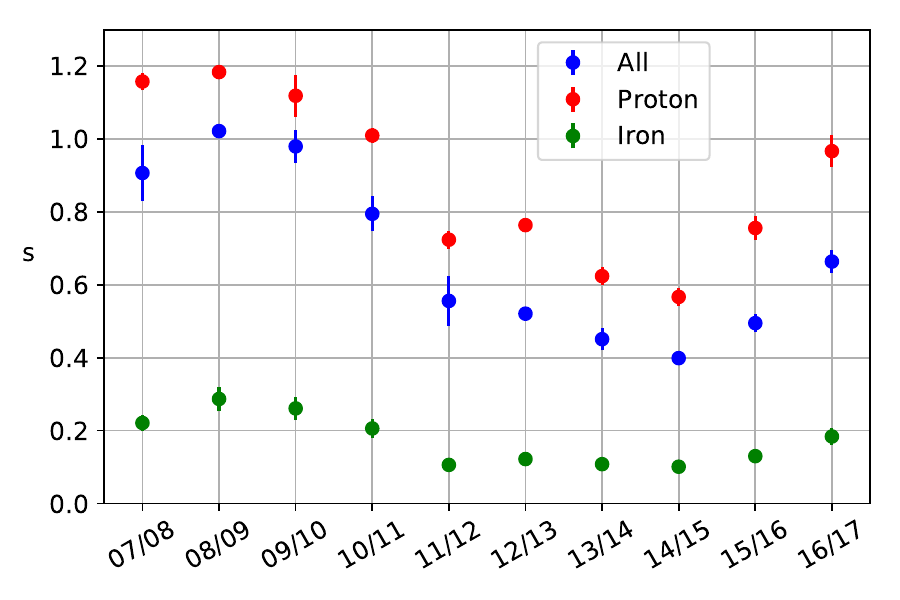}
  \end{center}
  \caption{Calculated Sun shadow size ratio in the years from 2007 through 2017 for pure proton, pure iron, and a mixed composition according to the HGm model.}
  \label{fig:s_all_proton_iron}
\end{figure}

In Fig.~\ref{fig:Dez15_40TeV_alle_teilchensorten}, the shadow for December 2015 can be seen at \mbox{$E=40$\,TeV} and for all different nuclei that were simulated.
As expected, the deflection of particles increases with increasing atomic number (decreasing rigidity).
In Fig.~\ref{fig:s_all_proton_iron} the ratio $s$ can be seen for pure proton, pure iron, and a mixed composition according to the HGm model for the years from 2007 through 2017.
Some obvious features can be seen from the figures:
First of all, as expected, the hydrogen shadow is much more focused and less fragmented as compared to the heavier components. 
Fragmentation and spread is increasing with increasing charge $Z$. 
The full shadow plots, displayed in the appendix in Figs.~\ref{fig:sunshadow_2007_to_2017_proton_only} and \ref{fig:sunshadow_2007_to_2017_iron_only}, also reveal that the measure for the shadow, $s$, is globally smaller for iron than for proton. 
In addition, the differences between the seasons are less pronounced for iron than for protons. 
The picture in the real world is expected to be proton-dominated up to PeV-energies, as the cosmic-ray spectrum itself is dominated by hydrogen \citep{wiebel_sooth1998}.
A detailed investigation of the ``true'' Sun shadow is presented in the following subsections.

\subsection{Sun shadow as a function of energy}
\label{sec:results:energy}
\begin{figure}[htbp]
  \begin{center}
    \includegraphics[width=0.95\linewidth]{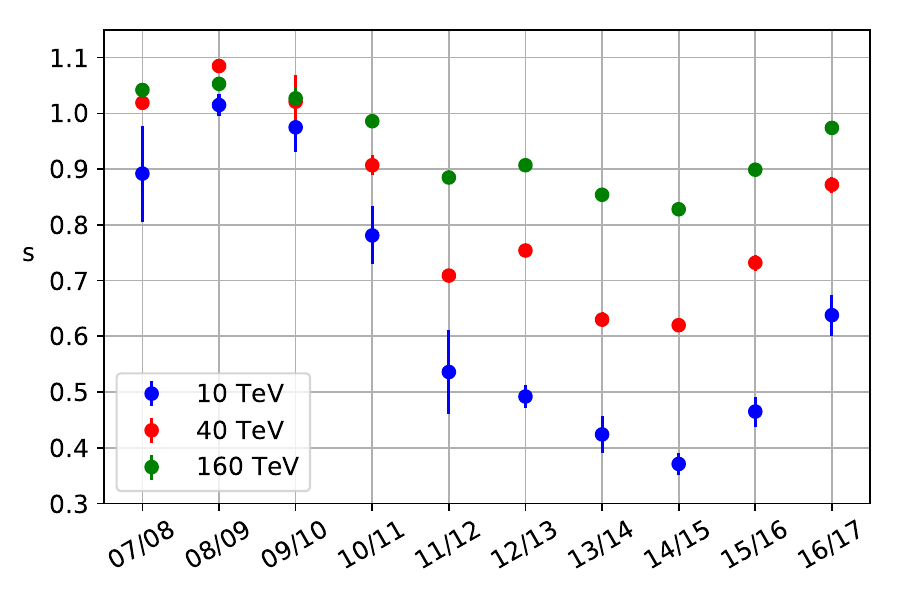}
  \end{center}
  \caption{Calculated Sun shadow in the years from 2007 through 2017 for the different energies with a mixed composition according to the HGm model.} 
  \label{fig:r_all_energies}
\end{figure}

\begin{figure}[htbp]
  \begin{center}
    \includegraphics[width=0.95\linewidth]{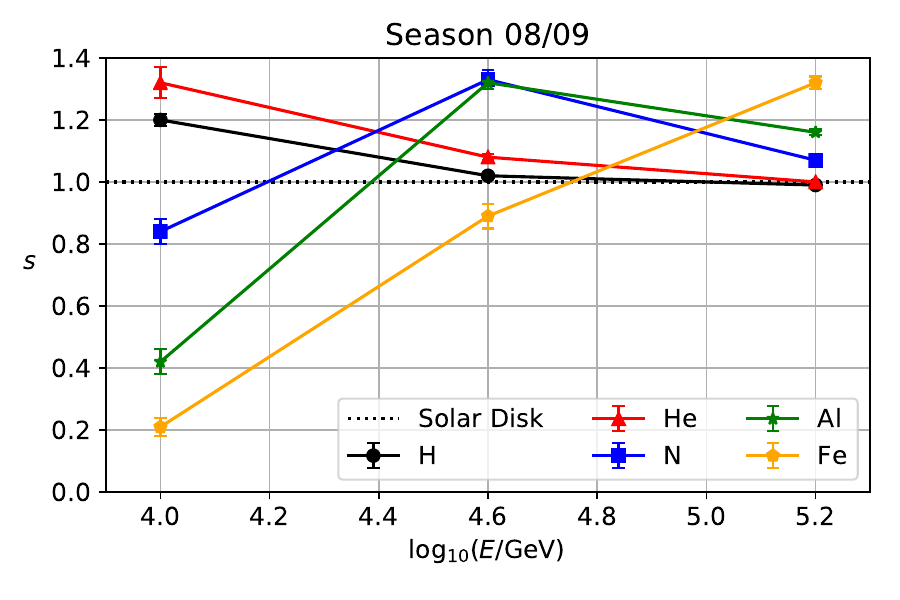}
  \end{center}
  \caption{Calculated Sun shadow in season 2008-09 for the different energies and mass numbers.
  Solar activity was comparatively low during this season (compare with Fig.~\ref{fig:SunSpots}).} 
  \label{fig:s_vs_E_8_9}
\end{figure}

\begin{figure}[h]
  \begin{center}
    \includegraphics[width=0.95\linewidth]{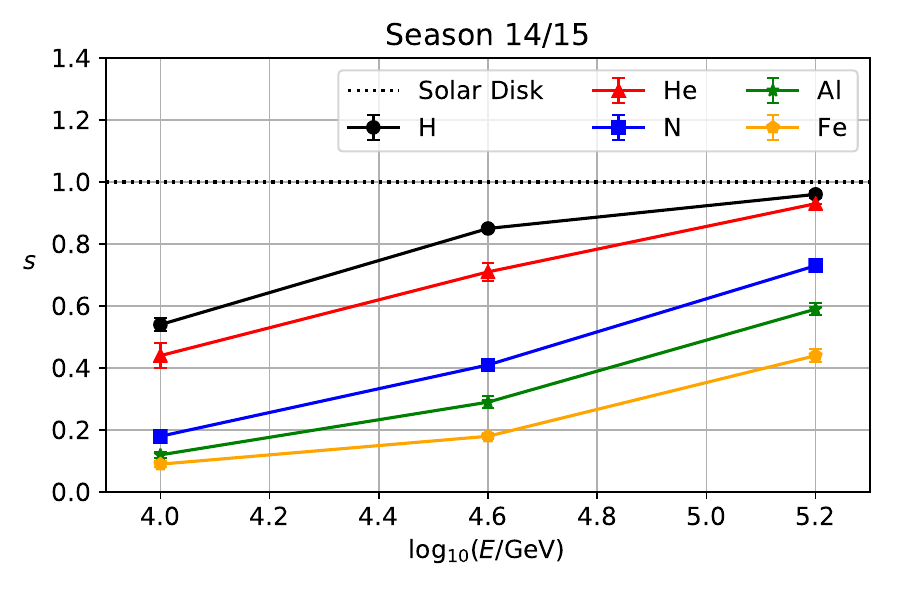}
  \end{center}
  \caption{Calculated Sun shadow in the season 2014-15 for the different energies and mass numbers.
  Solar activity was comparatively high during this season (compare with Fig.~\ref{fig:SunSpots}).} 
  \label{fig:s_vs_E_14_15}
\end{figure}

In Fig.~\ref{fig:r_all_energies}, the shadow depth $s$, calculated as described above, is displayed for the three different energies. 
Each data point contains contributions by all five elements that were simulated with a composition according to the HGm model.
As expected, the points for $E=10\,\mathrm{TeV}$ overall show the smallest $s$, indicating a weaker shadow, while the points for $E=158\,\mathrm{TeV}$ overall show the largest $s$, indicating a stronger shadow. 
In addition, the separation between the energies becomes stronger for seasons with high solar activity (season 2011-12 through season 2015-16).
This is also expected, as the effect of increasing magnetic activity and, in turn, a stronger magnetic field impacts low-energy particles the most. 
In Figs.~\ref{fig:sunshadow_2007_to_2017_10tev}--\ref{fig:sunshadow_2007_to_2017_160tev}, the full shadow plots can be seen for the different energies. 
In agreement with the numerical results, the shadow becomes more disk-like for higher energies, yielding $s$ values close to unity, while it shows a distinct temporal variation at lower energies, resulting in a smearing of the shadow in seasons with high solar activity.
In Figs.~\ref{fig:s_vs_E_8_9} and \ref{fig:s_vs_E_14_15}, the calculated $s$ is plotted for all 15 different combinations of energy and mass number for the seasons 2008-09 and 2014/15, respectively. 
These seasons were chosen as respective examples for periods with notably low and high solar activity (compare with Fig.~\ref{fig:SunSpots}). 
For season 2014-15 the behavior of $s$ as a function of energy and mass number is as expected: the shadow is stronger for high-energy particles with low atomic numbers and weaker for low-energy particles with high atomic numbers.
For season 2008-09, however, the picture is different.
While for iron nuclei, $s$ increases with increasing energy, there is a peak structure for aluminum and nitrogen nuclei, and $s$ even decreases for helium and hydrogen nuclei.
In order to better understand this effect, we reduce the particle properties to only one parameter in the following section. 
\begin{figure}[thbp]
  \begin{center}
    \includegraphics[width=\linewidth]{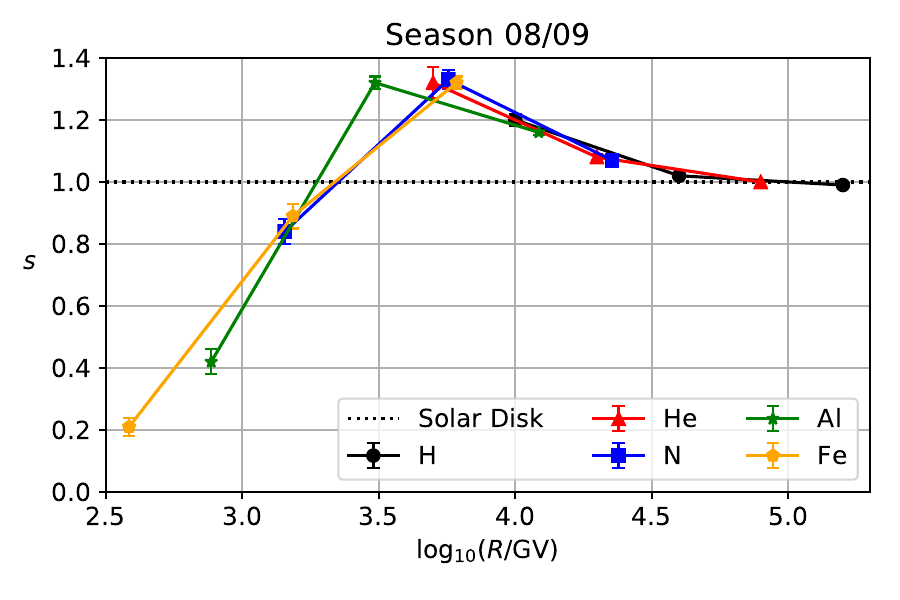}
  \end{center}
  \caption{Calculated Sun shadow in the season 2008/2009 as a function of rigidity.} 
  \label{fig:s_vs_R_8_9}
\end{figure}
\begin{figure}[h]
  \begin{center}
    \includegraphics[width=\linewidth]{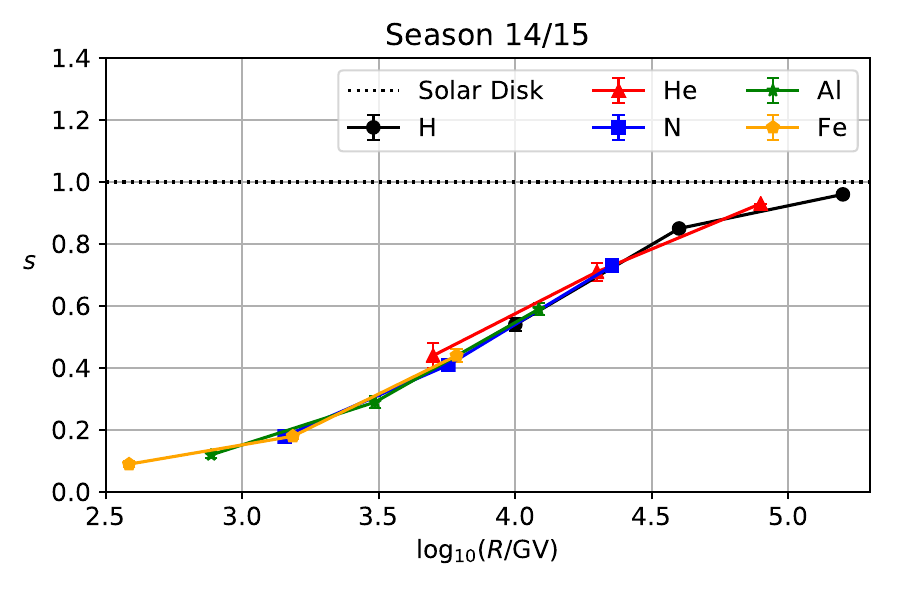}
  \end{center}
  \caption{Calculated Sun shadow in season 2014-15 as a function of rigidity.} 
  \label{fig:s_vs_R_14_15}
\end{figure}
\begin{figure}[h!]
  \begin{center}
    \includegraphics[width=\linewidth]{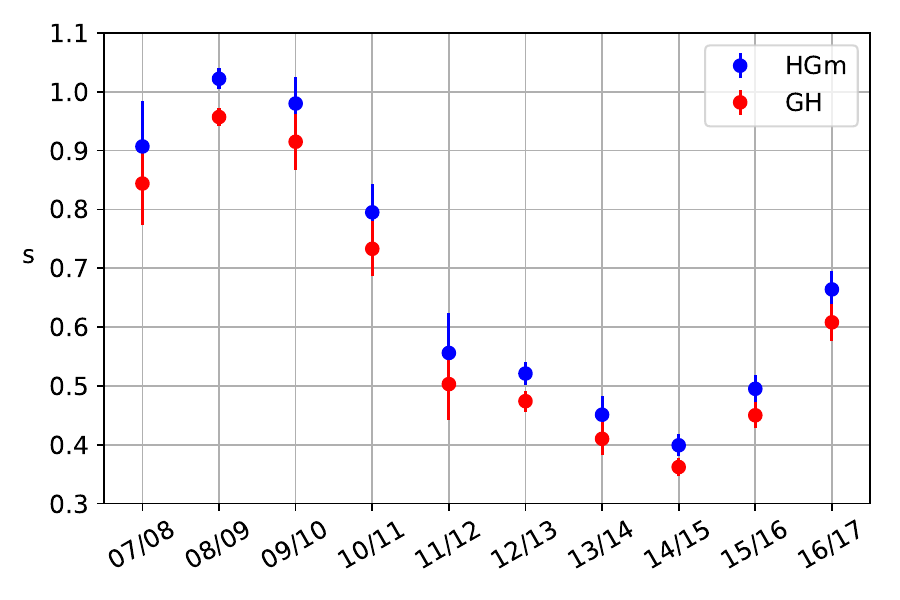}
  \end{center}
  \caption{Comparison of $s$ for the calculated Sun shadow in the years from 2007 through 2017 between the HGm and GH models.}
  \label{fig:r_2007_to_2017_HGm_GH}
\end{figure}

\begin{figure}[t!]
  \centering
    \includegraphics[trim = 0mm 14mm 4mm 3mm, clip, width=0.95\linewidth]{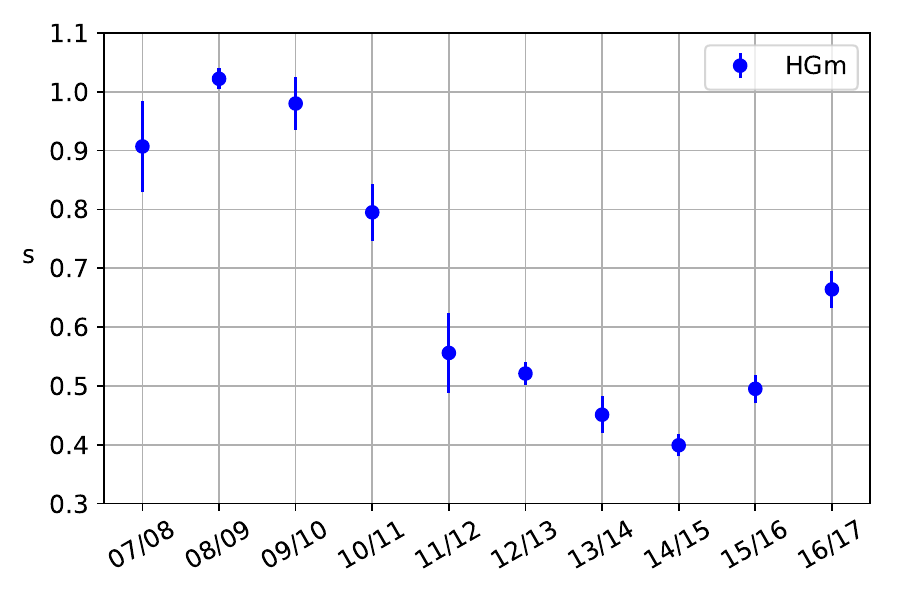} 
    \centering
    \includegraphics[trim = 4mm 4mm 4mm 3mm, clip, width=.95\linewidth]{sunspots.pdf}
  \caption{Temporal behavior of shadow ratio $s$ (assuming the HGm model) and the sunspot number in the years from 2007 through 2017.
  Sunspot numbers are taken from \cite{sidc}.}
  \label{fig:ratio_V_and Sun_Spot_Number_2007_to_2017}
\end{figure}
\begin{figure}[h!]
  \begin{center}
    \includegraphics[width=\linewidth]{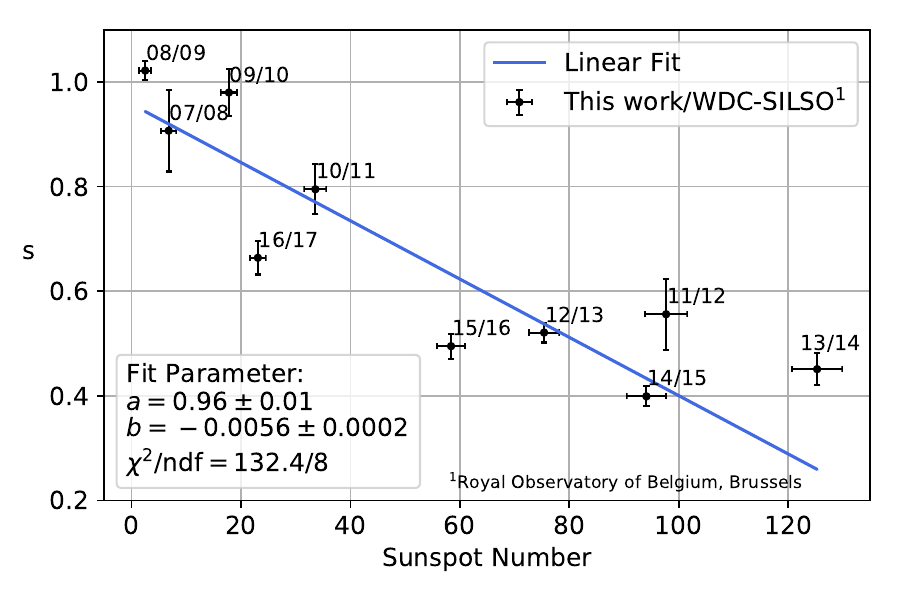}
  \end{center}
  \caption{Correlation of ratio $s$ and the sunspot number from 2007 through 2017.
  Sunspot numbers are taken from \cite{sidc}.}
  \label{fig:correlation_of_V_and_sunspot_number}
\end{figure}

\subsection{Sun shadow as a function of rigidity}
\label{sec:results:rigidity}
In this section, we use rigidity $R \sim E/Z$ in order to reduce the dependency of the shadow on energy and atomic number to one parameter.
In Figs.~\ref{fig:s_vs_R_8_9} and \ref{fig:s_vs_R_14_15}, the calculated $s$ is plotted as a function of rigidity for the respective seasons 2008-09 and 2014-15.
The expected equivalence of particles with the same rigidity is confirmed for both seasons.
As an example, the middle blue square in Fig.~\ref{fig:s_vs_R_14_15}, representing $\sim 10^{1.6}\,\mathrm{TeV}$ nitrogen nuclei, shows an $s$ comparable to that of $\sim 10^{2.2}\,\mathrm{TeV}$ iron nuclei, the rightmost orange pentagon. 
With rigidities of $\sim 5700\,\mathrm{GV}$ and $\sim 6100\,\mathrm{GV}$ respectively, these particles are expected to show a similar behavior with respect to their gyro-radii.
For the season 2008-09, however, $s$ does not increase monotonically, as it does in season 2014-15, but, rather, it peaks at a certain rigidity between $10^{3.5}$ and $10^{4.0}\,\mathrm{GV}$.
A possible explanation for this is that $s$ seems to be significantly larger than $1.0$ for years with low solar activity (compare Sect.~\ref{sec:results:mass}), that is, in years when the magnetic field is approximately described by a simple dipole.
It seems possible that particles above a certain rigidity no longer form the shadow typical for a dipole field anymore, but rather ``converge'' towards the shadow of an unmagnetized disk, which, by definition, would result in values of \mbox{$s \approx 1$}.
Particularly in view of the non-monotonous dependence of shadow size on rigidity that was found for the dipolar test case (see Sect.~\ref{sec:dipolar_test}) and illustrated in Fig.~\ref{fig:plot_shadowsize}, the transition between these two regions -- $s$ for a dipole field versus $s$ for an unmagnetized disk -- can possibly explain the peak visible in Fig.~\ref{fig:s_vs_E_8_9} as well.

Thus, our simulations indicate that an energy-dependent study of the Sun shadow by cosmic-ray observatories would be highly beneficial.
From these studies, we expect the observatories to see the dipole-like energy dependence, with a monotonic rise up to a peak energy.
In the peak, we expect the shadow effect to be up to $40$\% stronger than the geometrical-disk shadow.
At energies above this peak, we expect the shadow to converge to the size of the geometrical one.
In years of high solar activity, we predict a different behavior, which is a purely monotonic increase of the shadow depth, starting at low energies with a shadow much shallower than the geometric one.
For all energies, the shadow is expected to be shallower than the geometrical one. Toward high energies, it will converge to the geometrical shadow depth again.

\subsection{Sun shadow for different cosmic-ray flux models}
In Fig.~\ref{fig:r_2007_to_2017_HGm_GH}, the calculated $s$ is plotted for the HGm and GH models for the seasons 2007-08 to 2016-17. 
While there is a systematic offset in the total magnitude of $s$, its temporal variation is very similar for both models, suggesting that the shape of the temporal variation does not depend significantly on the primary cosmic-ray model used.
In Figs.~\ref{fig:sunshadow_2007_to_2017_HGm} and \ref{fig:sunshadow_2007_to_2017_GH} (see appendix), the calculated passing probabilities and full shadow plots for the HGm and GH models for seasons 2007-08 through 2016-17 can be seen.

\subsection{Sun shadow compared to solar activity}
Figure~\ref{fig:ratio_V_and Sun_Spot_Number_2007_to_2017} shows the temporal behavior of $s$ (assuming the HGm model) and the sunspot number averaged over the time periods considered in this work \citep[data taken from][]{sidc}).
We note that the time range of the Carrington rotations considered for a specific year and the time range in which the sunspot numbers are measured are slightly different since the duration of a Carrington rotation is always shorter than the duration of a calendar month.
Figure~\ref{fig:correlation_of_V_and_sunspot_number} shows the correlation of $s$ with the sunspot number, together with a linear fit of the form
\begin{align}
s = a + b \cdot \mathrm{SSN} .
\end{align}
While $a$ is reasonable close to $1.0$, which is expected as per the discussion in Sects.~\ref{sec:results:energy} and \ref{sec:results:rigidity}, there is currently no straightforward interpretation of $b$.
For the Spearman's rank correlation coefficient $\rho$ of $s$ and the sunspot number, we find 
$\rho_{\mathrm{Spearman}} = -0.86$.

\section{Summary}
\label{sec:conclusions}
In this paper, we investigate cosmic-ray transport across the Sun's corona in the energy range between \mbox{$\sim5$\,TeV} and \mbox{$\sim316$\,TeV} and for hydrogen, helium, nitrogen, aluminum, and iron nuclei. 
For that purpose, particles with three different energies were studied, namely $10^{1.0}$, $10^{1.6}$, and $10^{2.2}$~TeV (corresponding to $10$, $\sim 40$, and \mbox{$\sim158$~TeV}). 
We chose these supporting points based on their being equidistant in $\log{E_{\mathrm{CR}}}$.
The energy range was selected to reflect the range that most detectors measuring the cosmic-ray Sun shadow (such as IceCube, Tibet, ARGO-YBJ, HAWC, and others) are sensitive to. 
The data sample used by IceCube has a median energy of $\left<E_{\mathrm{CR}}\right>\sim 40$\,TeV, with 68\% of the events between $11$ and $200$\,TeV.
The mass numbers of the nuclei are chosen to reflect the most abundant mass groups, with nitrogen representing the CNO and aluminum representing the MgAlSi element group. 
By choosing these mass numbers and energies, we are able to compute a generic prediction for the expected time variability of the cosmic-ray Sun shadow that can be compared to measurements by the experiments mentioned above.
We associate the above-mentioned time variability with the time-changing solar magnetic field and its influence on the shape and depth of the Sun shadow of cosmic rays studied by ground-based cosmic-ray experiments such as IceCube, Tibet, ARGO-YBJ, and HAWC.
In the future, these observatories will be able to further improve their spatial and temporal resolution, possibly contributing to investigations of the solar cycle via indirect measurements of magnetic field properties. 

For the purpose mentioned above, we performed a numerical integration of particle trajectories in the solar magnetic field of the past ten years. 
Magnetogram data are obtained from GONG \citep{1996Sci...272.1284H} and interpreted in the PFSS model using the FDIPS code \citep{2016ascl.soft06011T}. 
For each year from 2007 through 2017, the respective four months, January, February, November, and December, are simulated. 
We chose to simulate these months as they correspond to the time period per year in which the Sun is visible from the South Pole. 
This way, our results are easily applicable to that of the IceCube detector, corresponding to the relevant energy range and observation times. 
It should be noted, however, that no detector effects, such as a limited acceptance or a limited angular resolution, have been taken into account.

%%%%%%%%

The variation of the cosmic-ray Sun shadow is verified in this simulation, and as expected, the effect is strongest at the lowest simulated energy of $10$~TeV. 
The variation of the shadow furthermore shows that in times of low solar activity, the shadow is relatively compact near the Sun's center. 
In times of high solar activity, however, the shadow tends to be washed out. 
It needs to be investigated by IceCube and other cosmic-ray experiments whether it is possible to detect such features of systematic deviation from a dipole structure despite the limitations associated with angular resolution. 
Such a measurement would make it possible to indirectly study the structure of the solar magnetic field and, thus, to contribute to its understanding. 
In this sense, our findings of the different energy behavior of the shadow for low activity (dipole-like) and high activity (monotonous) are very interesting, as they would pose another possibility for the detection of a dipole-like shadow behavior. 
We can make the clear prediction that in times of low solar activity, the energy-behavior should resemble the one of a pure dipole, with a shadow that becomes stronger with energy up to a maximum value, in which it exceeds the depth of the geometrical shadow. 
For the years studied here, the shadow is expected to become $40$\% deeper than the geometrical shadow. 
Above this peak energy, the shadow is expected to converge to the geometrical size.
This behavior stands in contrast to years of high activity. 
Here we find a monotonous increase of the shadow depth with energy, converging toward the geometrical size, never exceeding the geometrical shadow. 
Future studies will show if this is an effect of the unordered field of the Sun, which is much more significant in years of high solar activity as compared to the low activity years of the Sun.

We can further show that, while the shadow is still significantly distorted at $40$~TeV, the deviation from an undisturbed shadow becomes extremely small at $158$~TeV (compare with Sect.~\ref{sec:results:energy}). 
Thus, we expect effects of the inner heliospheric field on cosmic-ray propagation to become negligible with regard to the variation of the Sun shadow at these high energies. 
This result is in agreement with recent HAWC measurements \citep{hawc_aps2017}, which show a symmetric, deep shadow at energies around $100$~TeV.
Future investigations of several years of data with HAWC will help to further quantify these findings.

The absolute normalization of the shadow depth in our simulation is not directly comparable to IceCube data as we did not take into account detection effects like a limited acceptance or a limited angular resolution. 
In comparing the relative change in the deficit between the different years, however, our simulations show a similar temporal variation of the shadow depth as the data taken with IceCube \citep{ICECUBE_SUNSHADOW_PAPER}. 

\section{Outlook}
\label{sec:outlook}
We consider this paper as a first step toward achieving a detailed understanding of the effects of the solar magnetic field variation on the Sun's cosmic-ray shadow measured by different experiments at Earth.
In the future, there will be opportunities for further work, which either need improved data from the experiments or require modeling approaches that are computationally more involved.

\paragraph{Magnetic field modeling.}
One aspect concerns modeling the inner heliospheric magnetic field, which is done on the basis of available data. 
The data are collected by spacecraft, such as Helios \citep{schwenn2012physics}, which do not simultaneously provide data from the far side. 
We used the PFSS ansatz \citep{1969SoPh....6..442S, 1969SoPh....9..131A} in order to model the global coronal magnetic field, but it is desirable to compare these results to other models. 
The Tibet group, for instance, favors the CSSS model \citep{2013PhRvL.111a1101A} because in using this model, they found better agreement with the measured data. 
A similar comparison between these two models and IceCube data would be an asset for a better understanding of the solar coronal magnetic field.  
In addition to the existing studies by Tibet, a comparison of different coronal magnetic field models on the analysis level using IceCube data would be very useful for evaluating the validity of different models.
However, as these models both assume a purely radial magnetic field at a defined boundary sphere, no fundamental differences between these two are expected.
Ultimately, it will be desirable to probe more complex models like force-free field approximations \citep[see, e.g.,][]{force_free_field} or even full magnetohydrodynamic (MHD) simulations of the solar magnetic field. 

\paragraph{Quantitative comparison with data.}
In this paper, we compare our results regarding the time variability of the Sun shadow with current IceCube data, which is available for five years \citep[see][]{ICECUBE_SUNSHADOW_PAPER}.
In the future, IceCube will extend its data analysis to more years of data for the same solar cycle.
Also, HAWC will be able to provide information in a similar energy range. 
As a result, a comprehensive comparison of data and simulation will be possible:
first, by deducing the effects on the shadow caused by the magnetic field from simulations only, and, second, by comparing data and simulation on the detector level.
Future arrays like IceCube-Gen2 \citep{icecube_gen2} will improve such studies even further, possibly enabling an investigation of the properties of the magnetic field through the combination of simulation and detection of the cosmic-ray shadow with better angular and time resolution.

Integrating the particle propagation into a detector environment is necessary in order to test which magnetic field configuration fits the data best. 
This needs to be done for each instrument individually and it can only be done through the collaborations themselves.
Studies as those we present here, without including detector effects, are still desirable as this is the best way to find effects that arise from the magnetic field. 
The influence of coronal mass ejections (CMEs) on the cosmic-ray Sun shadow seen by different experiments is also something that can be studied, as shown in \cite{tibet_ecme}.

\paragraph{Toward a complete model of transport and interaction.}
Furthermore, adding particle interactions in the solar corona and in the Earth's atmosphere to the simulations will be another step that would increase our understanding of the phenomenon of cosmic-ray shadowing.

\begin{acknowledgements}
  We would like to thank Tobias Wiengarten for assistance with the magnetic field data, as well as Malina Reytemeier, Paul Evenson and the entire IceCube collaboration, as well as members of the \textit{Ruhr Astroparticle Plasma Physics (RAPP) Center} for helpful discussions on the subject. 
  We are further grateful for the support from the BMBF (05A14PC1) and from MERCUR (St-2014-0040). 
  This work utilizes data from the National Solar Observatory Integrated Synoptic Program, which is operated by the Association of Universities for Research in Astronomy, under a cooperative agreement with the National Science Foundation and with additional financial support from the National Oceanic and Atmospheric Administration, the National Aeronautics and Space Administration, and the United States Air Force. 
  The GONG network of instruments is hosted by the Big Bear Solar Observatory, High Altitude Observatory, Learmonth Solar Observatory, Udaipur Solar Observatory, Instituto de Astrof\'isica de Canarias, and Cerro Tololo Interamerican Observatory.
\end{acknowledgements}

\bibliographystyle{aa}
\bibliography{bibliography}

\begin{thebibliography}{43}
\expandafter\ifx\csname natexlab\endcsname\relax\def\natexlab#1{#1}\fi

\bibitem[{Aartsen {et~al.}(2019)Aartsen, Ackermann, Adams,
  {et~al.}}]{ICECUBE_SUNSHADOW_PAPER}
Aartsen, M.~G., Ackermann, M., Adams, J., {et~al.} 2019, Astrophys. J., 872,
  133

\bibitem[{{Abdo} {et~al.}(2011){Abdo}, {Ackermann}, {Ajello}, {Baldini},
  {Ballet}, {Barbiellini}, {Bastieri}, {Bechtol}, {Bellazzini}, {Berenji},
  {Bonamente}, {Borgland}, {Bouvier}, {Bregeon},
  {et~al.}}]{2011ApJ...734..116A}
{Abdo}, A.~A., {Ackermann}, M., {Ajello}, M., {et~al.} 2011, Astrophys. J.,
  734, 116

\bibitem[{{Altschuler} \& {Newkirk}(1969)}]{1969SoPh....9..131A}
{Altschuler}, M.~D. \& {Newkirk}, G. 1969, Sol. Phys., 9, 131

\bibitem[{{Alves Batista} {et~al.}(2013){Alves Batista}, {Erdmann}, {Evoli},
  {Kampert}, {Kuempel}, {M{\"u}ller}, {Schiffer}, {Sigl}, {van Vliet}, {Walz},
  \& {Winchen}}]{2013arXiv1307.2643A}
{Alves Batista}, R., {Erdmann}, M., {Evoli}, C., {et~al.} 2013, arXiv e-prints
  [\eprint[arXiv]{1307.2643}]

\bibitem[{{Amenomori} {et~al.}(2013){Amenomori}, {Bi}, {Chen},
  {et~al.}}]{2013PhRvL.111a1101A}
{Amenomori}, M., {Bi}, X.~J., {Chen}, D., {et~al.} 2013, Phys. Rev. Lett., 111,
  011101

\bibitem[{Amenomori {et~al.}(2018)Amenomori, Bi, Chen, {et~al.}}]{tibet_ecme}
Amenomori, M., Bi, X.~J., Chen, D., {et~al.} 2018, Astrophys. J., 860, 13

\bibitem[{{Arg{\"u}elles} {et~al.}(2017){Arg{\"u}elles}, {de Wasseige},
  {Fedynitch}, \& {Jones}}]{2017JCAP...07..024A}
{Arg{\"u}elles}, C.~A., {de Wasseige}, G., {Fedynitch}, A., \& {Jones},
  B.~J.~P. 2017, J. Cosm. Astropart. Phys., 7, 024

\bibitem[{Boris(1970)}]{boris}
Boris, J. 1970, Proc. of the 4th Conf. on Numerical Simulation of Plasmas
  (NRL), 3

\bibitem[{Bos(2017)}]{Bos2017}
Bos, F. 2017, PhD thesis, Ruhr-Universität Bochum

\bibitem[{Clark(1957)}]{clark}
Clark, G.~W. 1957, Phys. Rev., 108, 450

\bibitem[{{Edsj{\"o}} {et~al.}(2017){Edsj{\"o}}, {Elevant}, {Enberg}, \&
  {Niblaeus}}]{2017JCAP...06..033E}
{Edsj{\"o}}, J., {Elevant}, J., {Enberg}, R., \& {Niblaeus}, C. 2017, J. Cosm.
  Astropart. Phys., 6, 033

\bibitem[{Enriquez-Rivera \& Lara(2016)}]{Enriquez:2015nva}
Enriquez-Rivera, O. \& Lara, A. 2016, in {Proceedings, 34th International
  Cosmic Ray Conference (ICRC 2015): The Hague, The Netherlands, July 30-August
  6, 2015}, Vol. ICRC2015, 099

\bibitem[{Evoli {et~al.}(2008)Evoli, Gaggero, Grasso, \&
  Maccione}]{1475-7516-2008-10-018}
Evoli, C., Gaggero, D., Grasso, D., \& Maccione, L. 2008, J. Cosm. Astropart.
  Phys., 2008, 018

\bibitem[{{Gaisser}(2012)}]{HGm}
{Gaisser}, T.~K. 2012, Astropart. Phys., 35, 801

\bibitem[{{Gaisser} \& {Honda}(2002)}]{GaisserHonda}
{Gaisser}, T.~K. \& {Honda}, M. 2002, Ann. Rev. NPS, 52, 153

\bibitem[{Hapgood(1992)}]{hapgood}
Hapgood, M. 1992, Planet. Space Sci., 40, 711

\bibitem[{{Harvey} {et~al.}(1996){Harvey}, {Hill}, {Hubbard}, {Kennedy},
  {Leibacher}, {Pintar}, {Gilman}, {Noyes}, {Title}, {Toomre}, {Ulrich},
  {Bhatnagar}, {Kennewell}, {Marquette}, {Patron}, {Saa}, \&
  {Yasukawa}}]{1996Sci...272.1284H}
{Harvey}, J.~W., {Hill}, F., {Hubbard}, R.~P., {et~al.} 1996, Science, 272,
  1284

\bibitem[{{Hoyt} \& {Schatten}(1998)}]{1998SoPh..179..189H}
{Hoyt}, D.~V. \& {Schatten}, K.~H. 1998, Sol. Phys., 179, 189

\bibitem[{{IceCube-Gen2 Collaboration ({Aartsen}, M. G. et
  al.)}(2014)}]{icecube_gen2}
{IceCube-Gen2 Collaboration ({Aartsen}, M. G. et al.)}. 2014, arXiv e-prints,
  arXiv:1412.5106

\bibitem[{{Kissmann}(2014)}]{2014APh....55...37K}
{Kissmann}, R. 2014, Astropart. Phys., 55, 37

\bibitem[{{Merten} {et~al.}(2017){Merten}, {Becker Tjus}, {Fichtner},
  {Eichmann}, \& {Sigl}}]{2017JCAP...06..046M}
{Merten}, L., {Becker Tjus}, J., {Fichtner}, H., {Eichmann}, B., \& {Sigl}, G.
  2017, J. Cosm. Astropart. Phys., 6, 046

\bibitem[{Ng {et~al.}(2016)Ng, Beacom, Peter, \& Rott}]{Ng:2015gya}
Ng, K. C.~Y., Beacom, J.~F., Peter, A. H.~G., \& Rott, C. 2016, Phys. Rev.,
  D94, 023004

\bibitem[{{Ng} {et~al.}(2017){Ng}, {Beacom}, {Peter}, \&
  {Rott}}]{Ng+2017_PhRvD}
{Ng}, K. C.~Y., {Beacom}, J.~F., {Peter}, A. H.~G., \& {Rott}, C. 2017, \prd,
  96, 103006

\bibitem[{{Nisa} {et~al.}(2019){Nisa}, {Beacom}, {BenZvi}, {Leane}, {Linden},
  {Ng}, {Peter}, \& {Zhou}}]{whitepaper2020}
{Nisa}, M.~U., {Beacom}, J.~F., {BenZvi}, S.~Y., {et~al.} 2019, arXiv e-prints,
  arXiv:1903.06349

\bibitem[{{Nisa} {et~al.}(2017){Nisa}, {Hampel}, \& {HAWC
  Collaboration}}]{hawc_aps2017}
{Nisa}, M.~U., {Hampel}, Z., \& {HAWC Collaboration}. 2017, in APS Meeting
  Abstracts, Vol. 2017, APS April Meeting Abstracts, Y3.006

\bibitem[{Orlando \& Strong(2008)}]{Orlando:2008uk}
Orlando, E. \& Strong, A.~W. 2008, Astron. Astrophys., 480, 847

\bibitem[{{Parker}(1958)}]{parker}
{Parker}, E.~N. 1958, Astrophys. J., 128, 664

\bibitem[{Priest(1982)}]{Priest1982}
Priest, E. 1982, Solar Magnetohydrodynamics (Springer Netherlands)

\bibitem[{Qin {et~al.}(2013)Qin, Zhang, Xiao, Liu, Sun, \&
  Tang}]{why_is_boris_so_good}
Qin, H., Zhang, S., Xiao, J., {et~al.} 2013, Phys. Plasm., 20, 084503

\bibitem[{{Schatten} {et~al.}(1969){Schatten}, {Wilcox}, \&
  {Ness}}]{1969SoPh....6..442S}
{Schatten}, K.~H., {Wilcox}, J.~M., \& {Ness}, N.~F. 1969, Sol. Phys., 6, 442

\bibitem[{Schlickeiser(2002)}]{Schlickeiser2002}
Schlickeiser, R. 2002, Cosmic Ray Astrophysics (Springer)

\bibitem[{Schwenn \& Marsch(2012)}]{schwenn2012physics}
Schwenn, R. \& Marsch, E. 2012, Physics of the Inner Heliosphere I: Large-Scale
  Phenomena, Physics and Chemistry in Space (Springer Berlin Heidelberg)

\bibitem[{{Seckel} {et~al.}(1991){Seckel}, {Stanev}, \&
  {Gaisser}}]{1991ApJ...382..652S}
{Seckel}, D., {Stanev}, T., \& {Gaisser}, T.~K. 1991, Astrophys. J., 382, 652

\bibitem[{{SILSO World Data Center}(2007-2017)}]{sidc}
{SILSO World Data Center}. 2007-2017, International Sunspot Number Monthly
  Bulletin and online catalogue

\bibitem[{{Solanki} {et~al.}(2006){Solanki}, {Inhester}, \&
  {Sch{\"u}ssler}}]{2006RPPh...69..563S}
{Solanki}, S.~K., {Inhester}, B., \& {Sch{\"u}ssler}, M. 2006, Reports on
  Progress in Physics, 69, 563

\bibitem[{Strong \& Moskalenko(1998)}]{0004-637X-509-1-212}
Strong, A.~W. \& Moskalenko, I.~V. 1998, Astrophys. J., 509, 212

\bibitem[{Strong {et~al.}(2010)Strong, Porter, Digel, J{\'o}hannesson, Martin,
  Moskalenko, Murphy, \& Orlando}]{2041-8205-722-1-L58}
Strong, A.~W., Porter, T.~A., Digel, S.~W., {et~al.} 2010, Astrophys. J. Lett.,
  722, L58

\bibitem[{{Tadesse} {et~al.}(2014){Tadesse}, {Wiegelmann}, {MacNeice},
  {Inhester}, {Olson}, \& {Pevtsov}}]{force_free_field}
{Tadesse}, T., {Wiegelmann}, T., {MacNeice}, P.~J., {et~al.} 2014, Sol. Phys.,
  289, 831

\bibitem[{{Toth} {et~al.}(2016){Toth}, {van der Holst}, \&
  {Huang}}]{2016ascl.soft06011T}
{Toth}, G., {van der Holst}, B., \& {Huang}, Z. 2016, {FDIPS: Finite Difference
  Iterative Potential-field Solver}, Astrophysics Source Code Library

\bibitem[{{Vladimirov} {et~al.}(2011){Vladimirov}, {Digel}, {J{\'o}hannesson},
  {Michelson}, {Moskalenko}, {Nolan}, {Orlando}, {Porter}, \&
  {Strong}}]{2011CoPhC.182.1156V}
{Vladimirov}, A.~E., {Digel}, S.~W., {J{\'o}hannesson}, G., {et~al.} 2011,
  Comp. Phys. Comm., 182, 1156

\bibitem[{{Wiebel-Sooth} {et~al.}(1998){Wiebel-Sooth}, {Biermann}, \&
  {Meyer}}]{wiebel_sooth1998}
{Wiebel-Sooth}, B., {Biermann}, P.~L., \& {Meyer}, H. 1998, Astron. Astrophys.,
  330, 389

\bibitem[{{Wiengarten} {et~al.}(2014){Wiengarten}, {Kleimann}, {Fichtner},
  {K{\"u}hl}, {Kopp}, {Heber}, \& {Kissmann}}]{2014ApJ...788...80W}
{Wiengarten}, T., {Kleimann}, J., {Fichtner}, H., {et~al.} 2014, Astrophys. J.,
  788, 80

\bibitem[{Zhao \& Hoeksema(1995)}]{csss}
Zhao, X. \& Hoeksema, J.~T. 1995, J. Geophys. Res: Space Phys., 100, 19

\end{thebibliography}

\onecolumn
\appendix

\section{Shadow plots}
\label{sec:appendix:shadow}
\subsection{Sun shadow as a function of mass number}
\begin{figure*}[htbp]
  \centering
  \begin{tabular}{ccc} 
    \includegraphics[trim = 22mm 4mm 4mm 3mm, clip, width=0.32\linewidth]{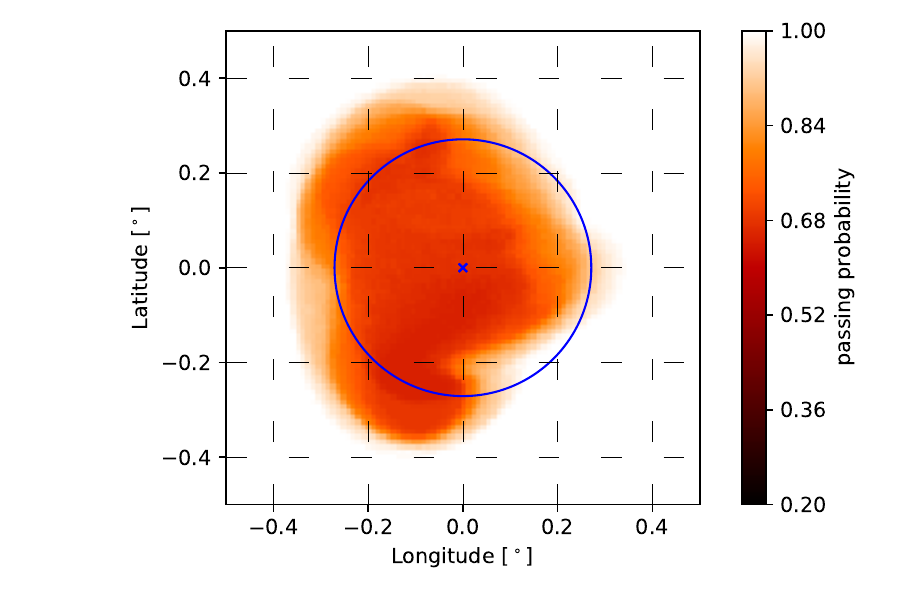} & 
    \includegraphics[trim = 22mm 4mm 4mm 3mm, clip, width=0.32\linewidth]{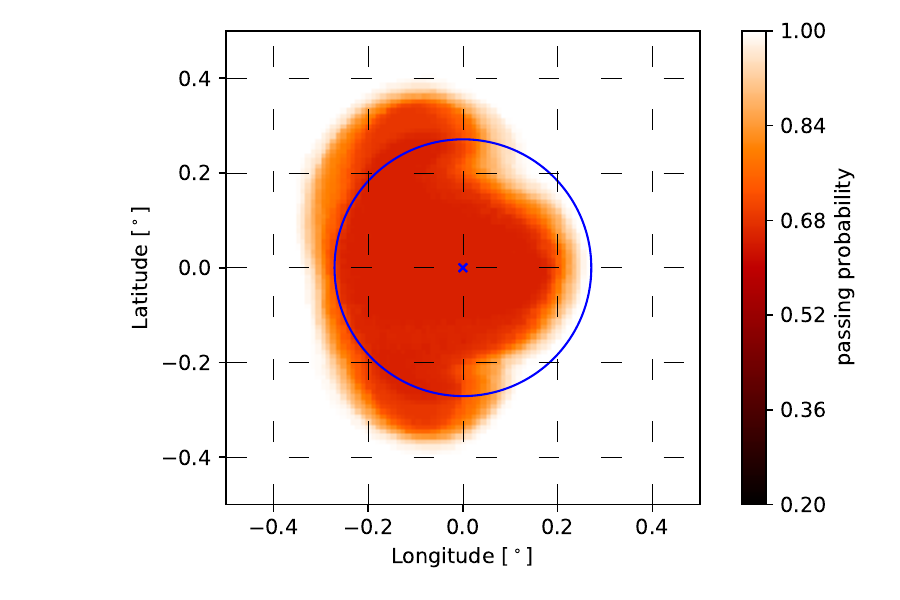} &
    \includegraphics[trim = 22mm 4mm 4mm 3mm, clip, width=0.32\linewidth]{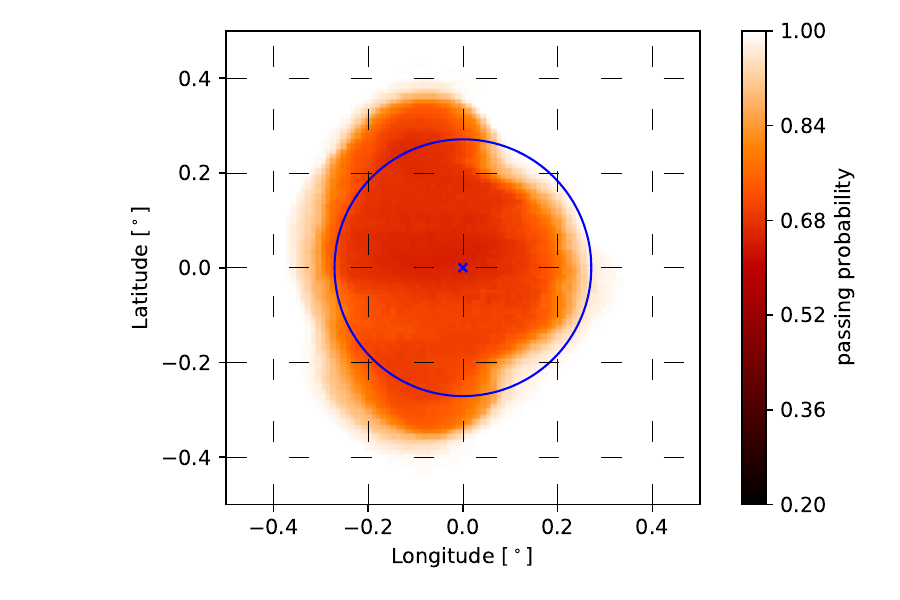} \\ 
    \includegraphics[trim = 22mm 4mm 4mm 3mm, clip, width=0.32\linewidth]{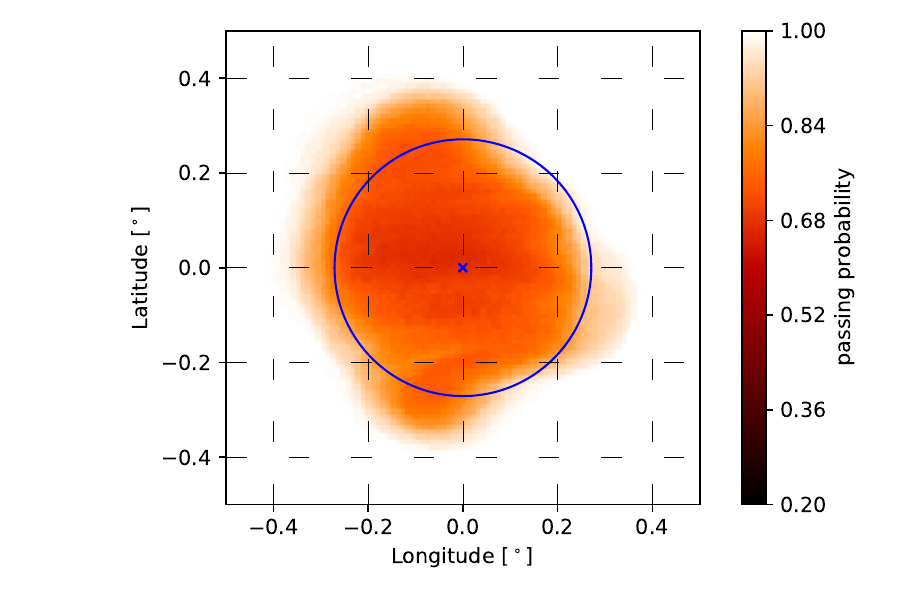} & 
    \includegraphics[trim = 22mm 4mm 4mm 3mm, clip, width=0.32\linewidth]{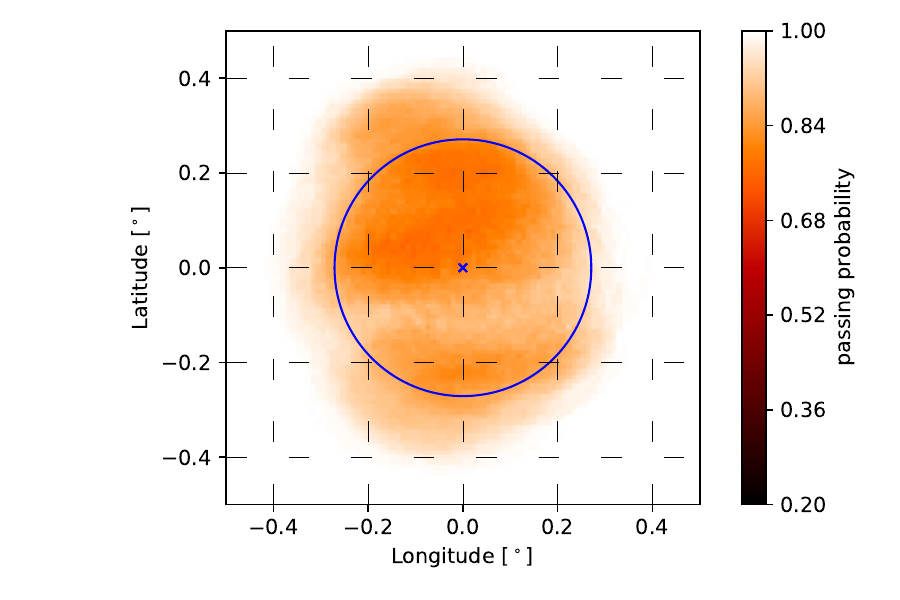} &
    \includegraphics[trim = 22mm 4mm 4mm 3mm, clip, width=0.32\linewidth]{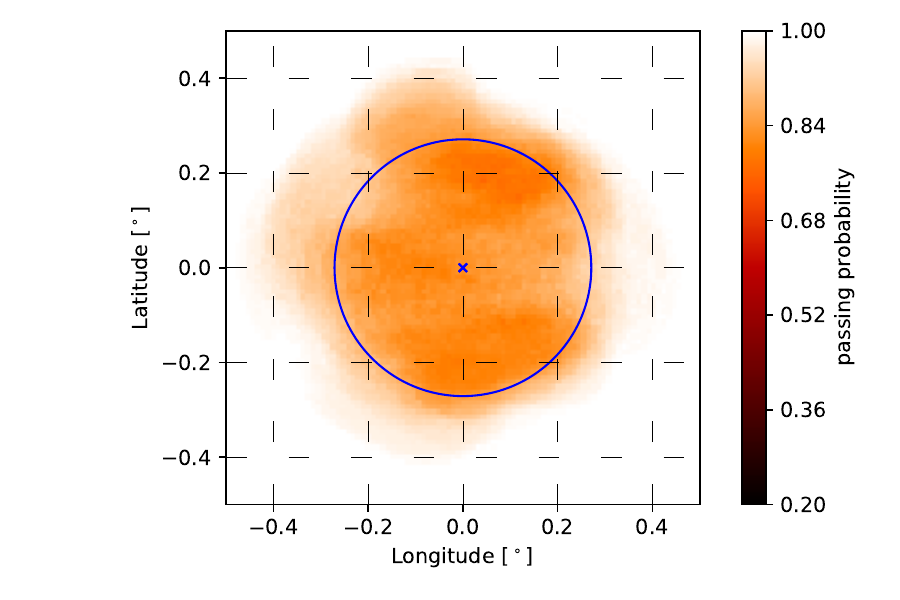} \\ 
    \includegraphics[trim = 22mm 4mm 4mm 3mm, clip, width=0.32\linewidth]{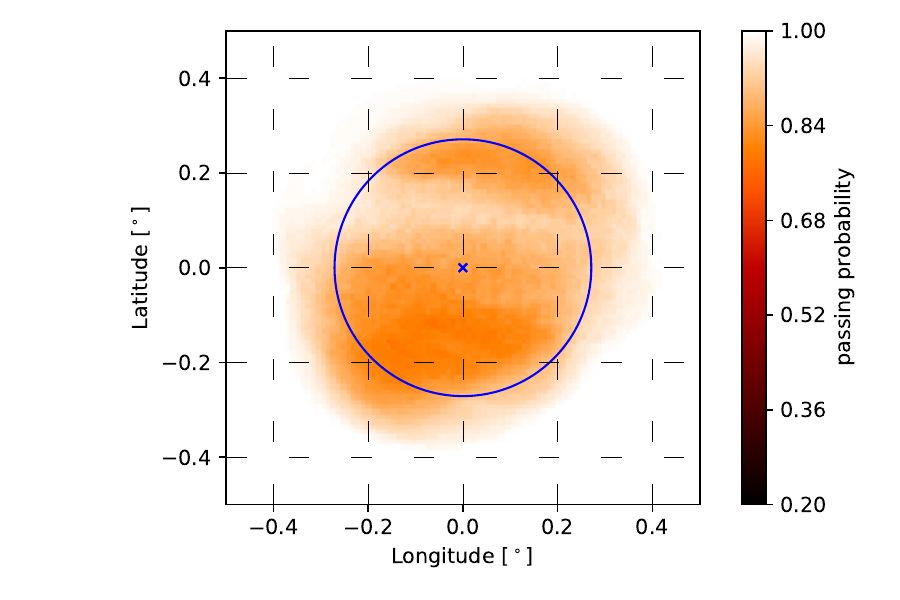} & 
    \includegraphics[trim = 22mm 4mm 4mm 3mm, clip, width=0.32\linewidth]{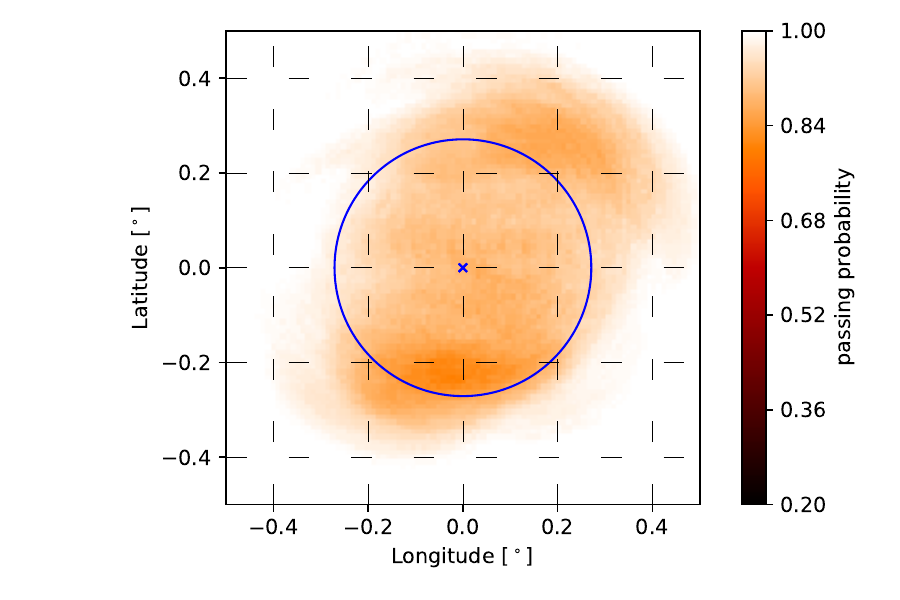} &
    \includegraphics[trim = 22mm 4mm 4mm 3mm, clip, width=0.32\linewidth]{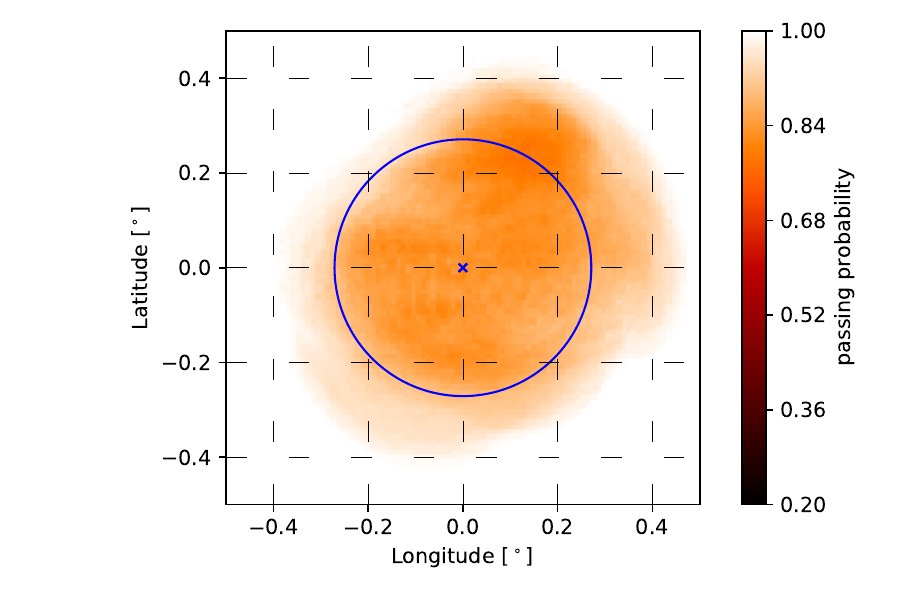} \\ & 
    \includegraphics[trim = 22mm 4mm 4mm 3mm, clip, width=0.32\linewidth]{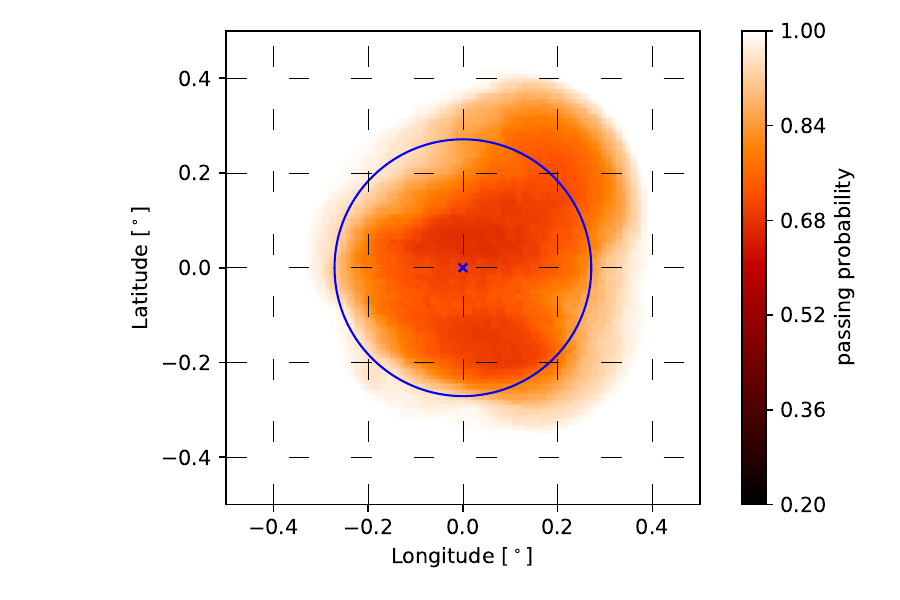} &
  \end{tabular}
  \caption{Calculated Sun shadow from 2007 through 2017 for protons only.}
  \label{fig:sunshadow_2007_to_2017_proton_only}
\end{figure*}

\begin{figure*}[htbp]
  \centering
  \begin{tabular}{ccc} 
    \includegraphics[trim = 22mm 4mm 4mm 3mm, clip, width=0.32\linewidth]{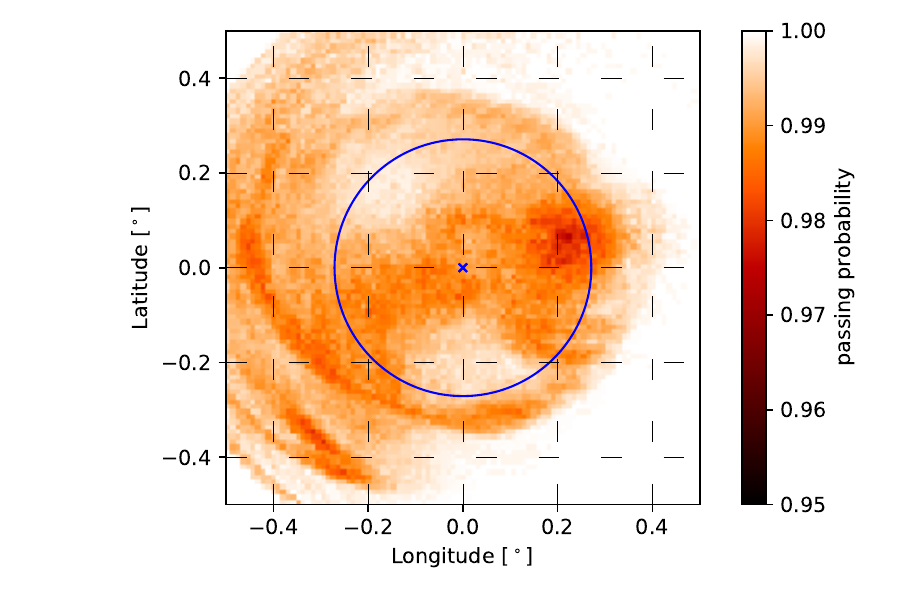} & 
    \includegraphics[trim = 22mm 4mm 4mm 3mm, clip, width=0.32\linewidth]{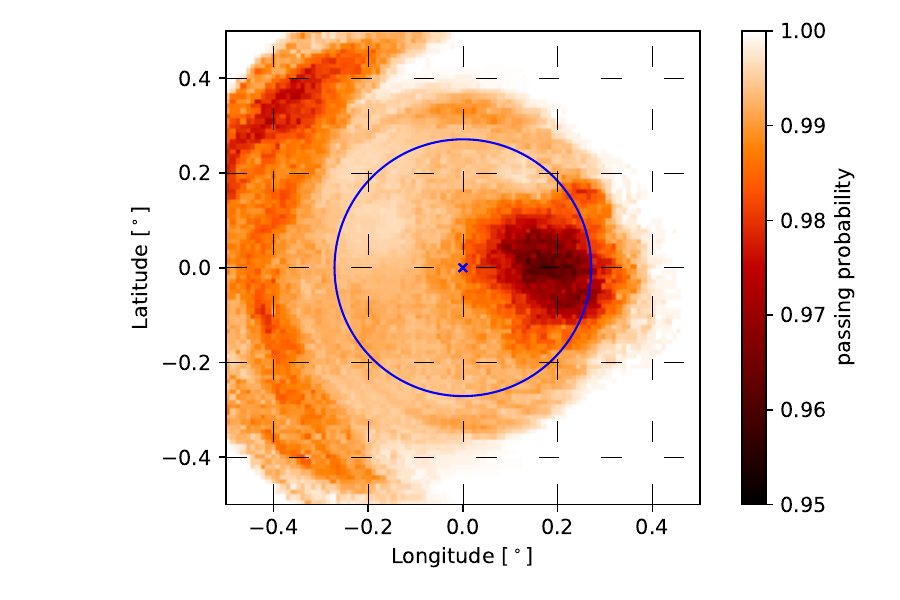} &
    \includegraphics[trim = 22mm 4mm 4mm 3mm, clip, width=0.32\linewidth]{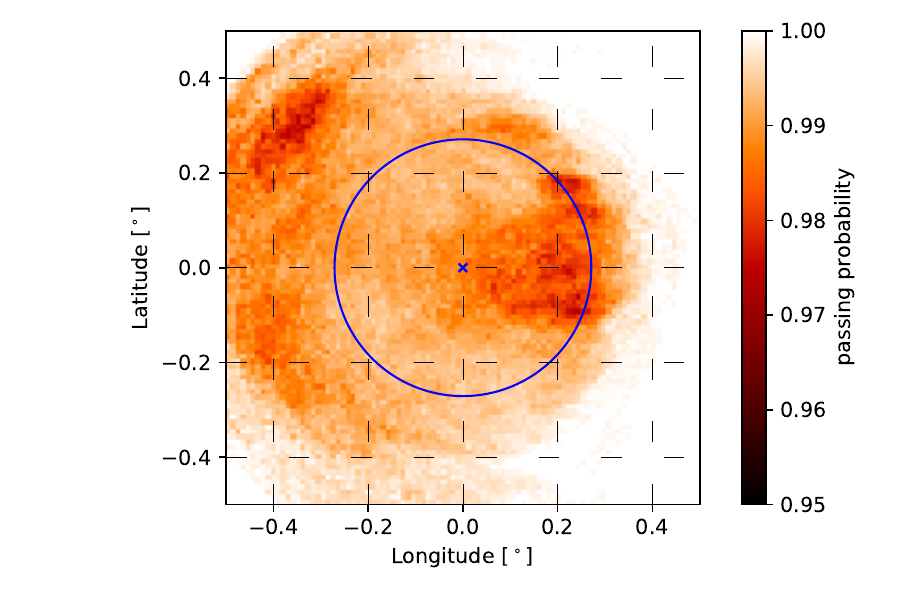} \\ 
    \includegraphics[trim = 22mm 4mm 4mm 3mm, clip, width=0.32\linewidth]{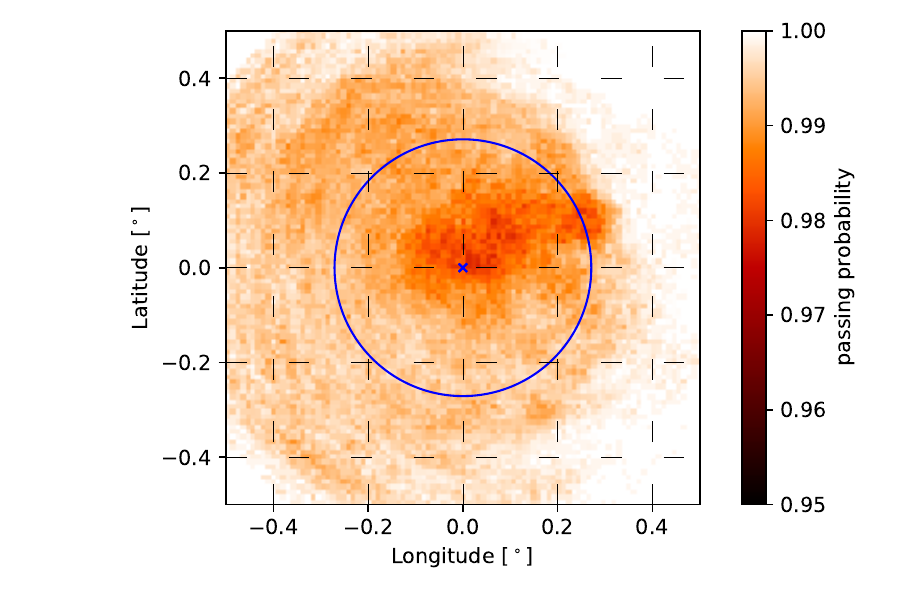} & 
    \includegraphics[trim = 22mm 4mm 4mm 3mm, clip, width=0.32\linewidth]{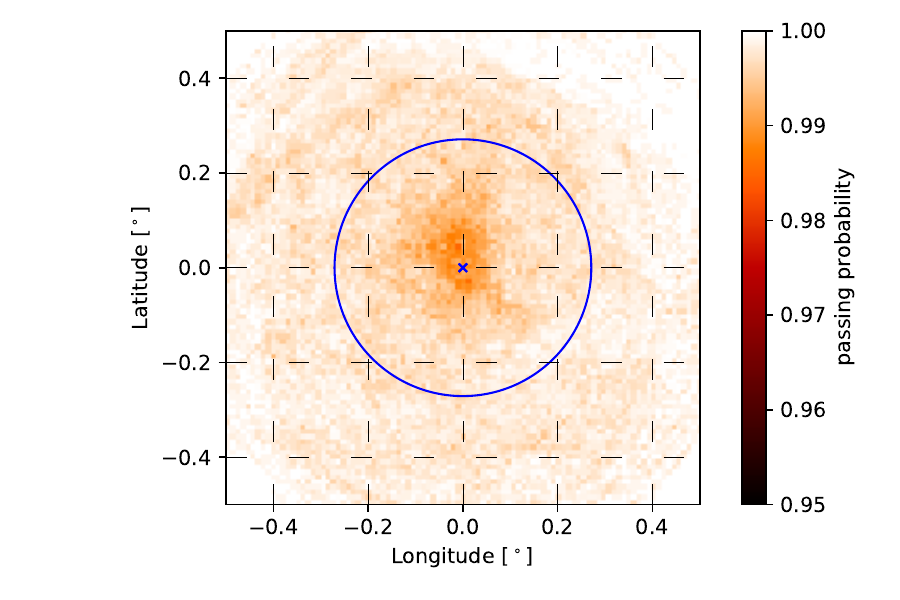} &
    \includegraphics[trim = 22mm 4mm 4mm 3mm, clip, width=0.32\linewidth]{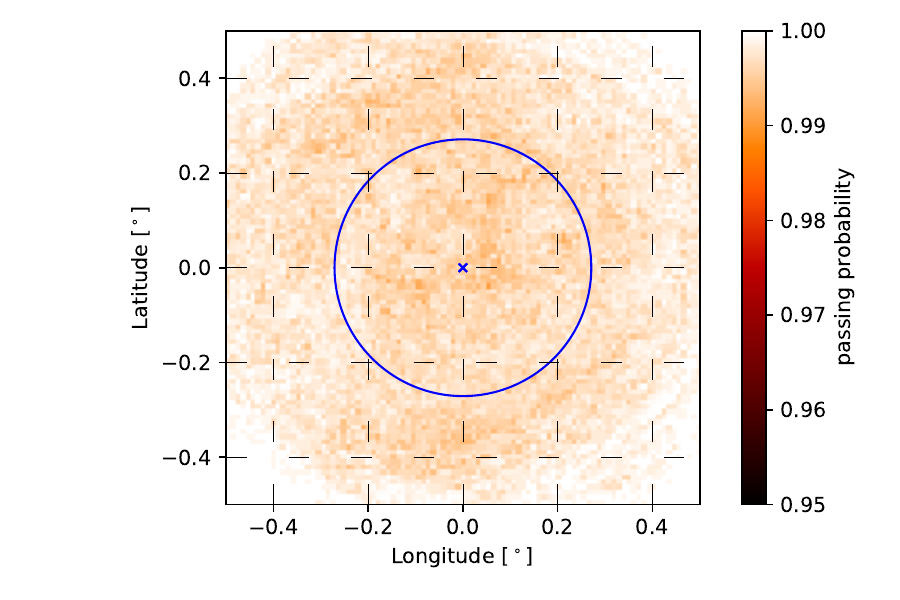} \\ 
    \includegraphics[trim = 22mm 4mm 4mm 3mm, clip, width=0.32\linewidth]{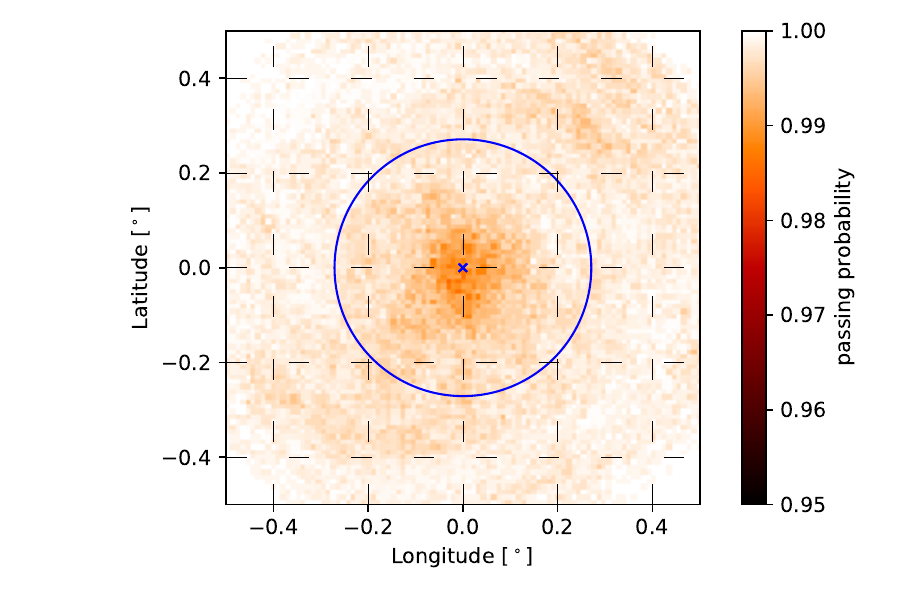} & 
    \includegraphics[trim = 22mm 4mm 4mm 3mm, clip, width=0.32\linewidth]{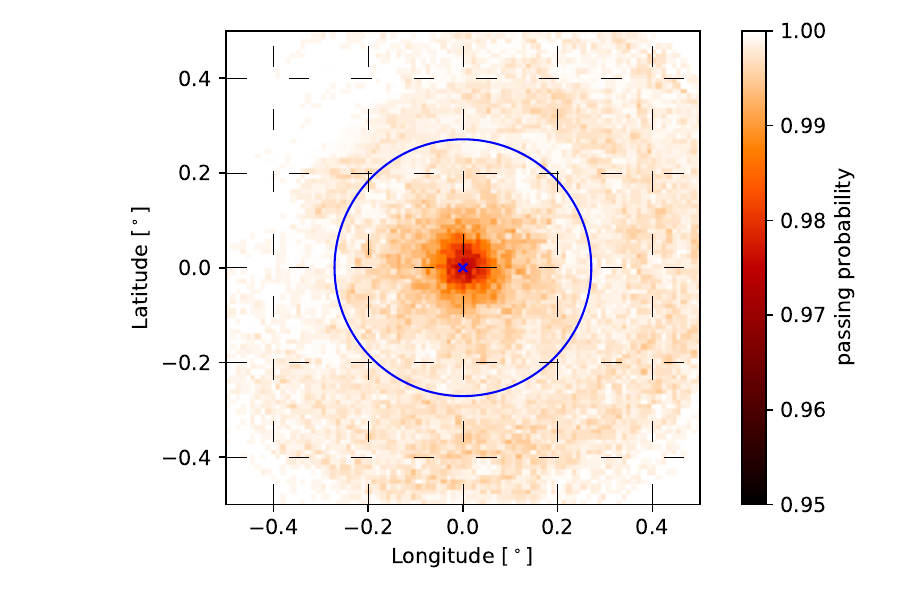} &
    \includegraphics[trim = 22mm 4mm 4mm 3mm, clip, width=0.32\linewidth]{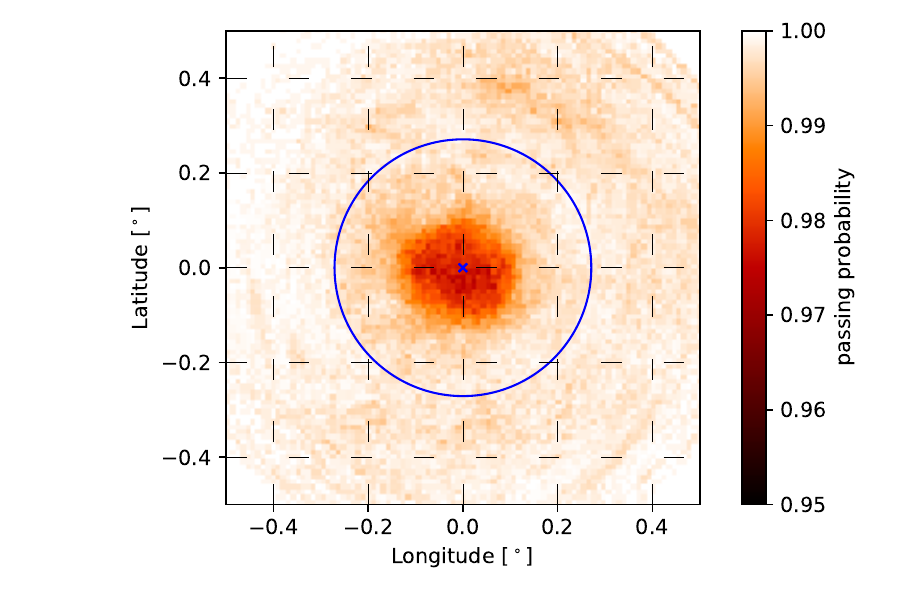} \\ & 
    \includegraphics[trim = 22mm 4mm 4mm 3mm, clip, width=0.32\linewidth]{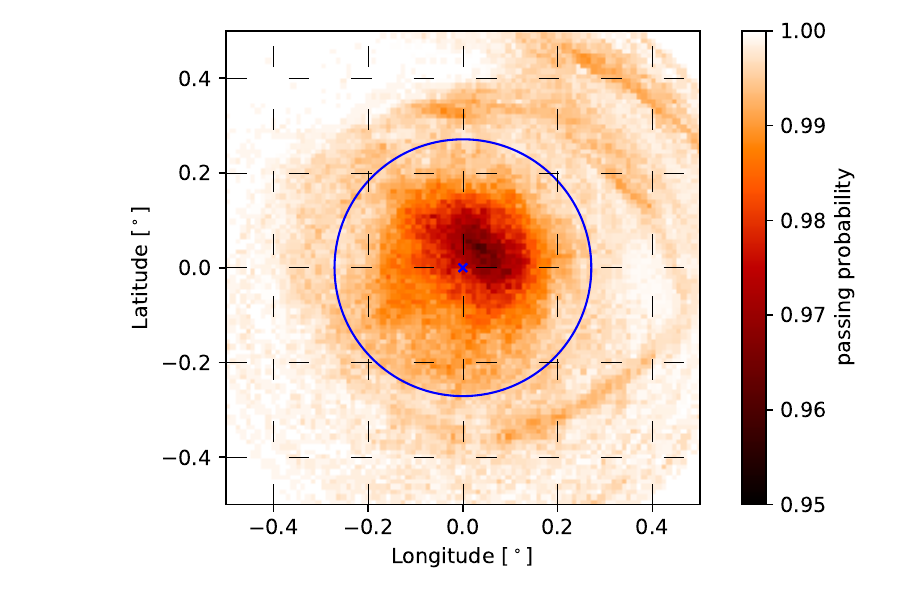} &
  \end{tabular}
  \caption{Calculated Sun shadow from 2007 through 2017 for iron nuclei only. We call attention to the narrow color scale.}
  \label{fig:sunshadow_2007_to_2017_iron_only}
\end{figure*}

In Figs.~\ref{fig:sunshadow_2007_to_2017_proton_only} and \ref{fig:sunshadow_2007_to_2017_iron_only}, the Sun shadow can be seen for proton and iron nuclei only. 

\clearpage
\subsection{Sun shadow as a function of energy}
\begin{figure*}[htbp]
  \centering
  \begin{tabular}{ccc} 
    \includegraphics[trim = 22mm 4mm 4mm 3mm, clip, width=0.32\linewidth]{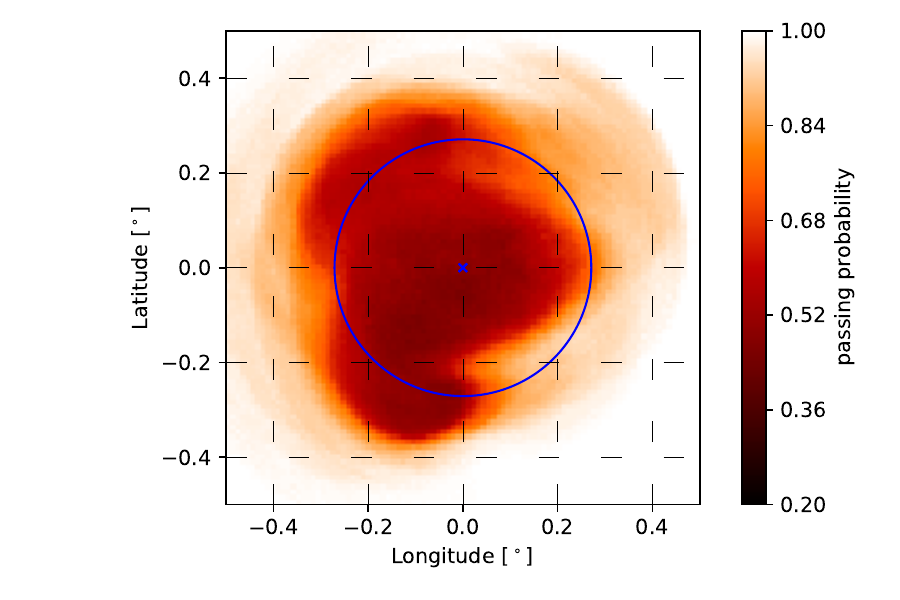} & 
    \includegraphics[trim = 22mm 4mm 4mm 3mm, clip, width=0.32\linewidth]{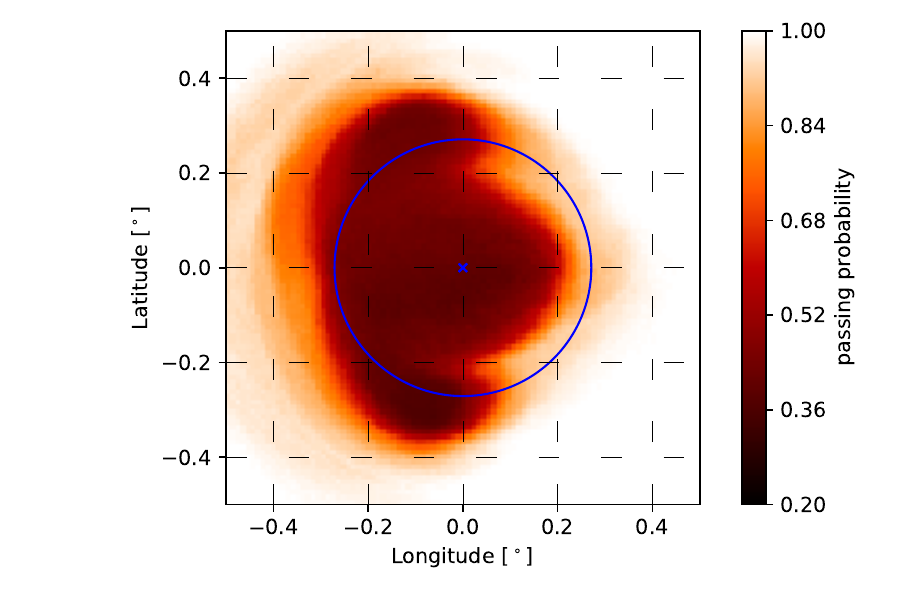} &
    \includegraphics[trim = 22mm 4mm 4mm 3mm, clip, width=0.32\linewidth]{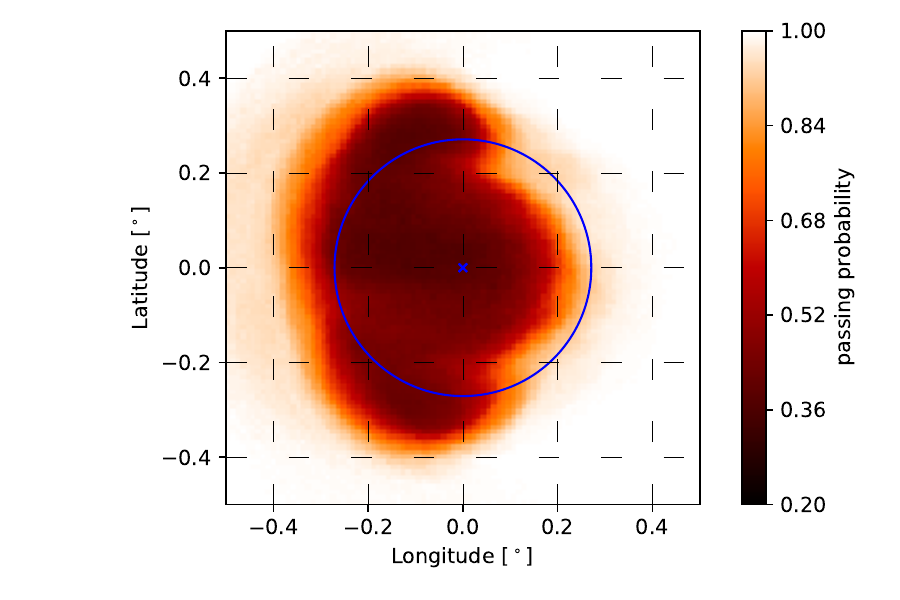} \\ 
    \includegraphics[trim = 22mm 4mm 4mm 3mm, clip, width=0.32\linewidth]{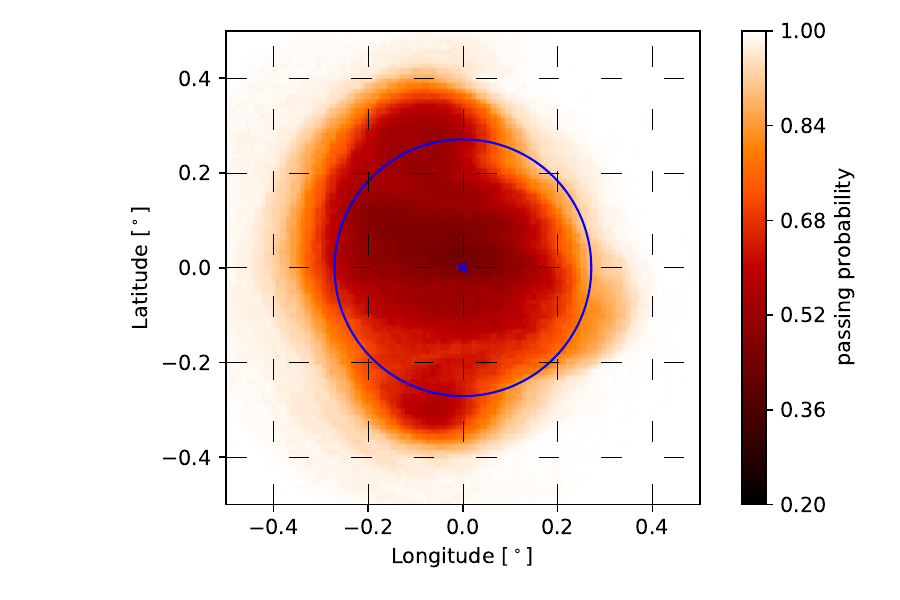} & 
    \includegraphics[trim = 22mm 4mm 4mm 3mm, clip, width=0.32\linewidth]{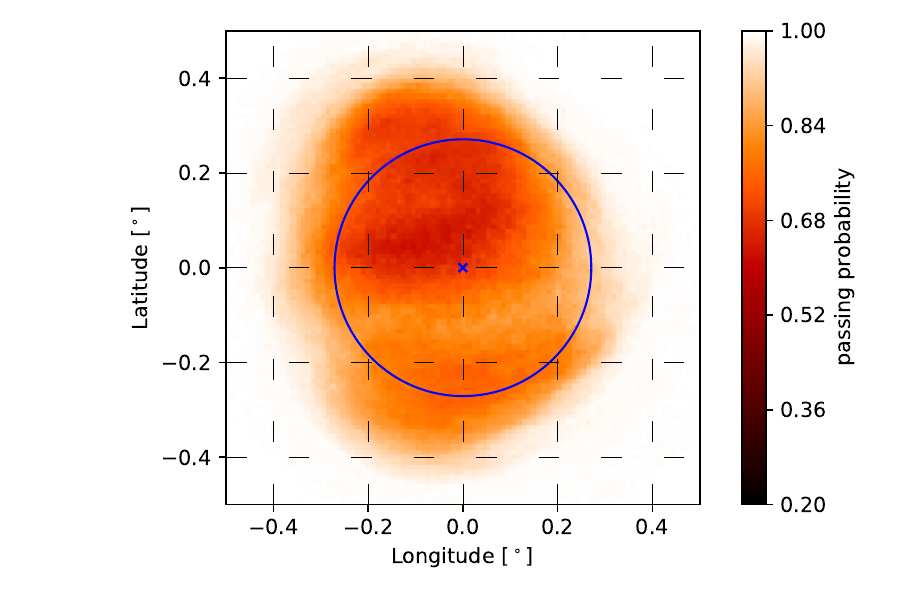} &
    \includegraphics[trim = 22mm 4mm 4mm 3mm, clip, width=0.32\linewidth]{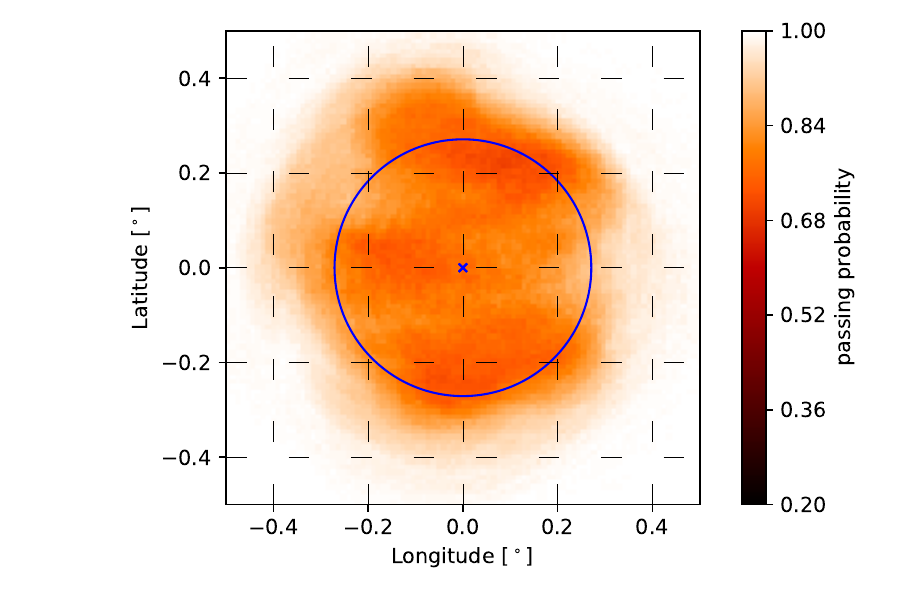} \\ 
    \includegraphics[trim = 22mm 4mm 4mm 3mm, clip, width=0.32\linewidth]{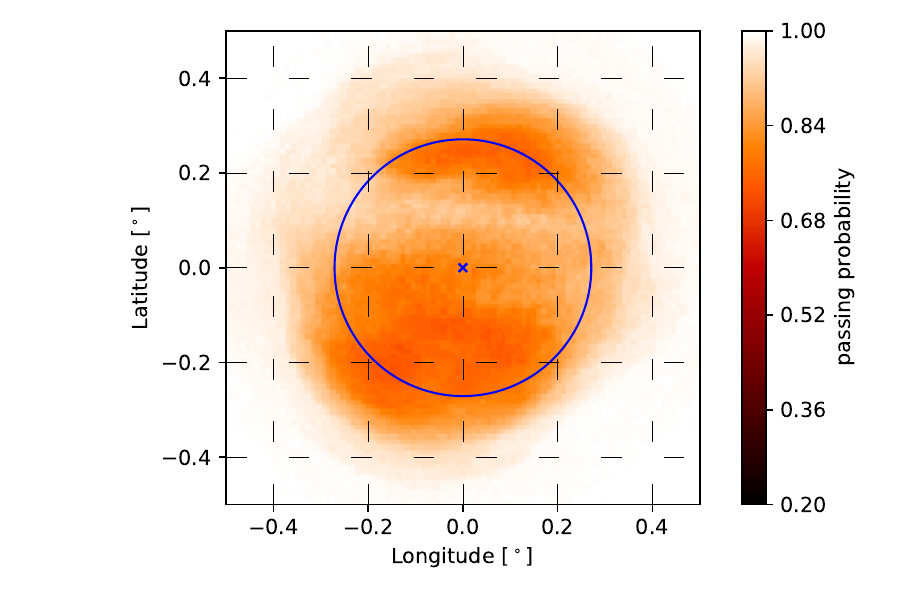} & 
    \includegraphics[trim = 22mm 4mm 4mm 3mm, clip, width=0.32\linewidth]{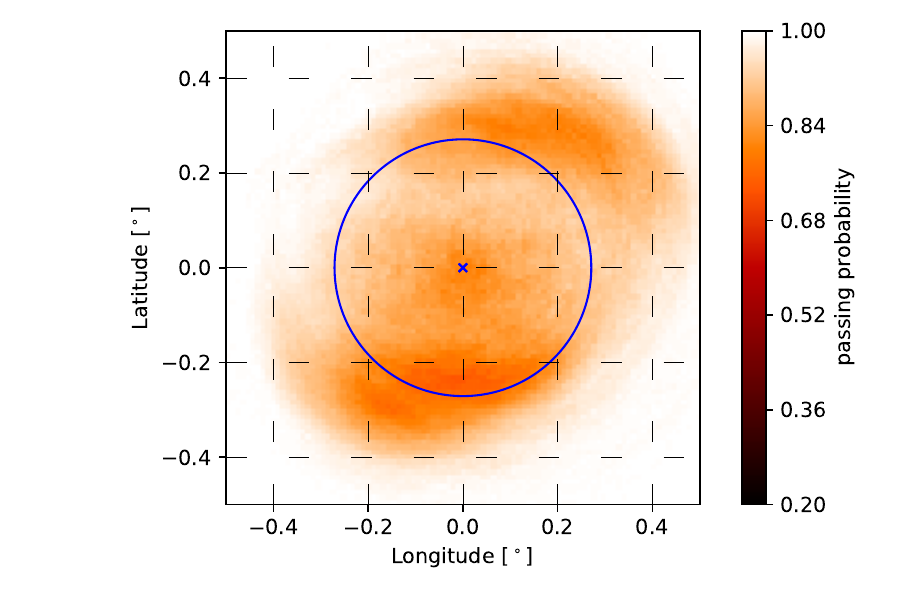} &
    \includegraphics[trim = 22mm 4mm 4mm 3mm, clip, width=0.32\linewidth]{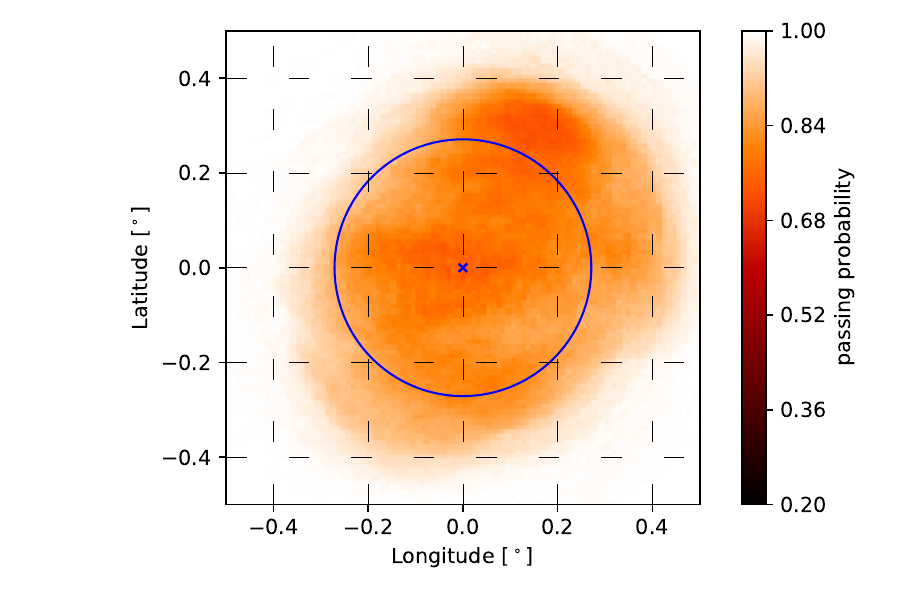} \\ & 
    \includegraphics[trim = 22mm 4mm 4mm 3mm, clip, width=0.32\linewidth]{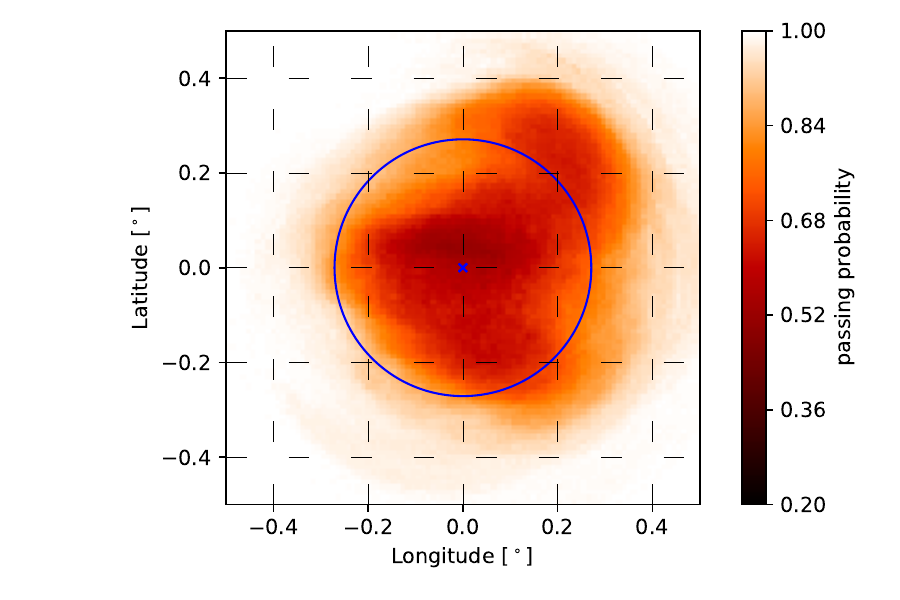} &
  \end{tabular}
  \caption{Calculated Sun shadow from 2007 through 2017 for $10\,\mathrm{TeV}$. Each plot contains contributions by all five elements that were simulated with a composition according to the HGm model.}
  \label{fig:sunshadow_2007_to_2017_10tev}
\end{figure*}

\begin{figure*}[htbp]
  \centering
  \begin{tabular}{ccc} 
    \includegraphics[trim = 22mm 4mm 4mm 3mm, clip, width=0.32\linewidth]{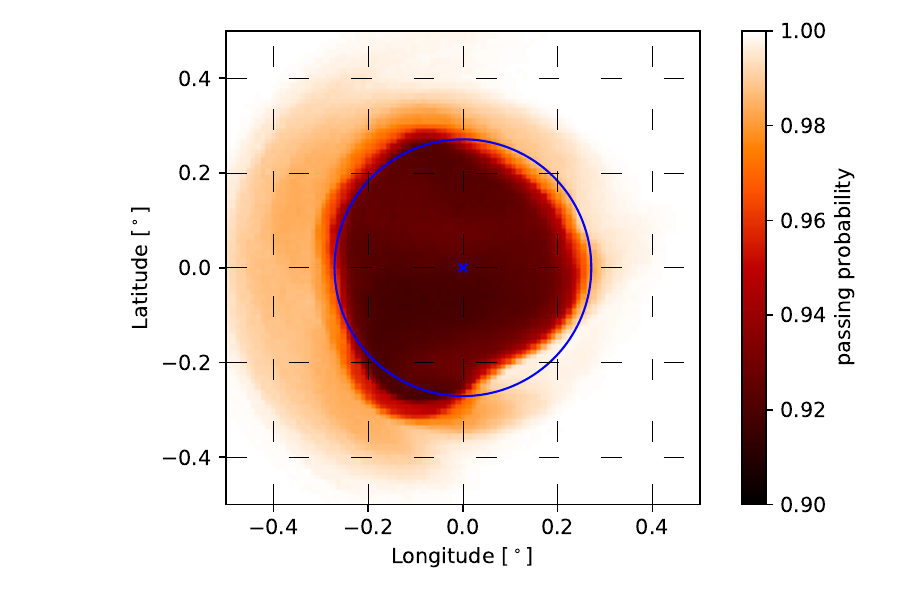} & 
    \includegraphics[trim = 22mm 4mm 4mm 3mm, clip, width=0.32\linewidth]{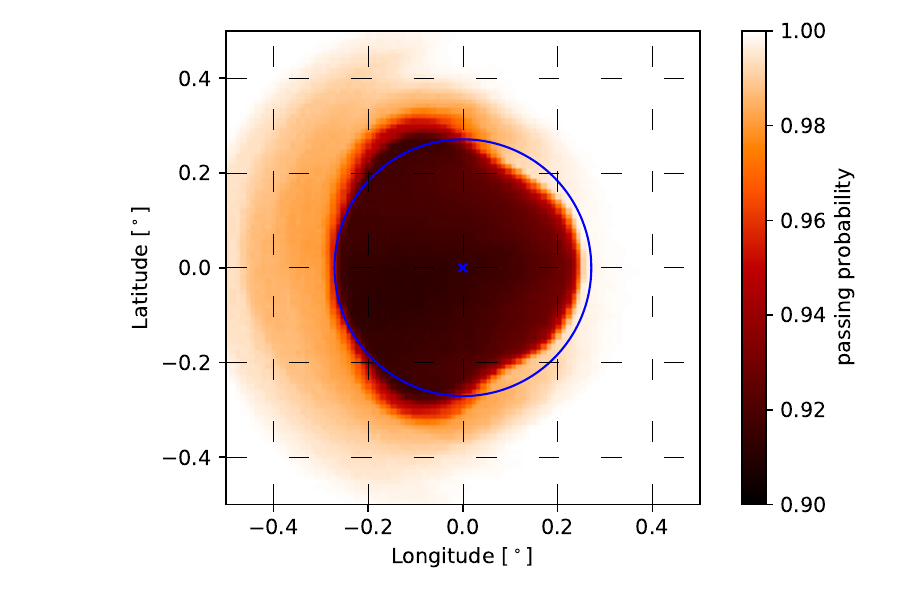} &
    \includegraphics[trim = 22mm 4mm 4mm 3mm, clip, width=0.32\linewidth]{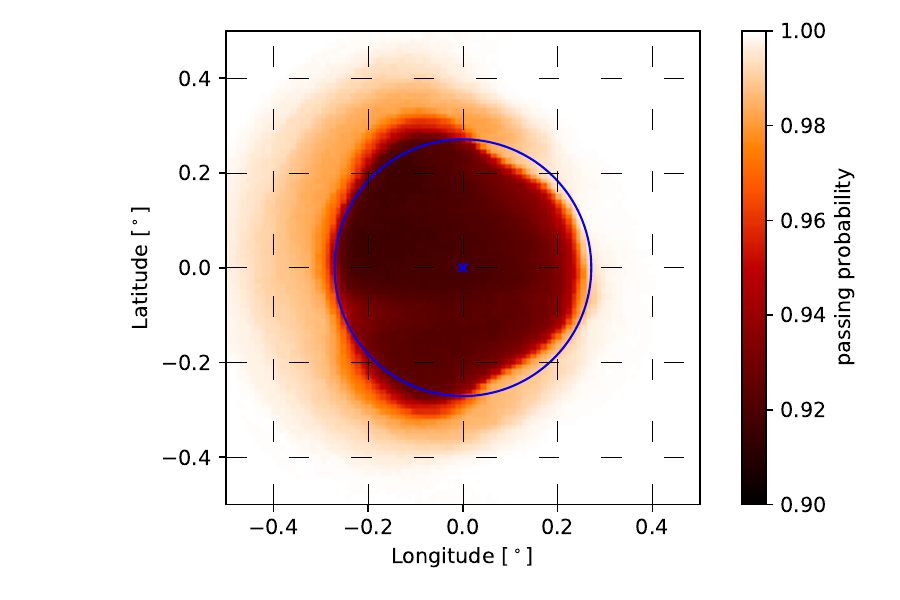} \\ 
    \includegraphics[trim = 22mm 4mm 4mm 3mm, clip, width=0.32\linewidth]{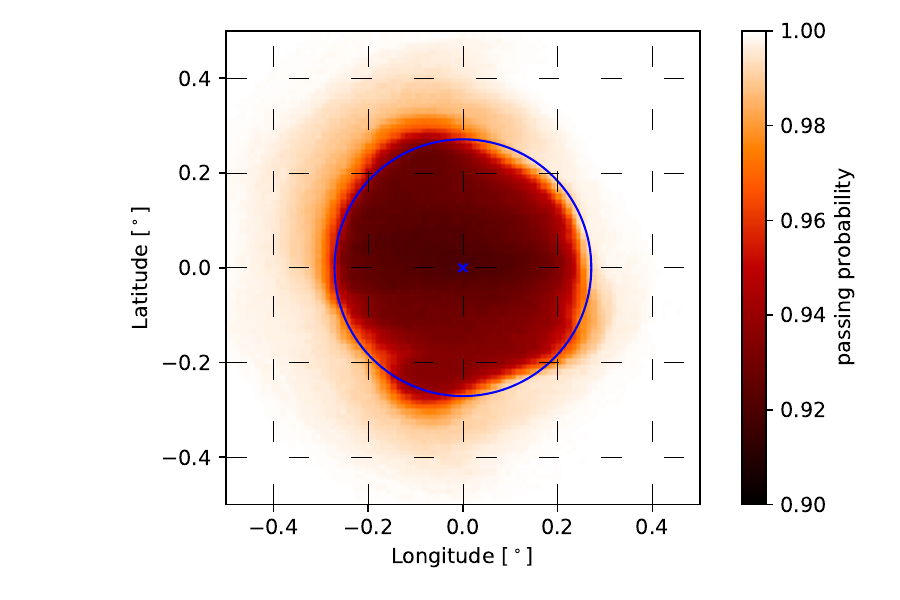} & 
    \includegraphics[trim = 22mm 4mm 4mm 3mm, clip, width=0.32\linewidth]{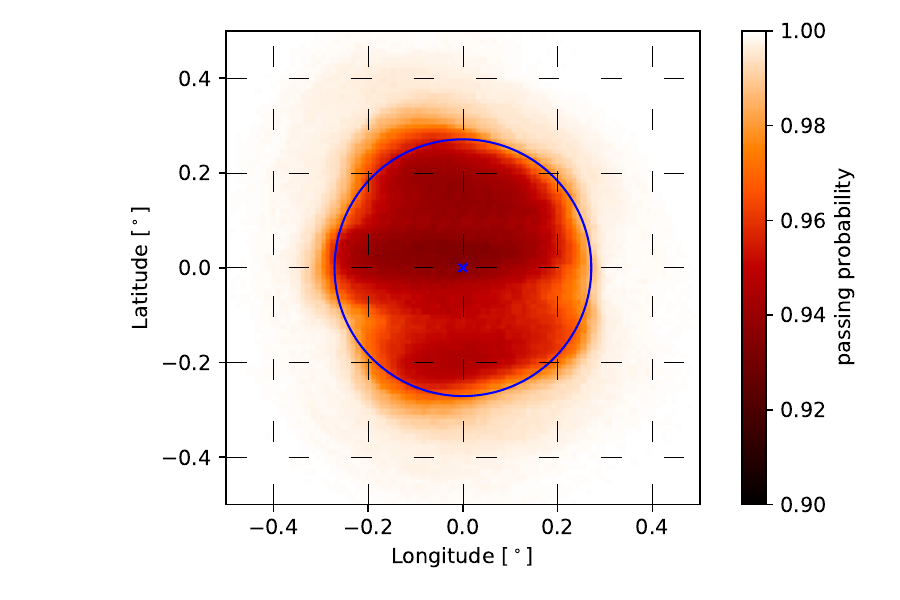} &
    \includegraphics[trim = 22mm 4mm 4mm 3mm, clip, width=0.32\linewidth]{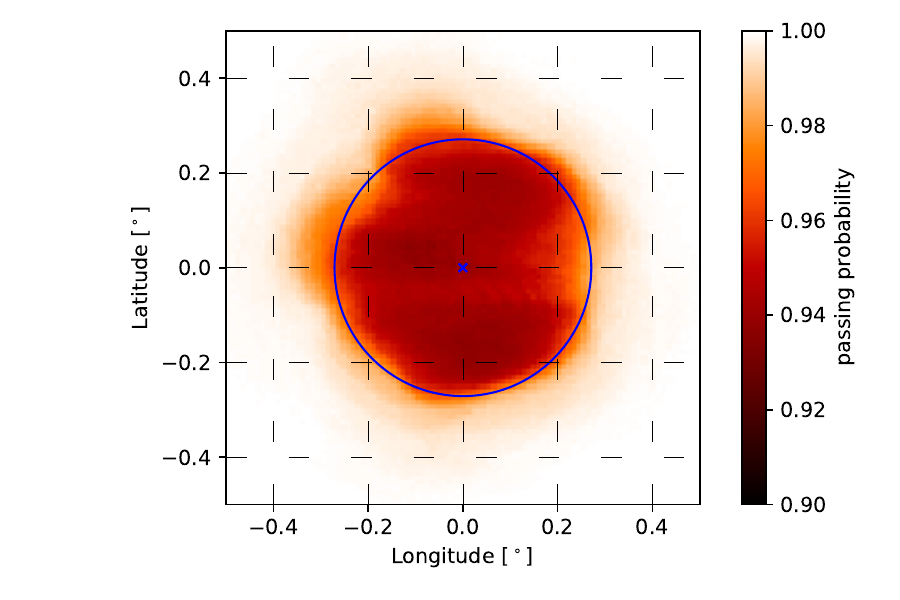} \\ 
    \includegraphics[trim = 22mm 4mm 4mm 3mm, clip, width=0.32\linewidth]{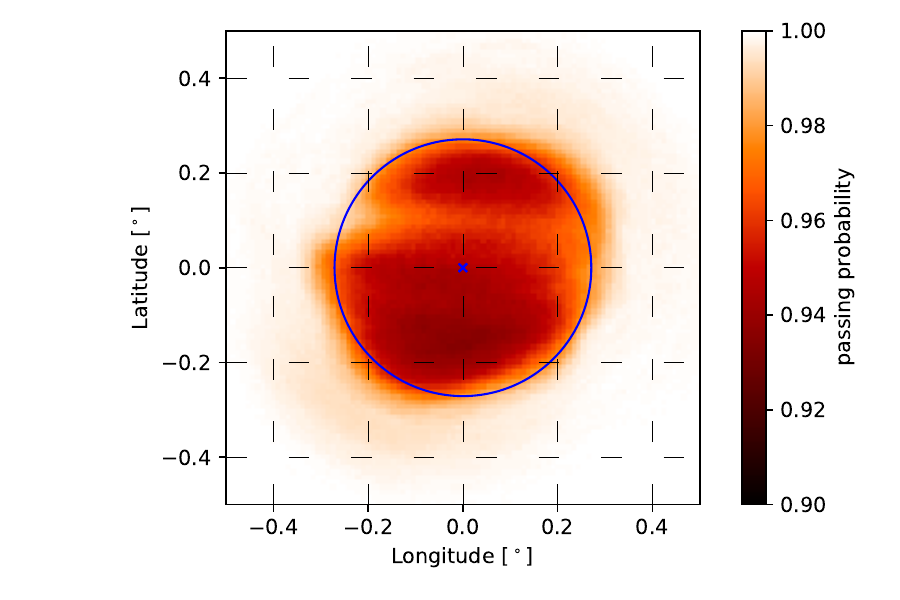} & 
    \includegraphics[trim = 22mm 4mm 4mm 3mm, clip, width=0.32\linewidth]{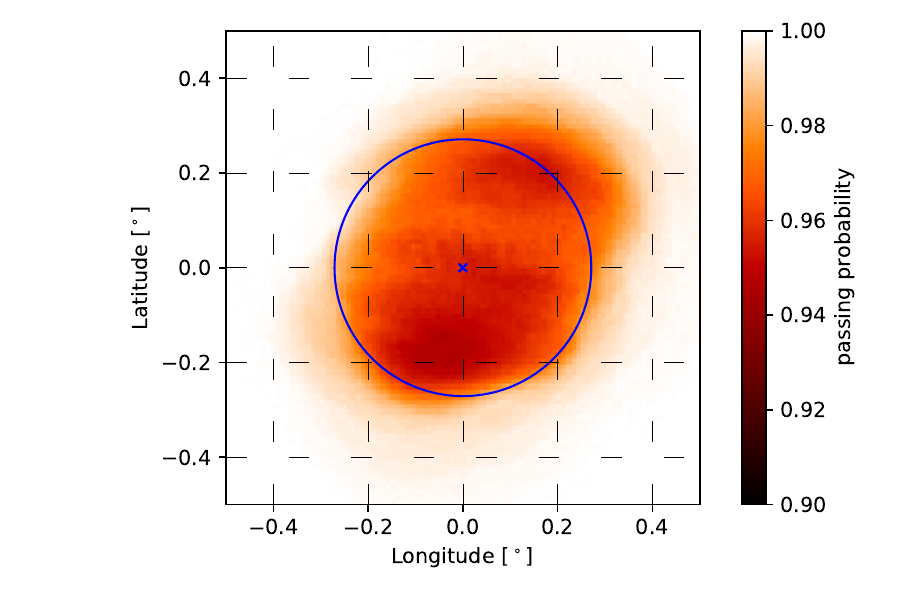} &
    \includegraphics[trim = 22mm 4mm 4mm 3mm, clip, width=0.32\linewidth]{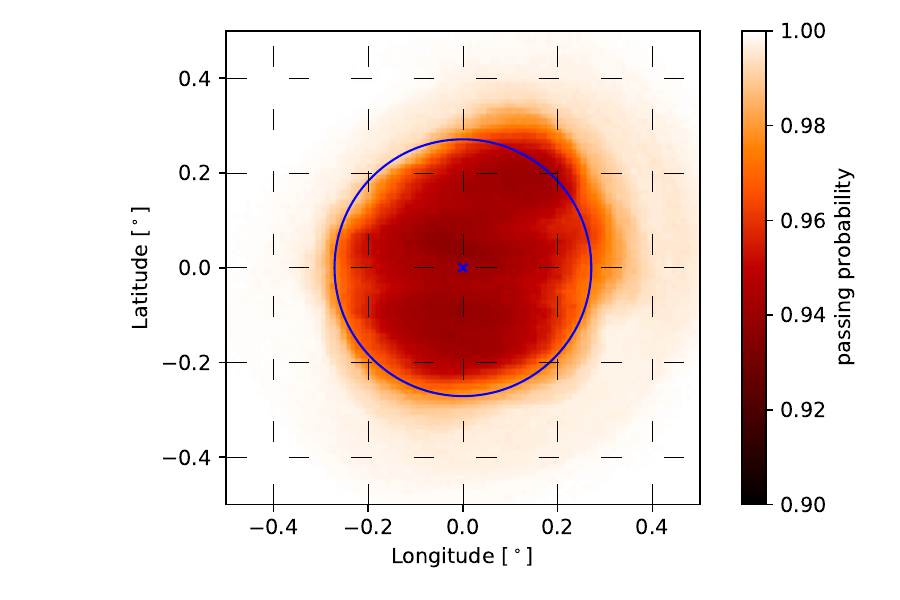} \\ & 
    \includegraphics[trim = 22mm 4mm 4mm 3mm, clip, width=0.32\linewidth]{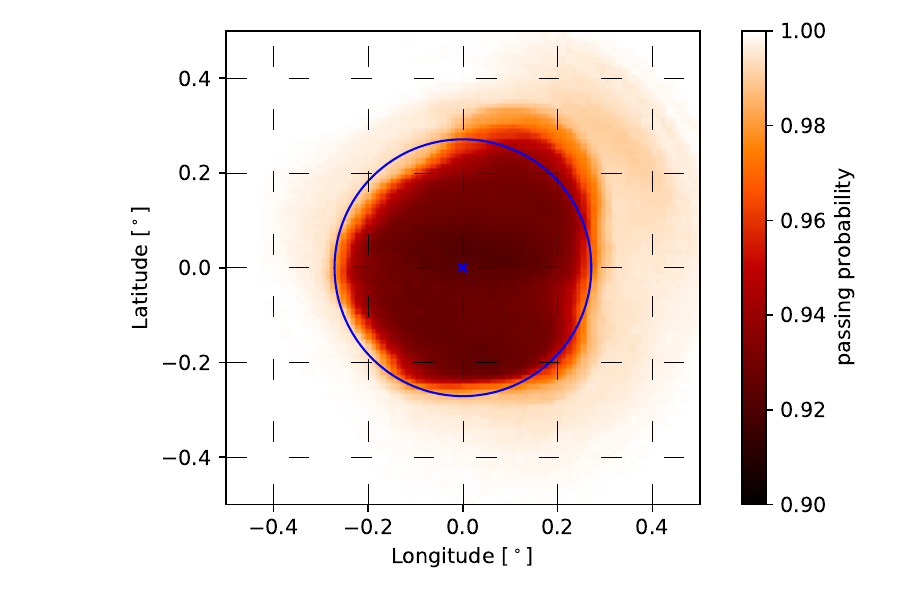} &
  \end{tabular}
  \caption{Calculated Sun shadow from 2007 through 2017 for $40\,\mathrm{TeV}$.
    Each plot contains contributions by all five elements that were simulated with a composition according to the HGm model.}
  \label{fig:sunshadow_2007_to_2017_40tev}
\end{figure*}

\begin{figure*}[htbp]
  \centering
  \begin{tabular}{ccc} 
    \includegraphics[trim = 22mm 4mm 4mm 3mm, clip, width=0.32\linewidth]{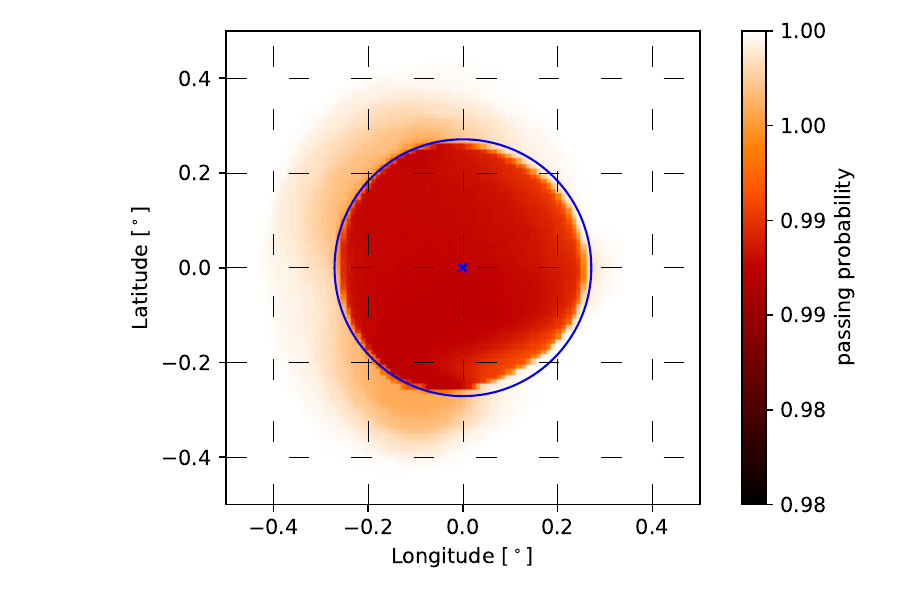} & 
    \includegraphics[trim = 22mm 4mm 4mm 3mm, clip, width=0.32\linewidth]{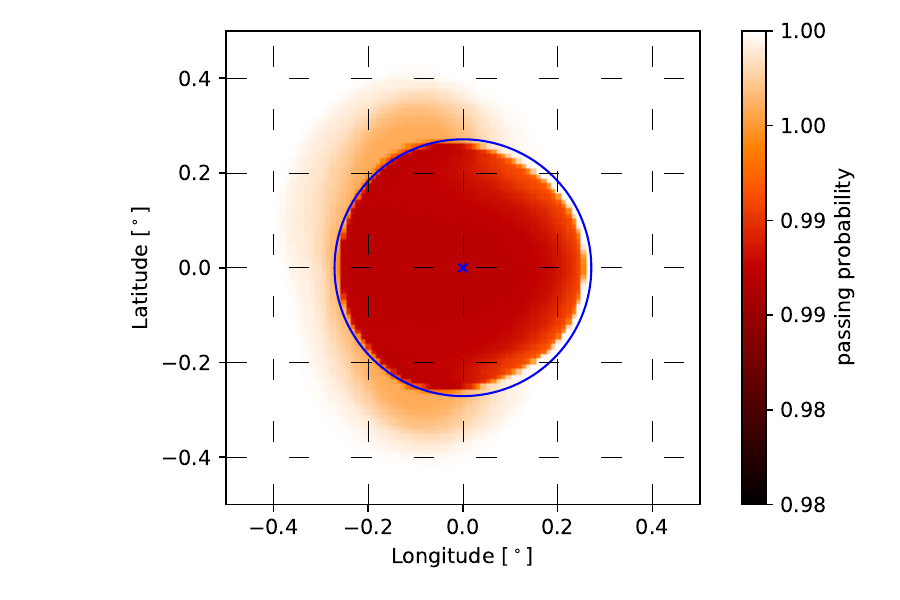} &
    \includegraphics[trim = 22mm 4mm 4mm 3mm, clip, width=0.32\linewidth]{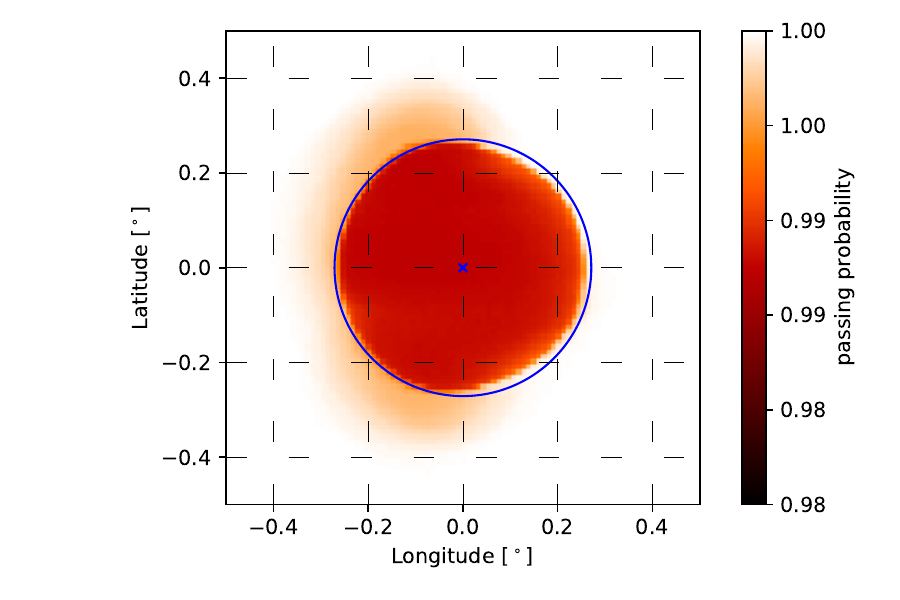} \\ 
    \includegraphics[trim = 22mm 4mm 4mm 3mm, clip, width=0.32\linewidth]{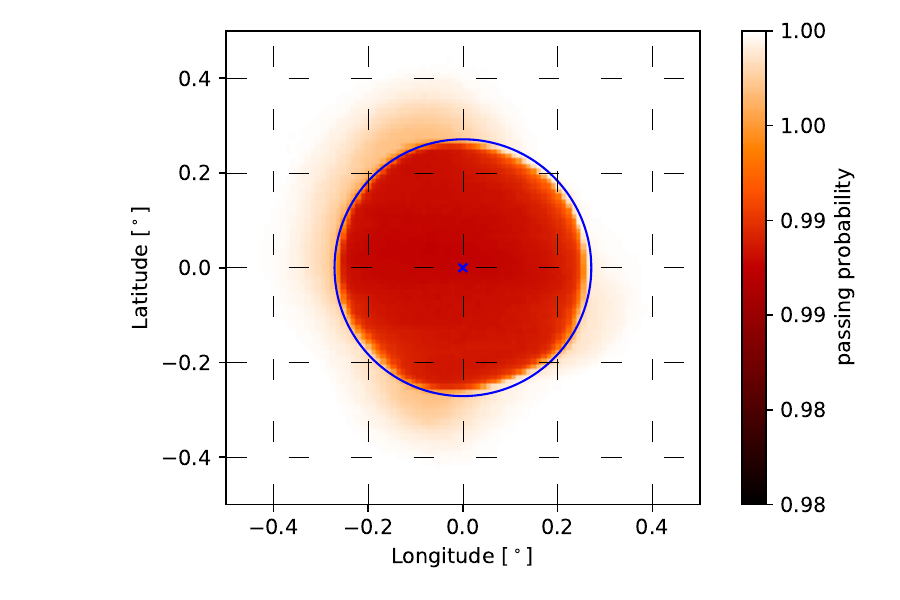} & 
    \includegraphics[trim = 22mm 4mm 4mm 3mm, clip, width=0.32\linewidth]{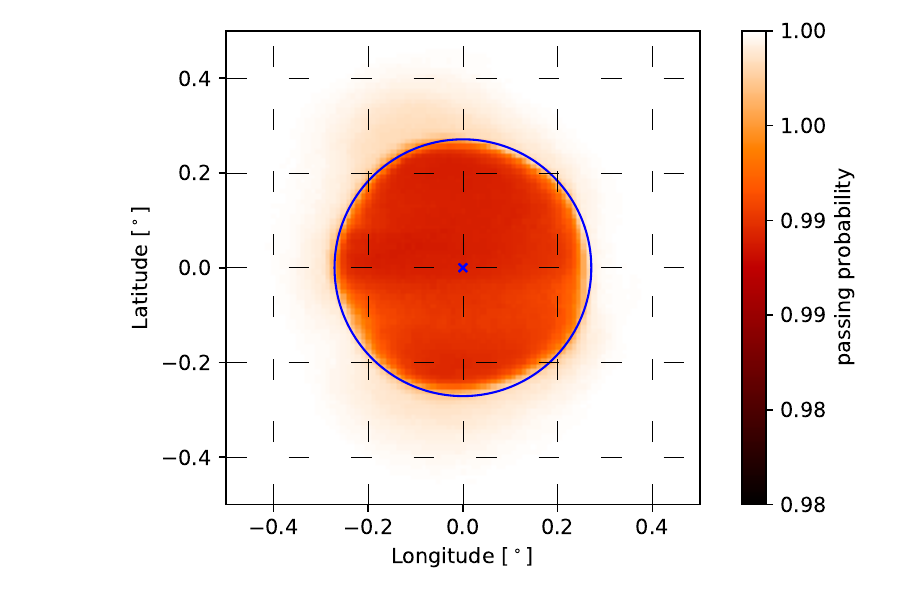} &
    \includegraphics[trim = 22mm 4mm 4mm 3mm, clip, width=0.32\linewidth]{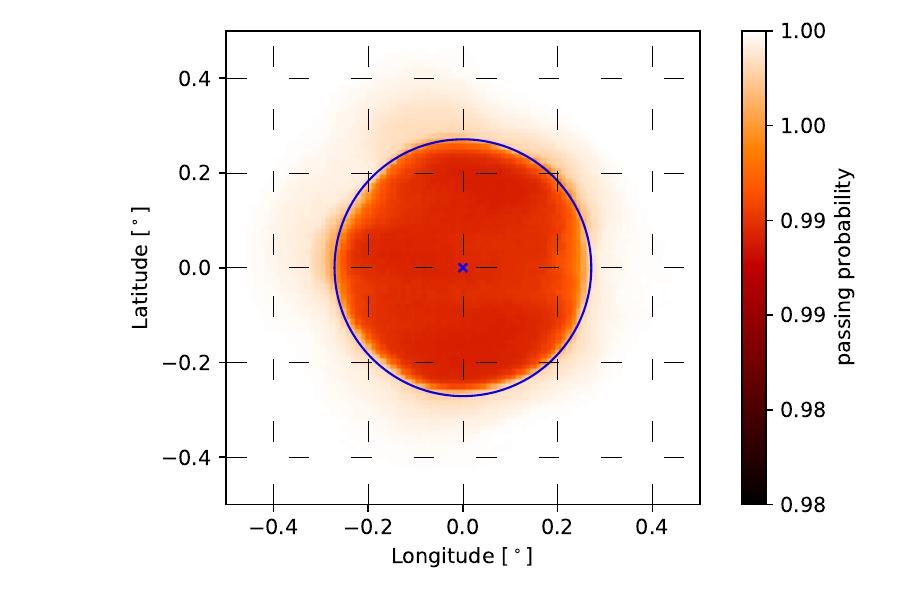} \\ 
    \includegraphics[trim = 22mm 4mm 4mm 3mm, clip, width=0.32\linewidth]{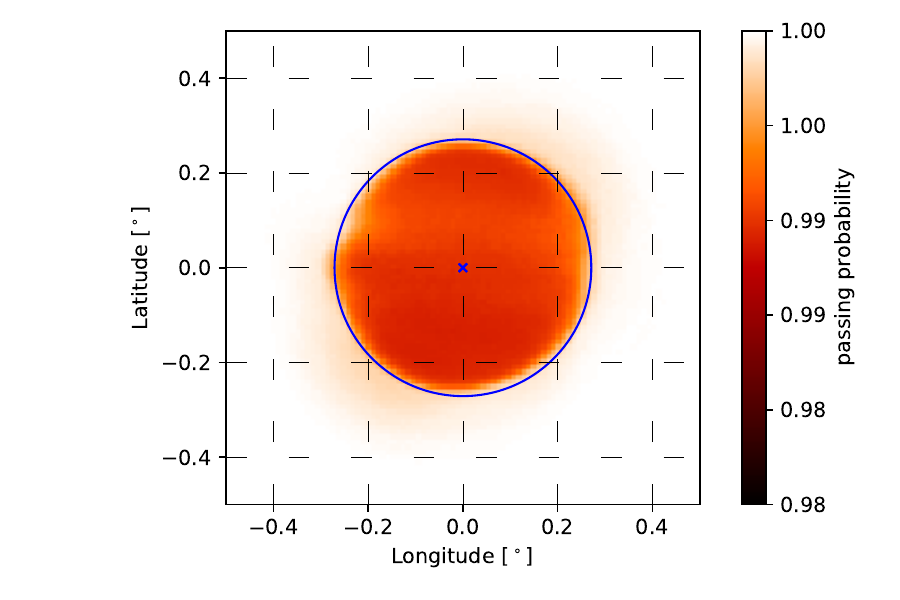} & 
    \includegraphics[trim = 22mm 4mm 4mm 3mm, clip, width=0.32\linewidth]{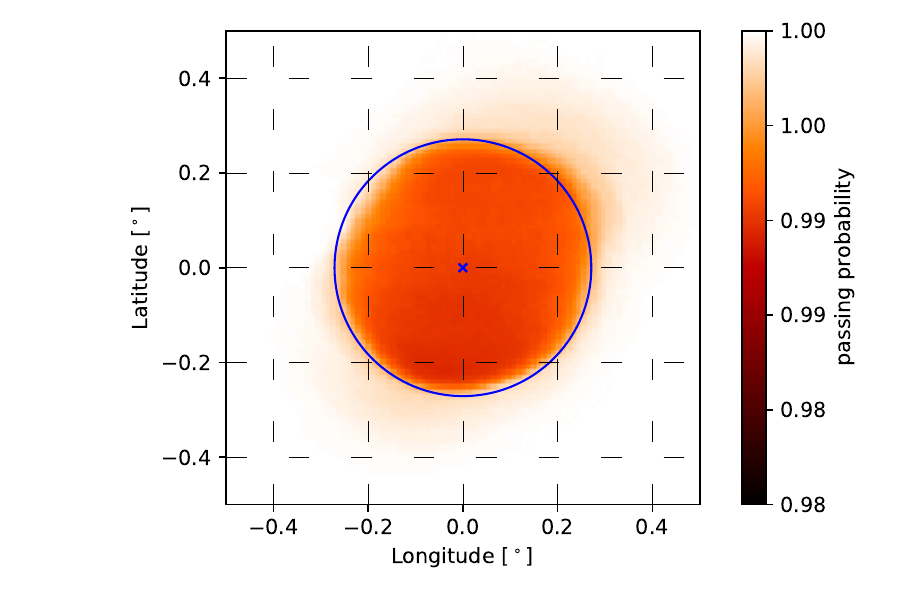} &
    \includegraphics[trim = 22mm 4mm 4mm 3mm, clip, width=0.32\linewidth]{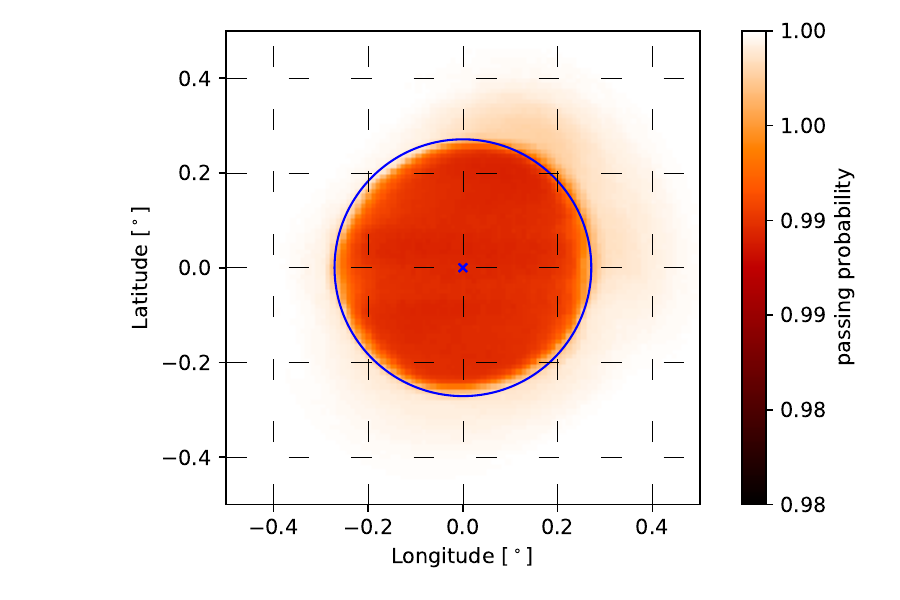} \\ & 
    \includegraphics[trim = 22mm 4mm 4mm 3mm, clip, width=0.32\linewidth]{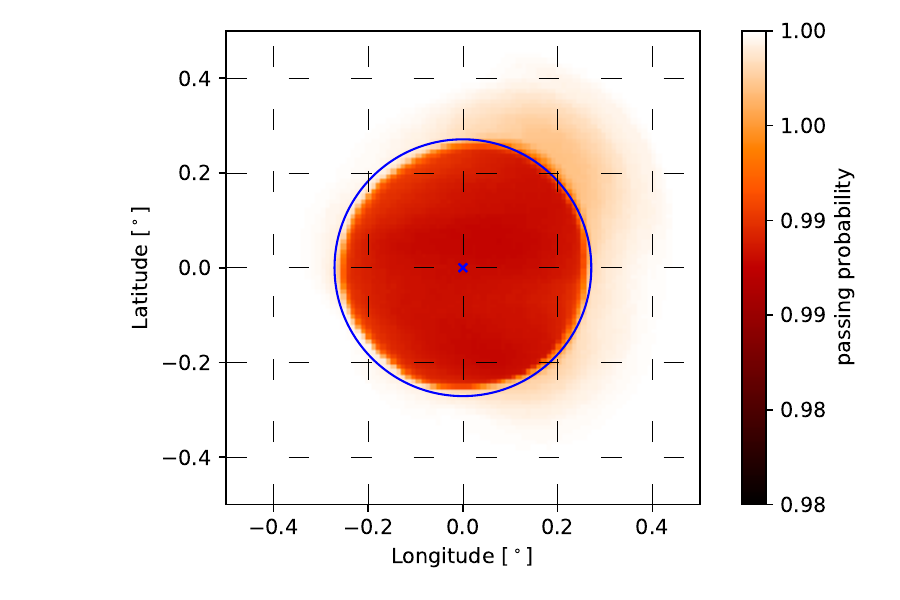} &
  \end{tabular}
  \caption{Calculated Sun shadow from 2007 through 2017 for $158\,\mathrm{TeV}$.
    Each plot contains contributions by all five elements that were simulated with a composition according to the HGm model.}
  \label{fig:sunshadow_2007_to_2017_160tev}
\end{figure*}

In Figs.~\ref{fig:sunshadow_2007_to_2017_10tev}--\ref{fig:sunshadow_2007_to_2017_160tev}, the Sun shadow can be seen at $10$, $40$ and $158\,\mathrm{TeV}$, respectively.

\clearpage
\subsection{Sun shadow for different cosmic-ray flux models}
\begin{figure*}[htbp]
  \centering
  \begin{tabular}{ccc} 
    \includegraphics[trim = 22mm 4mm 4mm 3mm, clip, width=0.32\linewidth]{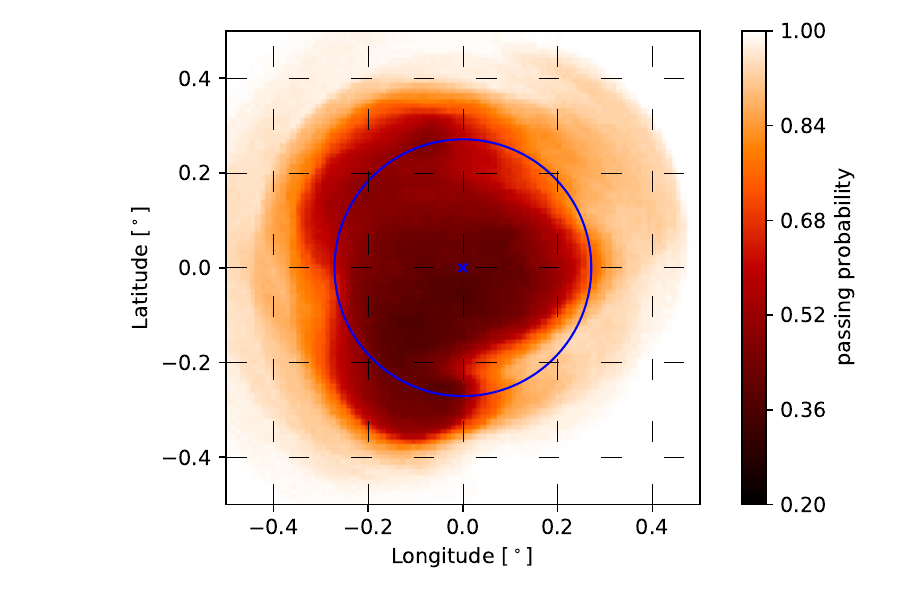} & 
    \includegraphics[trim = 22mm 4mm 4mm 3mm, clip, width=0.32\linewidth]{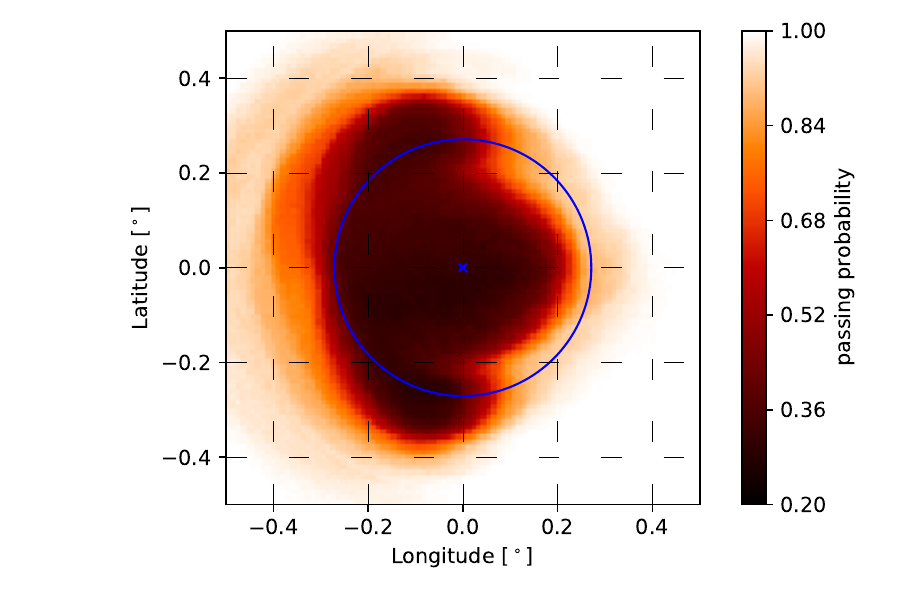} &
    \includegraphics[trim = 22mm 4mm 4mm 3mm, clip, width=0.32\linewidth]{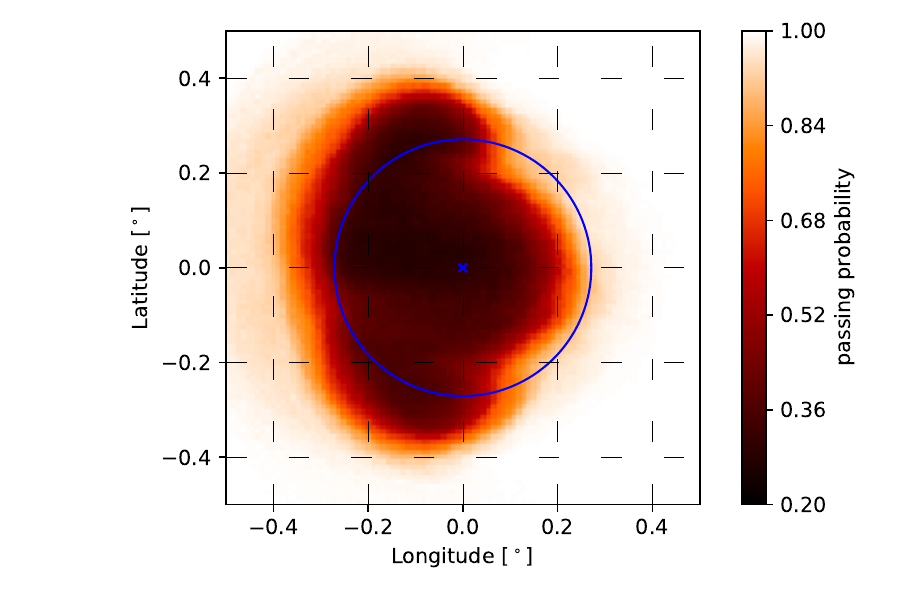} \\ 
    \includegraphics[trim = 22mm 4mm 4mm 3mm, clip, width=0.32\linewidth]{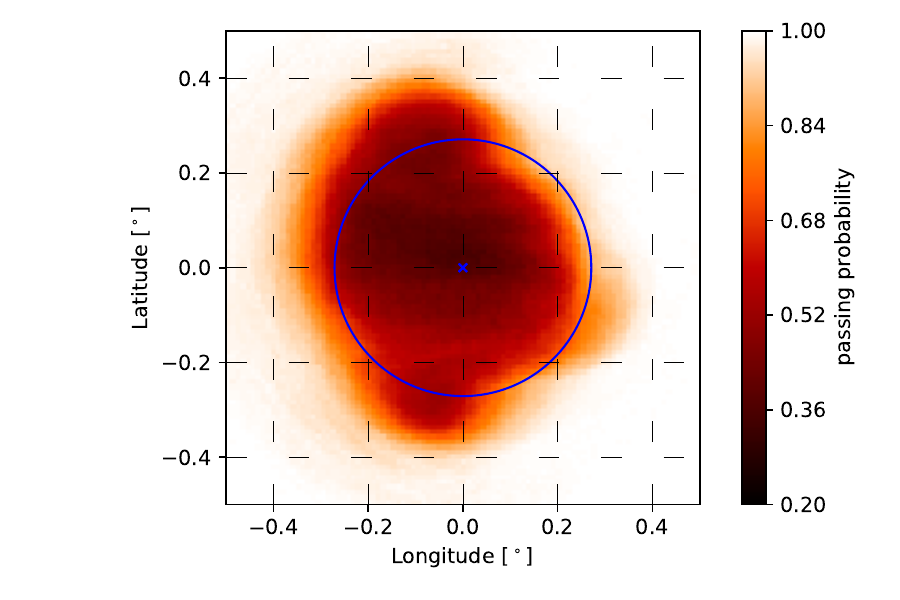} & 
    \includegraphics[trim = 22mm 4mm 4mm 3mm, clip, width=0.32\linewidth]{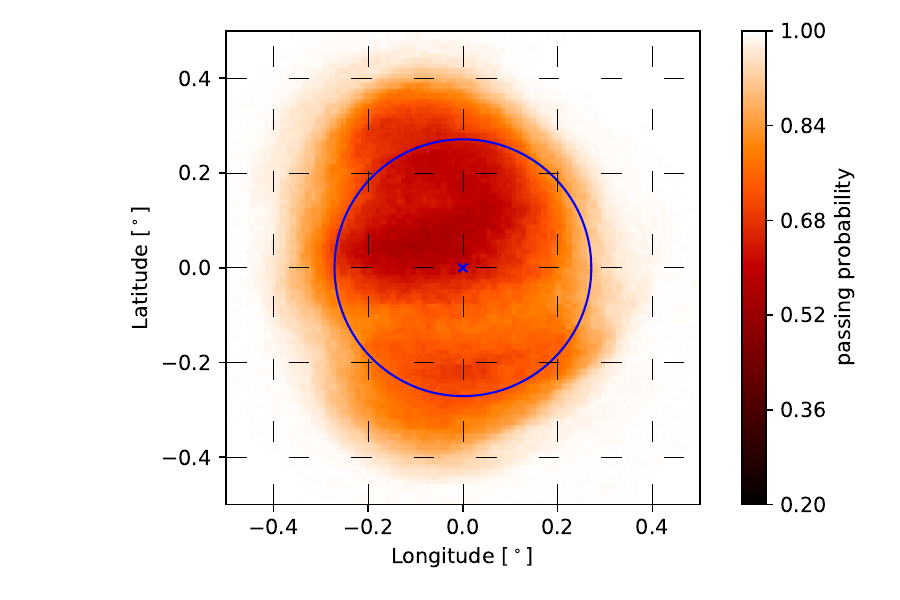} &
    \includegraphics[trim = 22mm 4mm 4mm 3mm, clip, width=0.32\linewidth]{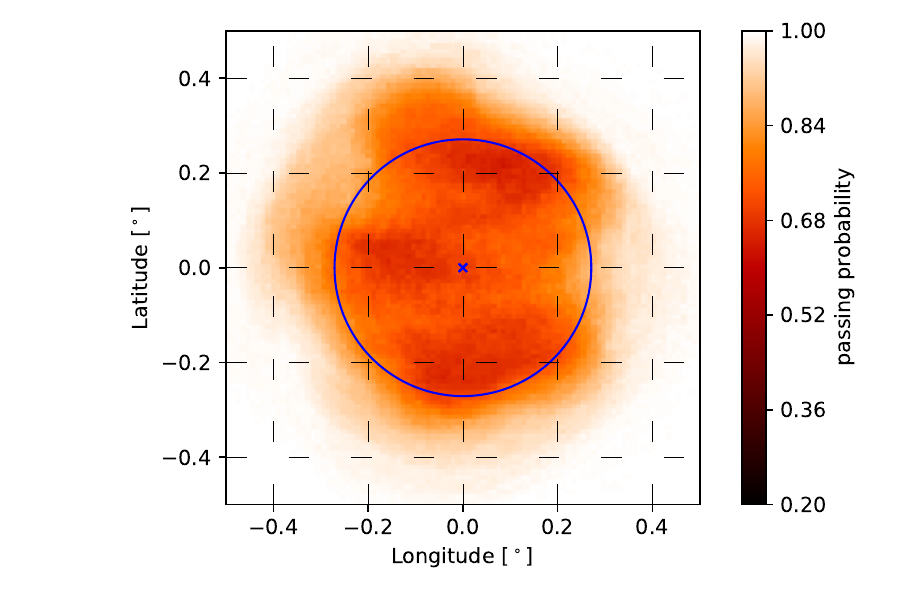} \\ 
    \includegraphics[trim = 22mm 4mm 4mm 3mm, clip, width=0.32\linewidth]{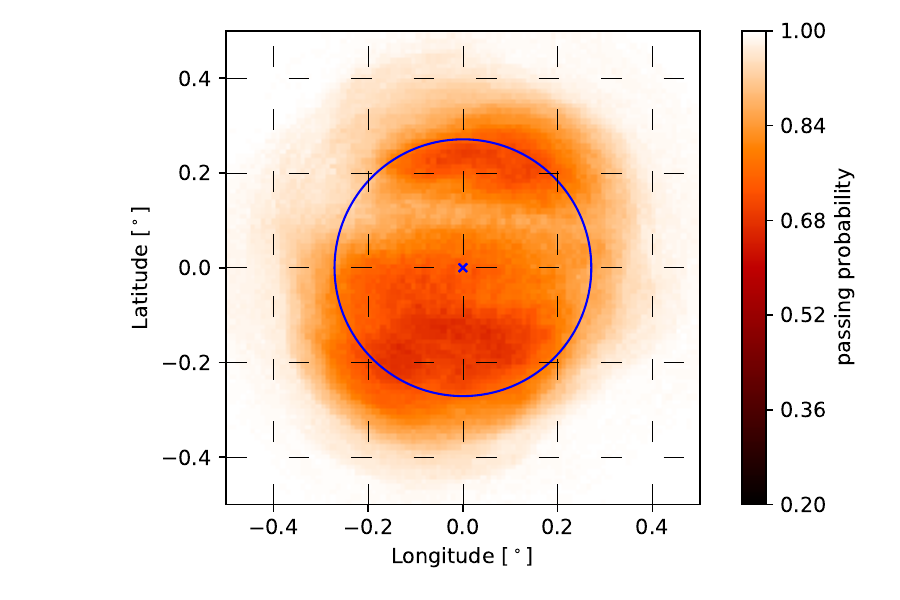} & 
    \includegraphics[trim = 22mm 4mm 4mm 3mm, clip, width=0.32\linewidth]{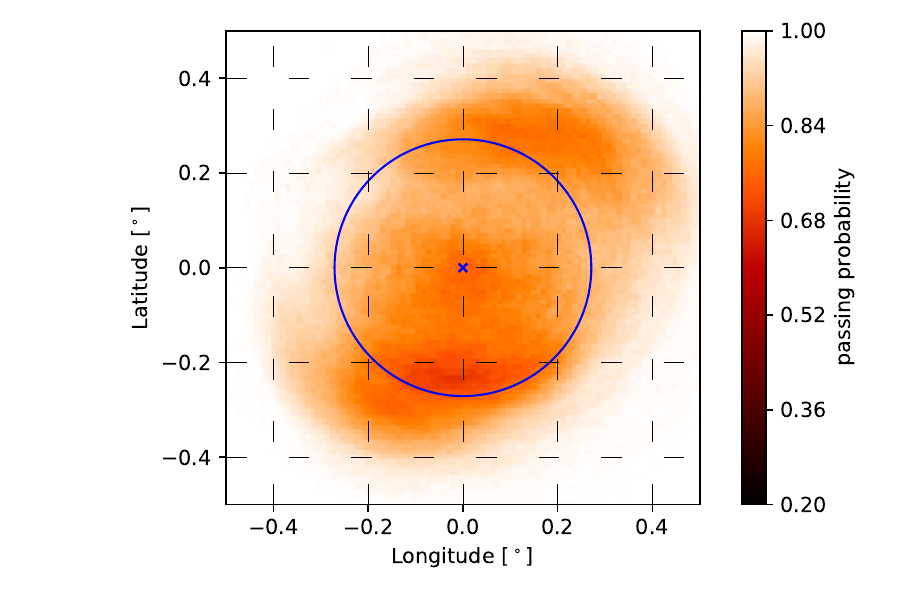} &
    \includegraphics[trim = 22mm 4mm 4mm 3mm, clip, width=0.32\linewidth]{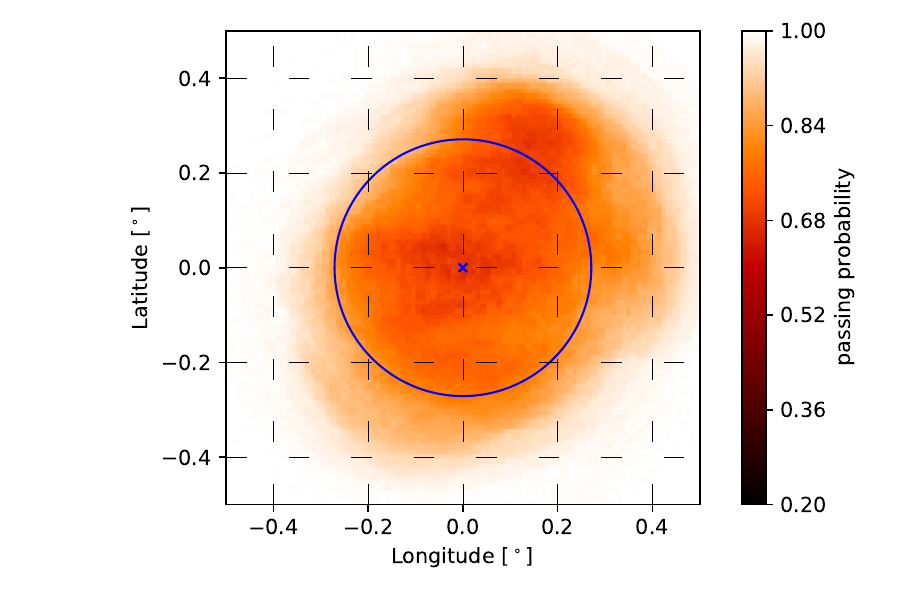} \\ & 
    \includegraphics[trim = 22mm 4mm 4mm 3mm, clip, width=0.32\linewidth]{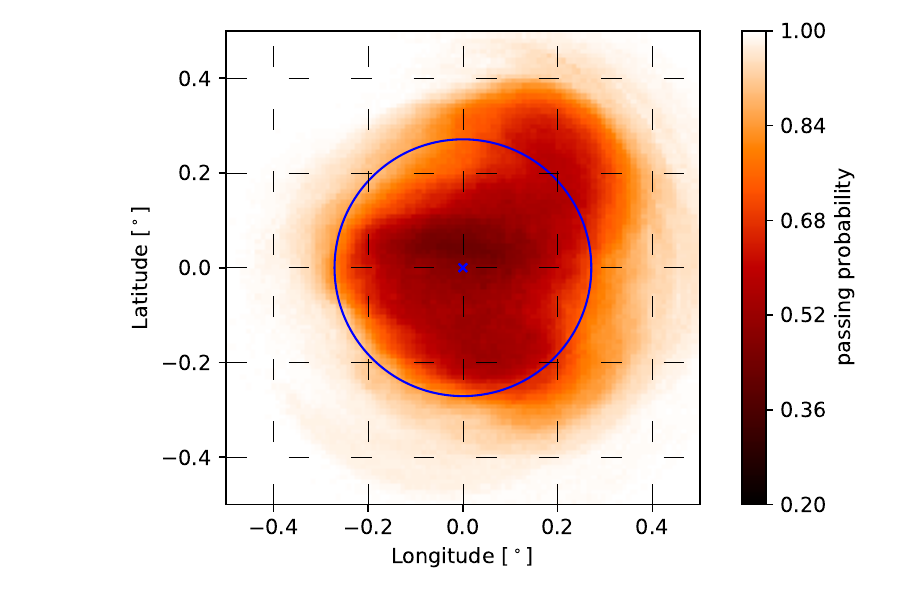} &
  \end{tabular}
  \caption{Calculated Sun shadow from 2007 through 2017 using the energy spectrum and composition according to the HGm model.}
  \label{fig:sunshadow_2007_to_2017_HGm}
\end{figure*}

\begin{figure*}[htbp]
  \centering
  \begin{tabular}{ccc} 
    \includegraphics[trim = 22mm 4mm 4mm 3mm, clip, width=0.32\linewidth]{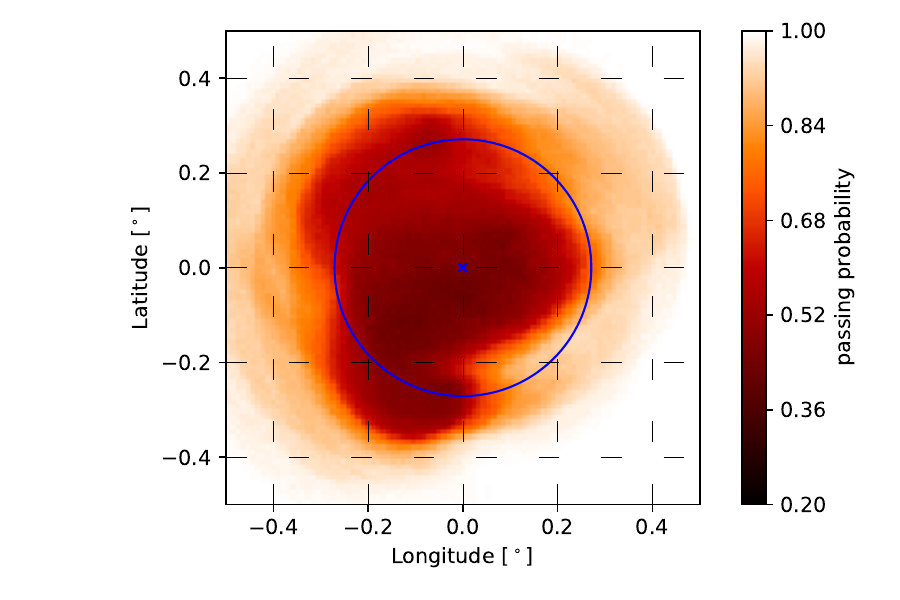} & 
    \includegraphics[trim = 22mm 4mm 4mm 3mm, clip, width=0.32\linewidth]{08_09_only_all_new_GH.pdf} &
    \includegraphics[trim = 22mm 4mm 4mm 3mm, clip, width=0.32\linewidth]{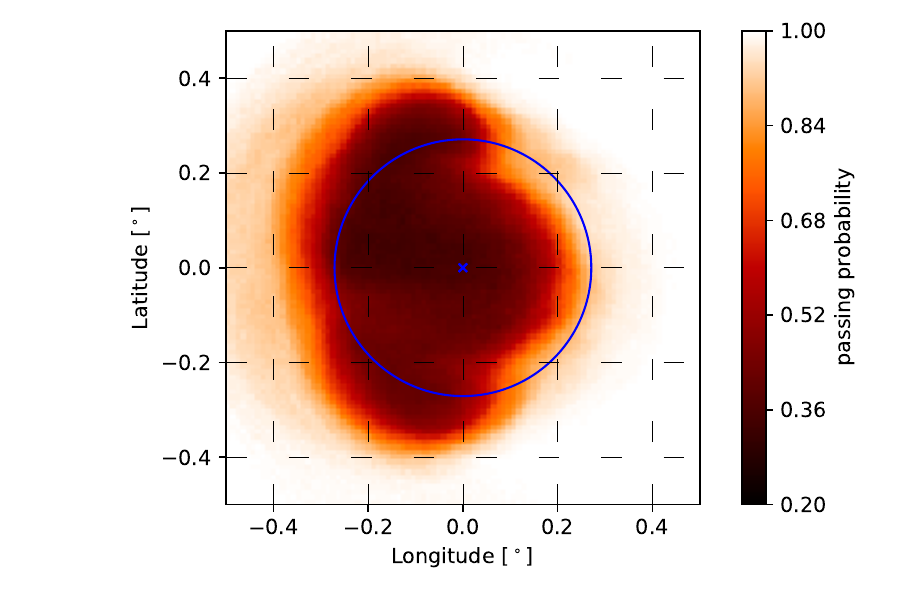} \\ 
    \includegraphics[trim = 22mm 4mm 4mm 3mm, clip, width=0.32\linewidth]{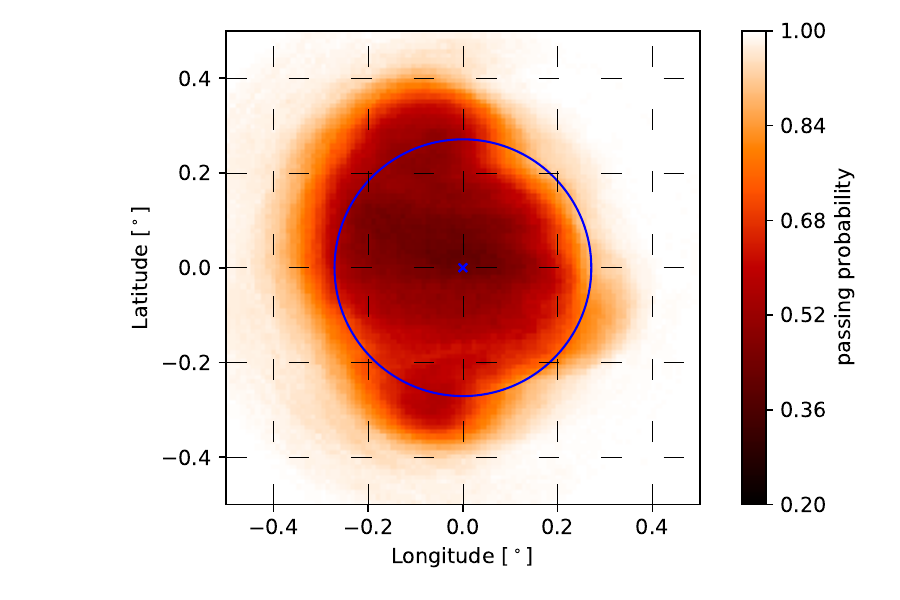} & 
    \includegraphics[trim = 22mm 4mm 4mm 3mm, clip, width=0.32\linewidth]{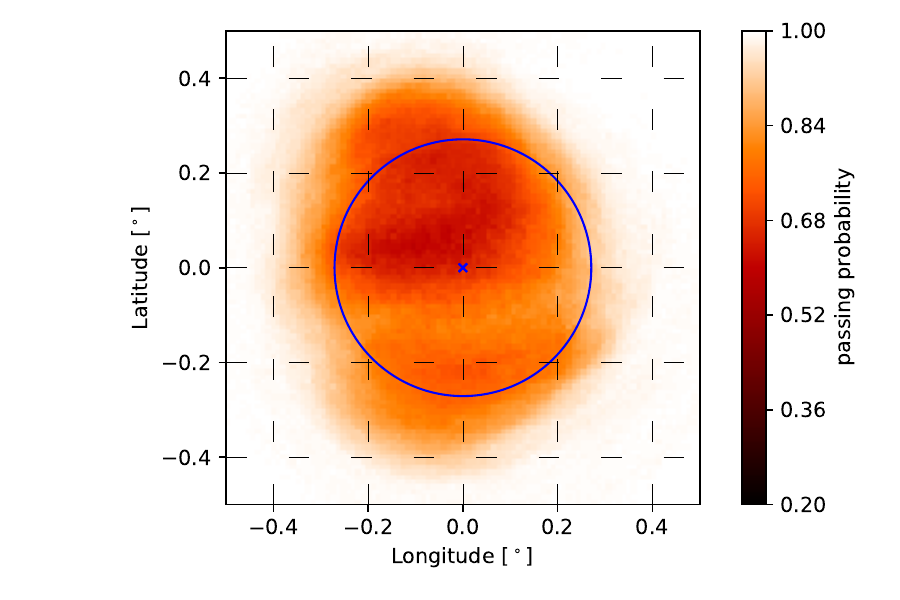} &
    \includegraphics[trim = 22mm 4mm 4mm 3mm, clip, width=0.32\linewidth]{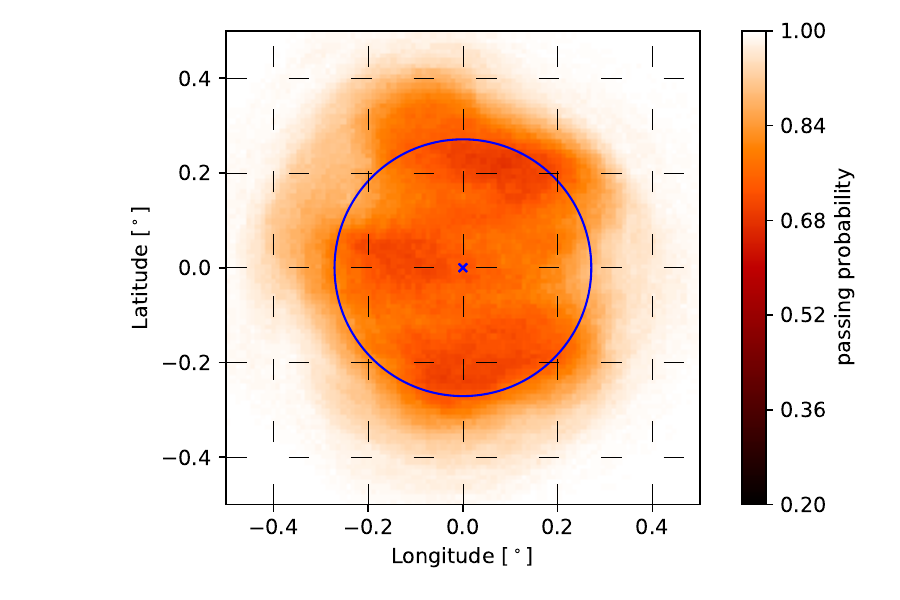} \\ 
    \includegraphics[trim = 22mm 4mm 4mm 3mm, clip, width=0.32\linewidth]{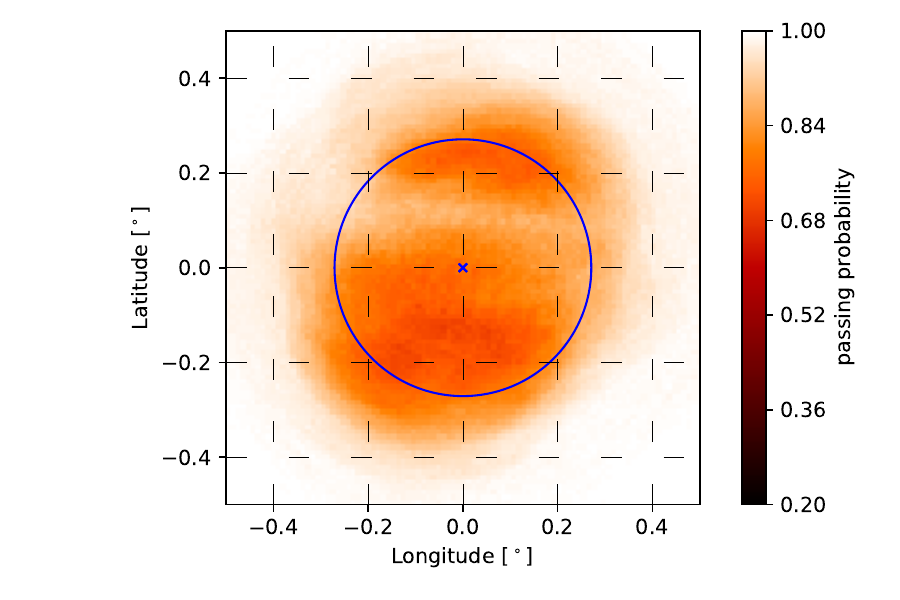} & 
    \includegraphics[trim = 22mm 4mm 4mm 3mm, clip, width=0.32\linewidth]{14_15_only_all_new_GH.pdf} &
    \includegraphics[trim = 22mm 4mm 4mm 3mm, clip, width=0.32\linewidth]{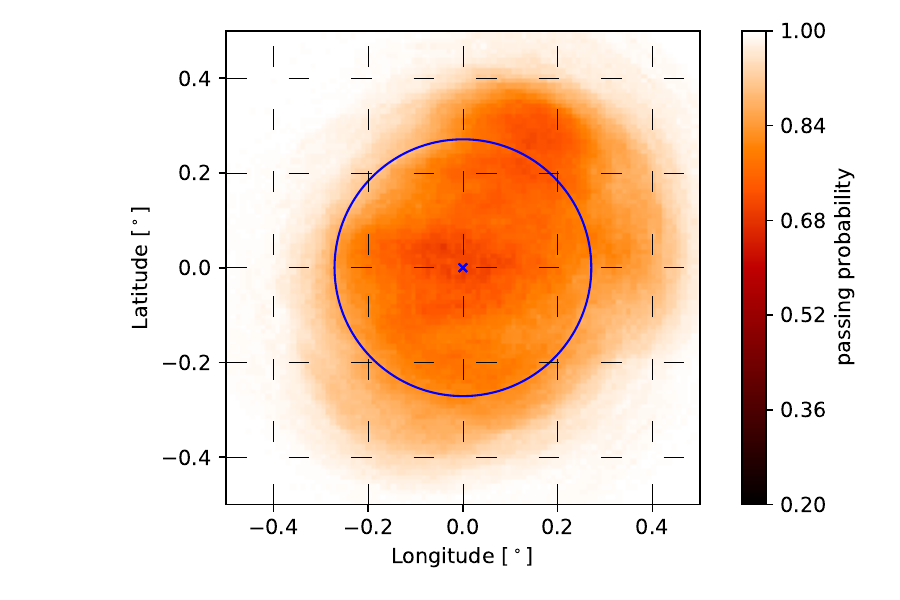} \\ & 
    \includegraphics[trim = 22mm 4mm 4mm 3mm, clip, width=0.32\linewidth]{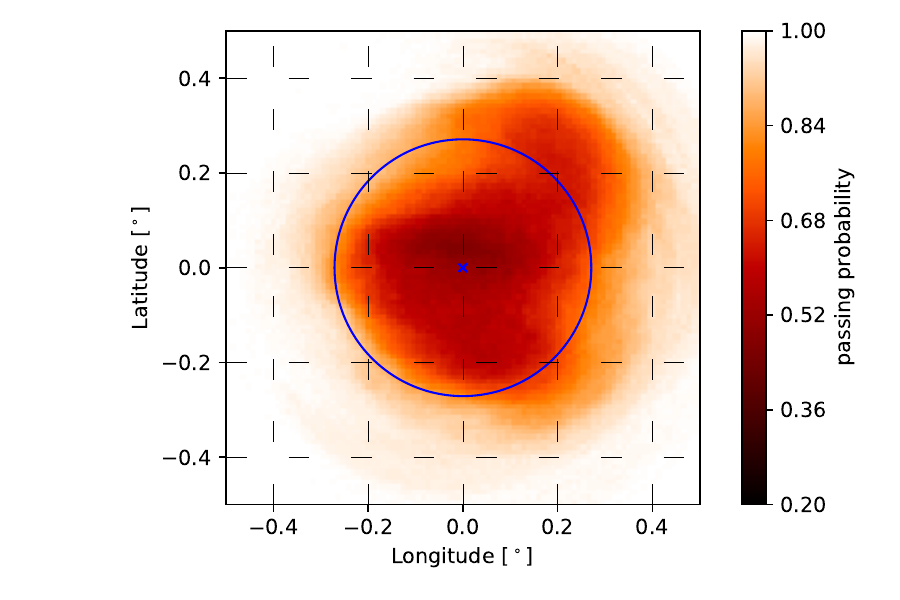} &
  \end{tabular}
  \caption{Calculated Sun shadow from 2007 through 2017 using the energy spectrum and composition according to the GH model.}
  \label{fig:sunshadow_2007_to_2017_GH}
\end{figure*}

In Figs.~\ref{fig:sunshadow_2007_to_2017_HGm} and \ref{fig:sunshadow_2007_to_2017_GH}, the Sun shadow can be seen for the HGm and GH model, respectively.

\end{document}